\documentclass[a4paper,fleqn,usenatbib]{mnras}

\usepackage{mathptmx}

\usepackage[T1]{fontenc}
\usepackage{ae,aecompl}

\usepackage{graphicx}	
\usepackage{amsmath}	
\usepackage{amssymb}	
\usepackage{booktabs}
\usepackage{longtable}
\usepackage{pdflscape}
\usepackage[draft]{todonotes}

\makeatletter
\def\input@path{{SelectionMethods/}{Colours/}{FluxComparison/}{Introduction/}{Discussion/}{TheData/}{Conclusion/}{AncillAndIdent/}{Clusters/}}
\makeatother
\graphicspath{{SelectionMethods/}{Colours/}{FluxComparison/}{Discussion/}{TheData/}{Conclusion/}{AncillAndIdent/}{Clusters/}}

\title[\textit{Planck} selected \textit{Herschel} proto-clusters]{Candidate high-z proto-clusters among the \textit{Planck} compact sources, as revealed by \textit{Herschel}-SPIRE}

\author[J. Greenslade et al.]{J. Greenslade$^{1}$\thanks{E-mail: J.Greenslade14@imperial.ac.uk},
D. L. Clements$^{1}$,
T. Cheng$^{1}$,
G. De Zotti$^{2, 3}$, 
D. Scott$^{4}$,\newauthor
E. Valiante$^{5}$,
S. Eales$^{5}$, 
M. N. Bremer$^{6}$,
H. Dannerbauer$^{7,8}$, 
M. Birkinshaw$^{6,9}$, \newauthor
D. Farrah$^{10}$, 
D. L. Harrison$^{11}$, 
M.~J.~Micha{\l}owski$^{12}$, 
I. Valtchanov$^{13}$,
I. Oteo$^{14}$, \newauthor 
M. Baes$^{15}$, 
A. Cooray$^{16}$ , 
M. Negrello$^{7}$, 
L. Wang$^{17, 18}$, 
P. van der Werf$^{19}$, \newauthor 
L. Dunne$^{7, 14}$, 
S. Dye$^{20}$ 
\\
$^{1}$Astrophysics Group, Imperial College, Blackett Laboratory, Prince Consort Road, London, UK \\
$^{2}$SISSA, Via Bonomea 265, 34136, Trieste, Italy\\
$^{3}$INAF-Osservatorio Astronomico di Padova, Vicolo dell'Osservatorio 5, I-35122 Padova, Italy\\
$^{4}$Department of Physics and Astronomy, University of British Columbia, Vancouver, BC V6T1Z1, Canada \\
$^{5}$School of Physics and Astronomy, Cardiff University, The Parade, Cardiff CF24 3AA, UK \\
$^{6}$HH Wills Physics Laboratory, University of Bristol, Tyndall Avenue, Bristol BS8 1TL, UK \\
$^{7}$Instituto de Astrof\'isica de Canarias (IAC), E-38205 La Laguna, Tenerife, Spain \\
$^{8}$Universidad de La Laguna, Dpto. Astrof\'isica, E-38206 La Laguna, Tenerife, Spain \\
$^{9}$Harvard-Smithsonian Center for Astrophysics, 60 Garden Street, Cambridge, MA 02138, USA \\
$^{10}$Department of Physics, Virginia Tech, Blacksburg, VA 24061, USA \\
$^{11}$Institute of Astronomy, University of Cambridge, Madingley Road,Cambridge CB3 0HA, UK \\
$^{12}$ Astronomical Observatory Institute, Faculty of Physics, Adam Mickiewicz University, ul.~S{\l}oneczna 36, 60-286 Pozna{\'n}, Poland \\
$^{13}$Herschel Science Centre, European Space Astronomy Centre, ESA, E-28691 Villanueva de la Ca ̃nada, Spain \\
$^{14}$Institute for Astronomy, University of Edinburgh, Royal Observatory, Blackford Hill, Edinburgh EH9 3HJ \\
$^{15}$Sterrenkundig Observatorium, Universiteit Gent, Krijgslaan 281 S9, B-9000 Gent, Belgium \\
$^{16}$Department of Physics and Astronomy, University of California, Irvine, CA 92697, USA \\
$^{17}$SRON Netherlands Institute for Space Research, Landleven 12, 9747 AD, Groningen, The Netherlands \\
$^{18}$Kapteyn Astronomical Institute, University of Groningen, Postbus 800, 9700 AV, Groningen, The Netherlands \\
$^{19}$Leiden Observatory, Leiden University, P.O. Box 9513, NL-2300 RA Leiden, The Netherlands \\
$^{20}$School of Physics and Astronomy, University of Nottingham, NG7 2RD, UK 
}

\date{Accepted XXX. Received YYY; in original form ZZZ}

\pubyear{2017}

\begin{document}
\label{firstpage}
\pagerange{\pageref{firstpage}--\pageref{lastpage}}
\maketitle

\begin{abstract}

By determining the nature of all the \textit{Planck} compact sources within $808.4$ deg$^2$ of large \textit{Herschel} surveys, we have identified 27 candidate proto-clusters of dusty star forming galaxies (DSFGs) that are at least 3$\sigma$ overdense in either 250, 350 or 500 $\mu$m sources.
We find roughly half of all the \textit{Planck} compact sources are resolved by \textit{Herschel} into multiple discrete objects, with the other half remaining unresolved by \textit{Herschel}.
We find a significant difference between versions of the \textit{Planck} catalogues, with earlier releases hosting a larger fraction of candidate proto-clusters and Galactic Cirrus than later releases, which we ascribe to a difference in the filters used in the creation of the three catalogues.
We find a surface density of DSFG candidate proto-clusters of ($3.3 \pm 0.7) \times10^{-2}$ sources deg$^{-2}$, in good agreement with previous similar studies. 
We find that a \textit{Planck} colour selection of $S_{857}/S_{545} < 2$ works well to select candidate proto-clusters, but can miss proto-clusters at $z < 2$. 
The \textit{Herschel} colours of individual candidate proto-cluster members indicate our candidate proto-clusters all likely all lie at z $>$ 1.
Our candidate proto-clusters are a factor of 5 times brighter at 353 GHz than expected from simulations, even in the most conservative estimates. Further observations are needed to confirm whether these candidate proto-clusters are physical clusters, multiple proto-clusters along the line of sight, or chance alignments of unassociated sources.

\end{abstract}




\section{Introduction}

The formation epoch of galaxy clusters remains a poorly constrained and understood component in galaxy formation and evolutionary theories. 
The masses and formation time of these structures in the early universe can not only place key constrains on cosmological theories and parameters \citep{Harrison2011}, but the elliptical galaxies in the cores of these massive clusters \citep{Kravatsov2012, Ma2015} are expected to go through an intense starburst phase at z $>$ 2, where a large portion of their stellar mass is rapidly built up over a timescale $< 1$ Gyr \citep{Eisenhardt2008, Hopkins2008, Petty2013, Granato2015}.
This starbursting phase  should be visible in the far-infrared (FIR) and sub-mm, where cool dust in the galaxies reemit absorbed UV photons.
At what point this takes place during the evolution of the cluster remains unknown, and the study and identification of clusters and proto-clusters at z $>$ 2 is important both for cosmology, and for understanding the evolutionary process within massive clusters and their members. 

However, few clusters or proto-clusters containing significant numbers of dusty star-bursting galaxies have been detected and confirmed at redshift z $> 2$ \citep{Daddi2009, Capak2011, Walter2012, Dannerbauer2014, Yuan2014, Casey2015}.
The rarity of proto-clusters, their large luminosity distance, and lack of an X-ray detectable intra-cluster medium or well formed red-sequence, makes traditional cluster selection techniques ineffective at selecting clusters in the earliest stage of their evolution.
The sub-mm, and to a lesser degree the FIR, also benefits from the negative k-correction, enabling reasonably easy identification of sources from redshift 2 to 8, at a fixed wavelength \citep{Blain2002, Casey2014}.

The dusty star forming galaxies (DSFGs), are thought to play a key role in the evolution of the massive ellipticals primarily seen today in the cores of local clusters \citep{Swinbank2006, Tacconi2008, Michalowski2009,  Stevens2010, Hickox2012, Casey2014, Toft2014, Simpson2014a, Dannerbauer2014, Ma2015, Wilkinson2016}.
The detection of a large number of physically associated DSFGs would be surprising, as the timescales on which they are expected to be sub-mm bright are only around 100 Myrs, so detecting several physically associated sources either implies some sort of large scale ($> 1$ Mpc) starburst triggering mechanism \citep[][]{Hung2016,  Oteo2017}, or that these sources are being externally re-fuelled, possibly by cosmic inflows \citep{Casey2016, Falgarone2017}.

However, several overdensities of sub-mm bright proto-clusters have already been discovered \citep[][]{Herranz2013, Ivison2013, Clements2014, Dannerbauer2014, Casey2016, MacKenzie2017,  Oteo2017, Oteo2017b}, some of which have spectroscopic redshifts and ALMA obsevations showing further sub-mm bright members \citep[][]{Ivison2013, Ivison2016,  Oteo2017}, implying that either a large scale triggering event ($>$10Mpc) ``activates'' the DSFGs simultaneously, or alternatively, that the duration of the starburst event is longer (0.5-0.7Gyrs) \citep{Granato2004, Lapi2011, Cai2013, Falgarone2017}.
Some evidence exists which suggests that the duty cycle of DSFGs in proto-clusters is indeed longer than those in the field \citep{Emonts2016, Dannerbauer2017}, with depletion timescales of several hundred Myrs.
Overall however, it is uncertain which of these scenarios is correct, and the discovery and study of further proto-clusters and their dusty components is needed, as measurements of the gas depletion timescale imply the former solution is correct, whereas the surface density of sub-mm bright proto-clusters implies the latter is correct.
Large field and all sky surveys in the sub-mm, such as \textit{Planck} \citep{Tauber2010, PlanckCollaboration2011} or \textit{Herschel} \citep{Pilbratt2010}, are ideal for selecting rare overdensities of DSFGs clustered together on the sky.

\citet{Negrello2005} studied the counts of extragalactic sources expected from low angular resolution surveys such as \textit{Planck}, and concluded that several luminous IR/sub-mm sources clustered on the scale of the instrument beam may appear as an unresolved or marginally resolved source.
The individual components that make up these sources could be chance projections along the line of sight, or physically associated.
Therefore many \textit{Planck} compact objects might resolve into high-$z$ clusters or proto-clusters of dusty sources when examined with a higher resolution instrument such as SPIRE \citep{Griffin2010} on the $\textit{Herschel}$ satellite.

\textit{Planck} has produced three catalogues of compact sources: The early release compact source catalogue \citep[ERCSC,][]{Ade2011}; The Planck catalogue of compact sources \citep[PCCS,][]{Ade2014}; and the second Planck catalogue of compact sources \citep[PCCS2,][]{PlanckCollaboration2015b}, based on 1.6, 2.6 and 5\footnote{For the highest frequency channels only.} full surveys of the sky.
In each catalogue, the compact \textit{Planck} sources were compiled into nine separate sub-catalogues, one for each \textit{Planck} channel, ranging from 30 to 857 GHz.
The beamsizes vary both between channel and between catalogues, but are generally around 4 to 5 arcminutes for the 217, 353, 545 and 857 GHz channels we use here, corresponding to 2 to 2.5 Mpc at $z=2$.
\textit{Herschel}-SPIRE's 350 $\mu$m band is matched to \textit{Planck}'s 857 GHz channel, while SPIRE's 500 $\mu$m channel has a similar wavelength and passband to \textit{Planck}'s 545 GHz band (500 $\mu$m against 550 $\mu$m).

\textit{Herschel} performed several wide surveys over large areas of the extragalactic sky. 
SPIRE operated at 250, 350 and 500 $\mu$m, with beam FWHMs of 17.9, 24.2, and 35.4 arcseconds respectively.\footnote{SPIRE Handbook,  Version 3.1, February 8, 2017, \url{http://herschel.esac.esa.int/Docs/SPIRE/spire_handbook.pdf} .}
The largest extragalactic surveys are the \textit{Herschel}-ATLAS \citep[H-ATLAS, 616 deg$^2$,][]{Eales2010}, the \textit{Herschel} multi-tiered extragalactic survey \citep[HerMES, including the HerMES large mode survey (HELMS), $\sim$ 370 deg$^2$, ][]{Oliver2010, Oliver2012} and the \textit{Herschel} Stripe 82 survey \citep[HerS, 79 deg$^2$, ][]{Viero2014}.
This provides over 1000 deg$^2$ of sky with approximate flux density 1$\sigma$ limits of 7-10 mJy in the three SPIRE bands.
By cross-matching the higher resolution \textit{Herschel} map with the catalogues of \textit{Planck} sources, the nature of all the \textit{Planck} compact sources in the \textit{Herschel} fields can be determined.

Several authors have already found plausible high redshift clusters using the \textit{Planck} data \citep{Herranz2013, Clements2014, Baes2014, PlanckCollaboration2015c, PlanckCollaboration2015d}, with a variety of approaches. 
Both \citet{Herranz2013} and \citet{Clements2014} performed similar cross-matches between \textit{Herschel} and \textit{Planck} in order to search for clusters of DSFGs.
\citet{Herranz2013} used 134 deg$^{2}$ of preliminary H-ATLAS Phase 1 data and the ERCSC and discovered a redshift 3.26 candidate cluster/proto-cluster of sub-mm sources surrounding the lensed source H12-00 \citep{Fu2012, Clements2016}.
\citet{Clements2014} meanwhile, cross-matched the ERCSC with the HerMES survey, and found evidence for four further candidate proto-clusters of DSFGs, with each candidate proto-cluster having total SFRs > 1000M$_{\odot}$yr$^{-1}$.

Here we set out to investigate and characterise the nature of all the \textit{Planck} compact sources that fall within any of the major \textit{Herschel} fields, using H-ATLAS, HerMES and HerS with the aim of searching for further rare cluster/proto-cluster candidates and potentially other rare and unexpected sources.
{In general, since we are unable to confirm whether our detected clusters / proto-clusters contain a well developed intracluster medium, and since they generally span scales on the order of arcminutes, we will refer to them as proto-clusters rather than clusters unless otherwise stated. 
This is prudent given our uncertainties about the evolutionary state of these systems, but we do allow for the possibility that some of our proto-clusters are actually physically evolved clusters.

The rest of this paper is organised as follows: In Section \ref{Sec:Data}, we describe the data sets used in this paper.
In Section \ref{Sec:SelectionMethods} we outline the methodology used to cross-match with \textit{Herschel}, and present the matches we found between \textit{Planck} and \textit{Herschel}.
In Section \ref{Sec:Photometry} we verify the photometry of our sources from the \textit{Planck} and \textit{Herschel} observations.
In Section \ref{Sec:Colours} we examine the colours of the \textit{Planck}-detected sources, and discuss the likely nature of the reddest sources discovered, whilst in Section \ref{Sec:Clusters} we further characterise our candidate proto-clusters.
In Sections \ref{Sec:Discussion} and \ref{Sec:Conclusion} we discuss the implications of our findings and summarise our results. 
Throughout this paper, we assume a standard cosmology, with $H_0 = 67.7$km s$^{-1}$ Mpc$^{-1}$, $\Omega_M = 0.3$ and $\Omega_{\Lambda} = 0.7$.

\section{Data Sets}

\label{Sec:Data}

In this section, we provide a brief overview of the construction of the three \textit{Planck} and 17 \text{Herschel} catalogues used in this paper, as well as the limits of each catalogue and any key differences between them.
A summary of the \textit{Herschel} field properties is given in Table \ref{Table:Surveysizes}, and a map of their location on the sky is given in Fig. \ref{Fig:AllSky}. 

\begin{figure*}
\includegraphics[width=\linewidth]{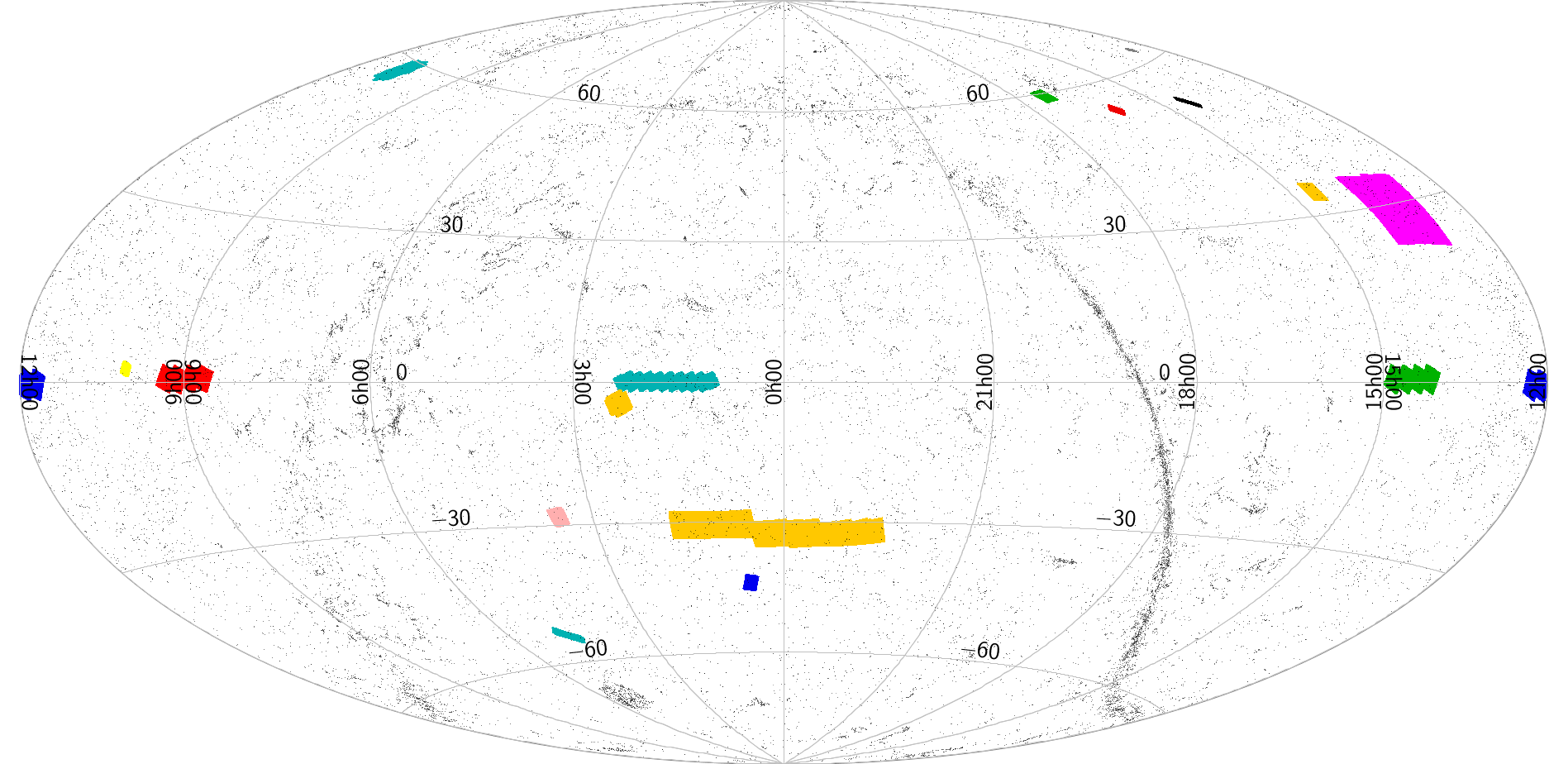}
\centering{\caption{All-sky map showing the 17 \textit{Herschel} fields under examination here (coloured polygons) and all 38,260 \textit{Planck} compact sources (black dots). Some of the major fields include the NGP (Magenta, Dec of 30$^{o}$), the SGP (Yellow, Dec of -30$^{o}$), HeRS (Turquoise, centre) and the three GAMA fields centred at a Dec of 0$^{o}$ and RA of 09$^{h}$ (Red), 12$^{h}$ (Blue) and 15$^{h}$ (Green). The Milky Way is indicated by the thick band of \textit{Planck} sources stretching across the sky.}\label{Fig:AllSky}}
\end{figure*}

\subsection{The \textit{Planck} compact catalogues of sources}
The ERCSC used $SExtractor$ \citep{Bertin1996} on the \textit{Planck} maps to identify sources in each band; this is based on extracting a number of connected bright pixels that are some threshold above a background measurement.
The PCCS and PCCS2, on the other hand, divided the maps into multiple patches, and convolved these patches with a second-order Mexican-hat wavelet that had been locally optimised to detect point sources \citep{Lopez-Caniego2006}.
Peaks $> 5\sigma$ in the resulting convolved map were then classified as detections \citep{Ade2014}.

We focus on the High Frequency Instrument's (HFI) 857 GHz (350$\mu$m) and 545 GHz (550$\mu$m) channels \citep{PlanckHFITeam2010}, since the peak of dust emission in galaxies (around 100$\mu$m) will be redshifted into these bands between $z = 1$ and $5$.
The quoted FWHM beam-size varies between catalogue releases, between 4.23 to 4.63 arcminutes in the 857 GHz band, and between 4.47 and 4.83 in the 545 GHz band due to improvements in calibration and beam information \citep{PlanckCollaboration2015b}.
The 90\% flux completeness level for the 857 GHz band is given as 680 and 790 mJy\footnote{This is higher than for the PCCS1 \citep[see section 3.2.3 of ][]{PlanckCollaboration2015b}.} at Galactic latitudes |b| $ > $ 30 for the PCCS and PCCS2 respectively.
The ERCSC does not provide a 90\% completeness level, but the faintest source detected at  |b| $ > $ 30$^o$ is 655 mJy, with the flux density of the faintest 10$\sigma$ source at  |b| $ > $ 30$^o$ being 813 mJy, demonstrating that the limits of the three catalogues at 857 GHz are all typically around 700 to 800 mJy. 
We use the aperture photometry flux density estimate in the \textit{Planck} catalogues, as it performs best when compared to \textit{Herschel} \citep[See table 12 of ][]{PlanckCollaboration2015b}, is likely to correctly capture emission from extended structures, and is available in all 3 catalogues.

\subsection{H-ATLAS}

H-ATLAS surveyed five fields: The Northern Galactic Pole (NGP, 170 deg$^2$), the Southern Galactic Pole (SGP 285 deg$^2$), and three smaller fields that lie along the equatorial plane at RAs of approximately 9, 12 and 15 hours, referred to as GAMA09, GAMA12 and GAMA15 (around 54 deg$^2$ each) which correspond to 3 of the fields surveyed by the Galaxy and Mass Assembly (GAMA) project \citep{Driver2011}.
Maps were produced with the \textit{Herschel} Interactive Pipeline Environment \citep[$HIPE$, ][]{Ott2010}, and the typical 1$\sigma$ total noise per \textit{Herschel} beam (confusion plus instrumental) in the final background-subtracted and filtered maps is 7.4, 9.4 and 10.2 mJy for the 250, 350 and 500 $\mu$m bands respectively \citep[][Maddox et al. in preparation, Smith et al. in preparation]{Valiante2016}.
Sources were extracted using the Multi-band Algorithm for source detection and extraction ($MADX$, Maddox, in preparation).

\subsection{HerMES}

HerMES field sizes varied from 0.4 deg$^2$ for GOODS-North, up to 280 deg$^2$ for the HELMS field.
The majority of the fields have 1$\sigma$ total noises of 6.2 - 6.8, 7.1 - 7.5 and 8.2 - 8.9 mJy for the 250, 350 and 500 $\mu$m bands respectively, with the exception of FLS, ADFS, ELAIS-N1, ELAIS-S1, BOOTES and XMM-LSS, which have 1$\sigma$ noise levels of 7.9, 8.2 and 10.1 mJy \citep{Nguyen2010}.
We exclude the HELMS field from further study, since no publicly released, formally verified catalogue of detected sources is yet available for cross-matching, and the field is strongly contaminated with Galactic cirrus. 
We do examine the \textit{Planck} compact sources present in HELMS using a private catalogue in section \ref{Sec:Discussion}, but do not include them in our final results.
The maps used in this paper were produced using the SPIRE-HerMES Iterative Mapper \citep{Levenson2010}, and the catalogues we used were the DR4 xID250 catalogues \citep{Wang2014}.

\subsection{HerS}

The \textit{Herschel} Stripe 82 Survey \citep[HerS, ][]{Viero2014}, is a 79 deg$^2$ survey taken along the SDSS Stripe 82 region with the SPIRE instrument on \textit{Herschel}.
Sources were extracted from the 250 $\mu$m map using \textit{STARFINDER} requiring S/N $>3$, after filtering the maps with a high pass filter to remove extended emission.
Flux estimates were then extracted from the 350 and 500 $\mu$m maps, using the 250 $\mu$m source positions as a prior.
The 1$\sigma$ median total noise is 7.1, 7.1 and 8.4 mJy for the 250, 350 and 500 $\mu$m bands respectively.

\begin{table}
\centering
\begin{tabular}{lllll}
\toprule
 & & \multicolumn{2}{c}{\textit{Planck} source count} \\
\toprule
Field &  Area [deg$^2$] & 857GHz & 545GHz \\
\midrule
NGP     &     170.0 & 82 & 21 \\
SGP     &     285.0 & 91 & 35  \\
GAMA09 &    53.4 & 26 & 13 \\
GAMA12    &         53.6 & 15 & 5 \\
GAMA15             &         54.6 & 16 & 13\\
ADFS             &    7.5 & 3 & 3 \\
BOOTES            &    11.3 & 11 & 2 \\
CDFS-SWIRE           &    10.9 & 5 & 1\\
COSMOS HerMES         &       4.4 & 2 & 1 \\
EGS HerMES          &     2.7 & 1 & 0 \\
ELAIS N1 SWIRE          &     12.3 & 6 & 2 \\
ELAIS S1 SWIRE       &    7.9 & 2 & 0 \\
FLS       &      6.7 & 5 & 3 \\
GOODS-North & 13.5 & 0 & 0 \\
LOCKMAN-SWIRE            &      16.1 & 6 & 3 \\
XMM-LSS-SWIRE            &    18.9 & 4 & 2 \\
HERS          &     79.0 & 38 & 14 \\
\midrule
\textbf{Total} & 808.4 & 313 & 118 \\
\bottomrule
\end{tabular}

\caption{The 17 \textit{Planck}/\textit{Herschel} Fields under consideration in this paper, their areas, and the number of unique \textit{Planck} sources detected within them across all three compact source catalogues. \label{Table:Surveysizes} }
\end{table}

\section{Selection Methods}
\label{Sec:SelectionMethods}


At 857 and 545 GHz, the \textit{Planck} beam physically corresponds to a size of a few hundred kpc at a redshift of 0.1, and around 2.5 Mpc at redshifts 1 to 3.
Therefore, most sources will not be resolved in the \textit{Planck} maps, since only local (z $\ll$ 1) extragalactic sources, extended cirrus, or galaxy clusters larger than 2.5 Mpc could have emission extended on larger angular scales.
By visually inspecting the \textit{Herschel} maps at the positions of the \textit{Planck} sources, the nature of the \textit{Planck} sources in these regions can be studied.

\begin{figure}
   \centering
   \includegraphics[width=\linewidth]{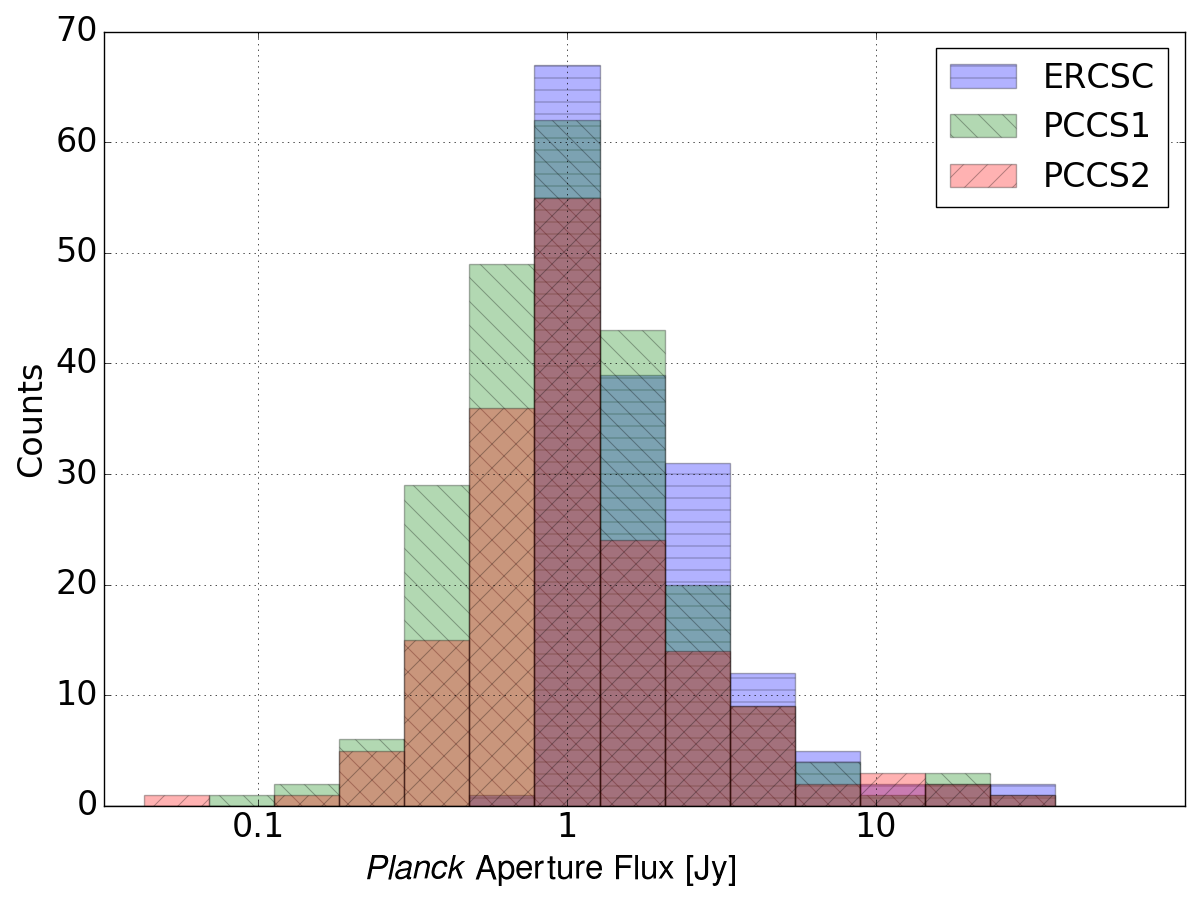}
   \caption{Aperture flux distribution of \textit{Planck} sources that lie in one of our \textit{Herschel} fields from the ERCSC (Blue) , PCCS1 (Green) and  PCCS2 (Red).} 
        \label{Fig:DistofFluxes}
\end{figure}

\subsection{Creation of the \textit{Planck}-\textit{Herschel} catalogue}

\label{Sec:Catalogue}

As different detection pipelines used in the creation of the ERCSC, PCCS and PCCS2 could be sensitive to different source populations, we include all three as part of our analysis.
We crossmatch each \textit{Planck} catalogue with the 17 catalogues of \textit{Herschel} sources.
We use a search radius equal to the \textit{Planck} FWHM at 857 GHz in the PCCS2, which used the most up to date calibration and beam information (4.63 arcminutes).
We varied this search radius between 4.00 and 5.00 arcminutes to check for consistency, as the \textit{Planck} beam FWHM varies not only with channel, but also with \textit{Planck} catalogue, typically between 4.2 and 4.8 arcminutes.
With the exception of some minor changes in the number of \textit{Herschel} sources detected in each \textit{Planck} source, our conclusions remained consistent. 
The \textit{Herschel} source density is high enough that there are always multiple \textit{Herschel} sources per \textit{Planck} beam, typically $> 10$.
For the 857 GHz - 350 $\mu$m match, there are 160 \textit{Planck} sources in the \textit{Herschel} fields from the ERCSC, 229 from the PCCS1 and 168 from the PCCS2.
The 545 GHz - 500 $\mu$m match finds 50 \textit{Planck} sources from the ERCSC, 99 from the PCCS1 and 60 from the PCCS2. 

In Fig. \ref{Fig:DistofFluxes}, we plot the aperture flux values for our \textit{Planck} sources.
While the PCCS and PCCS2 appear to be similar in terms of their flux distribution, the ERCSC distribution is skewed towards higher flux values. 
The ERCSC, using \textit{SExtractor}, requires isolated, bright, connected pixels in order to flag a detection, with the minimum flux found for the whole ERCSC being 655 mJy at 857 GHz.
The PCCS and PCCS2, on the other hand, require a single local peak in the \textit{Planck} map after convolution with the filter, and so can contain $>5\sigma$ sources with aperture fluxes as low as 69 mJy in our catalogue.
However, for bright sources detected in all three catalogues, the distributions should be similar, and above an aperture flux of $\sim$ 750 mJy, we find a much better match between the source flux densities.



Cross-matching the three versions of the catalogues together to find the total number of \textit{Planck} compact sources detected in at least one of the three catalogues, we find 313 \textit{Planck} sources in the \textit{Herschel} fields from the 857 GHz band and 118 unique \textit{Planck} sources from the 545 GHz band.
Combining these two catalogues together to search for unique objects, we find a total of 354 sources detected across 808.4 deg$^2$



We also create a catalogue of \textit{Herschel} sources that fall within 4.63 arcminutes of each \textit{Planck} source.
We created two uniform catalogues of \textit{Herschel} sources for the 857 GHz and 545 GHz data by selecting \textit{Herschel} sources using a minimum flux density limit of 25.4 mJy at 250, 350 or 500 $\mu$m (i.e. a source must be at least 25.4 mJy in one of the three SPIRE bands).
This is approximately 3 times the highest median total error seen in any of the \textit{Herschel} fields.
This results in 3,709 individual sources with S$_{350} >$ 25.4 mJy ($\gtrsim 3 \sigma$) that lie within 4.63 arcminutes of a \textit{Planck} 736 GHz source, and 693 \textit{Herschel} sources with S$_{500} >$ 25.4 mJy that lie within 4.63 arcminutes of a \textit{Planck} 545 GHz source.

Finally, we cross-matched our catalogue of unique compact \textit{Planck} objects with the \textit{Planck} Sunyaev-Zel'dovich Galaxy Cluster Catalogue \citep{PlanckCollaboration2015a}, the \textit{Planck} Galactic cold clump catalogue \citep{PlanckCollaboration2015e}, and the \textit{Planck} High Z catalogue \citep[PHZ, ][]{PlanckCollaboration2015c}.
We found no matches with the SZ catalogue, a single match with the Galactic cold cores catalogue, PLCKERC857 G339.76-85.56, and four matches in the PHZ, PCCS1 545 G160.59-56.75, PCCS1 545 G084.81+46.34, PLCKERC545 G007.56-64.14 and PCCS1 545 G012.89-66.24.

\subsection{The nature of the \textit{Planck} sources}

\label{Sec:Ancil}



\label{Sec:Cross}

We visually inspected each source in the \textit{Herschel} 350 $\mu$m maps at the position of the \textit{Planck} objects, to identify the nature of each \textit{Planck} source.
A summary of our results is presented in Table \ref{Table:Id}, a full table of identifications is available in Appendix \ref{Sec:Apen}, and images of the 324 that lie on the maps and away from the edge are available in Appendix \ref{Sec:Im}.

Most local ( z $\ll$ 1 ) galaxies can be identified by their bright, point source or extended emission in the \textit{Herschel} maps.
Cross-matching these with the NASA Extragalactic Database (NED) identifies 192 local galaxies, two QSOs and eight lens candidates that have known H-ATLAS identifications.
Four times, single bright sources with S$_{350} > 50$ mJy are found to have no optical or other known counterparts in NED or elsewhere.
These we assign as additional lens candidates, though these could also easily be examples of hyper luminous infrared galaxies, with L$_{fir}> 10^{13}L_\odot$, and are note necessarily lensed
Sources were also cross-matched with SIMBAD \citep{Wenger2000}, and three stars were identified this way.
Fourteen of the \textit{Planck} sources lie just outside the map coverage, and these are included in Table \ref{Table:Id} but not considered further.

For the remaining 131 sources, as well as examining the \textit{Herschel} maps, we examined the Improved Reprocessing of the IRAS Survey \citep[IRIS][]{MivilleDeschenes2005} maps at the positions of the \textit{Planck} sources to search for bright emission at 100 $\mu$m, which will be present for Galactic cirrus but not for proto-clusters of DSFGs at redshifts $\gtrsim 1$.
\textit{Planck} objects with structures in the 100 $\mu$m map were conservatively catagorised as Galactic cirrus, 43 in total.
This left 88 regions without an identification.

To search for proto-clusters amongst these 88, we counted the number of 250, 350 and 500 $\mu$m sources with fluxes $>$ 25.4 mJy that lie within 4.63 arcminutes of the \textit{Planck} position, with the flux limit chosen to compare to published number counts.
Assuming our sources are Poisson distributed, number counts from \citet{Clements2010} and \citet{Valiante2016} suggest that the expected number of 250, 350 and 500 $\mu$m sources are 16.5 $\pm $ 4.1, 9.1 $\pm$ 3.0 and 2.7 $\pm$ 1.7 per \textit{Planck} beam.
Any objects that show a 3$\sigma$ overdensity in any of the three SPIRE bands (at least 31, 19 or 9 sources\footnote{Using Poisson statistics} in the 250, 350 or 500 $\mu$m bands, respectively) are classed as candidate proto-clusters of galaxies, 27 in total are found in this way.
These over-densities are not necessarily physical associated proto-clusters, as they could also be line of sight effects of unrelated sources, multiple clusters / proto-clusters along the same line of sight \citep{Flores-Cacho2016, Negrello2017}, or they might be explained by differences in the actual distributions of the number of \textit{Herschel} sources in the tail of the distribution compared to Poisson.
The assumption of Poisson oversimplifies the complex distribution of galaxies, so in order to justify our assumption, we simulate 10,000 \textit{Planck} beams (circles of radius 4.63 arcminutes) at random positions on the NGP 350 $\mu$m map, and count the number of \textit{Herschel} sources with $S_{350} > 25.4$ mJy.
We then and compare this to our Poisson assumption that 19 or more sources indicates an overdensity.
Only 16 of 10,000 of the random positions contain at least 19 \textit{Herschel} sources with $S_{350} > 25.4$ mJy, with an average of $8.8 \pm 2.9$ per \textit{Planck} beam, in good agreement with the estimates from \citet{Valiante2016}.
Interpreting the 16 out of 10,000 as a probability, and converting this to an equivalent $\sigma$ value in the normal distribution, this corresponds to a $2.94\sigma$ overdensity, in excellent agreement with our choice of assuming these sources are Poisson distributed.
We find similar results for the 250 and 500 $\mu$m bands.
Therefore, these 27 \textit{Planck} sources are clearly overdense in \textit{Herschel} sources, and we assign them as candidate proto-clusters, though we retain the possibility that these are line of sight effects or multiple clusters / proto-clusters along the line of sight remains.
The remaining 61 sources in our maps remain unclassified, as we cannot reliably determine their nature.
We thus have a total of 340 unique \textit{Planck} compact sources across both the 857 GHz and 545 GHz bands, including local galaxies, galactic cirrus, proto-cluster candidates, lensed sources, stars, QSO's and sources we were unable to assign a classification.




\begin{table}
\centering
\caption{Identifications of all the \textit{Planck} objects that fall within one of the \textit{Herschel} survey fields under consideration here.}
\label{Table:Id}
\begin{tabular}{l|lll}
\toprule
Type & 857 GHz & 545 GHz & Unique \\
\midrule
Local galaxies & 187 & 54 & 192\\
Galactic cirrus & 37 & 18 & 43 \\
Proto-cluster candidates & 21 & 10 & 27\\
Lensed sources & 12 & 2 & 12 \\
Stars & 3 & 2 & 3\\
QSOs & 2 & 1 & 2\\
Off Map & 13 & 3 & 14 \\
No Assignment Given & 38 & 28 & 61 \\
\midrule
Total & 313 & 118 & 354 \\
\bottomrule
\end{tabular}
\end{table}

\subsection{Properties of our catalogues}

\subsubsection{Nature of the unassaigned sources}

We cannot reliably assign a catagory for several of our sources.
These could be false detections by \textit{Planck} or a series of fainter sources which we do not detect in \textit{Herschel}.
Significant differences at low S/N were seen from preliminary versions of the catalogues, which were created from preliminary versions of the \textit{Planck} maps. (D. Harrison, private communication).
Keeping the parameters for the \textit{Planck} catalogue creation the same, sources near the detection threshold would appear / disappear, depending on the preliminary version of the map used in the creation of the \textit{Planck} catalogue. 
For sources detected at a high S/N, this was very rare, whilst for sources near the detection thresholds, this was more common.

In our catalogue, for sources not assigned a counterpart, the median detection level in the PCCS and PCCS2 is 5.4 $\pm$ 0.5$\sigma$, near the detection threshold of 5$\sigma$ (for our proto-clusters, this is similar at 5.4 $\pm$ 0.3$\sigma$).
However, ten of the 65 are detected in multiple catalogues (six of these were detected in both the ERCSC and either the PCCS or PCCS2, thus using different detection methods).
This is unlikely if these ten sources are false detections.
Of these ten, five have colours that would be selected as a high redshift candidate by the Planck High Z collaboration (PHZ) in their analysis of candidate high-$z$ sources in \textit{Planck} \citep{PlanckCollaboration2015c}.
This could indicate an overdensity of red compact sources, too faint to be included in our analysis.
This conclusion was also hinted at when varying our search radius between 4.00 and 5.00 arcminutes; several of our unassigned sources became classified as candidate proto-clusters, and several candidate proto-clusters became unassigned.
In all cases we found roughly 30 candidate proto-clusters, with the exact number depending both on our choice of search radius, and flux density limit.
It is therefore likely that some of the unassigned sources are proto-clusters of DSFGs, but for the specific values we have chosen they do not pass our threshold test.

\begin{figure}
   \centering
   \includegraphics[width=\linewidth]{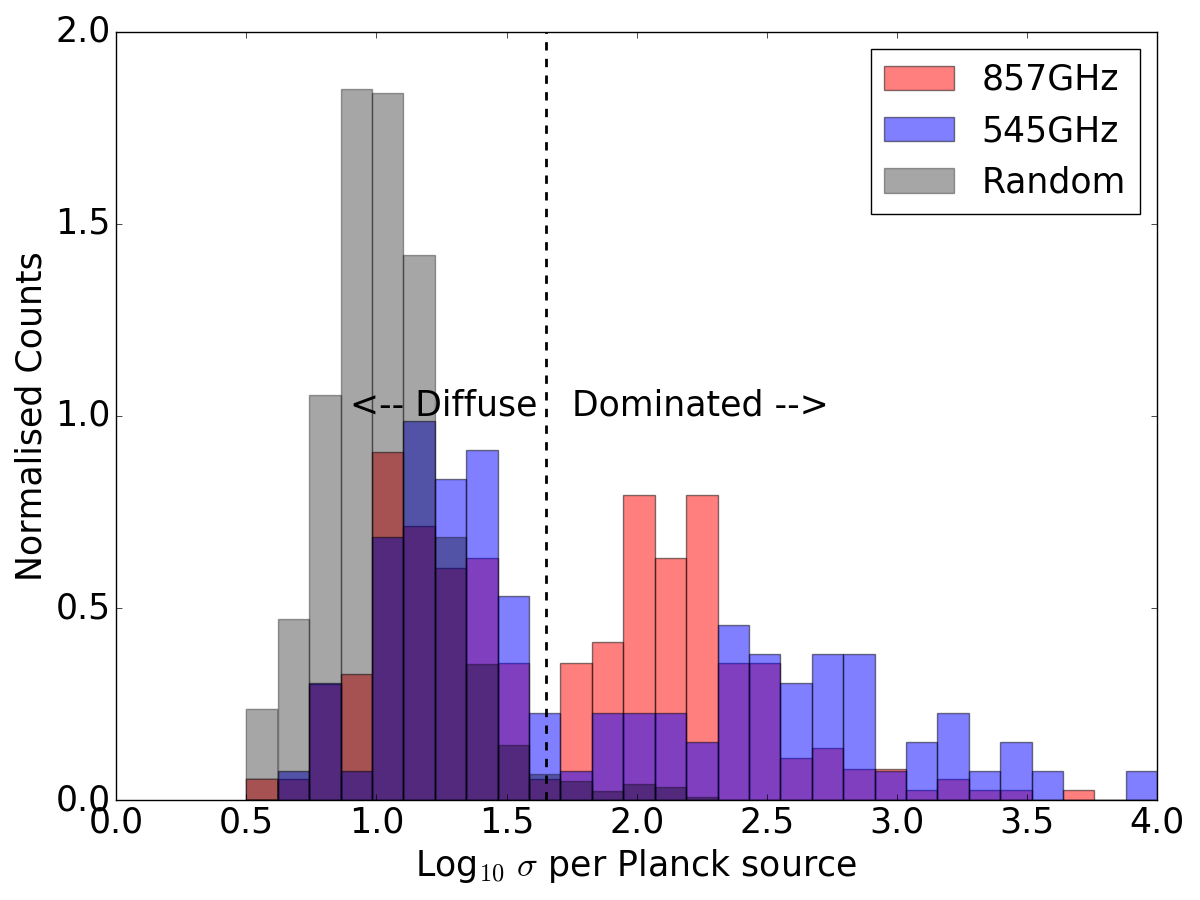}
   \caption{Log of the dispersion at 350 $\mu$m for the \textit{Herschel} sources contained within a \textit{Planck} object in the 857 (Red) and the 545 (Blue) GHz catalogues. The vertical black dashed line indicates the selected division between ``diffuse'' and ``dominated" sources. In grey is the result from taking 1,000 random positions in the NGP field, showing very few sources in the ``dominated'' region.} 
\label{Fig:CountsHerschel}
\end{figure}

\subsubsection{Diffuse and Dominated Sources}
\label{Sec:DifandDom}



Given we are searching for proto-clusters, we take all the \textit{Herschel} sources associated with a \textit{Planck} 857 GHz object, and calculate the standard deviation of their \textit{Herschel} S$_{350}$ flux densities, $\sigma_{350}$.
A large value of $\sigma_{350}$ is likely due to singular bright sources, whereas a small value indicates either of multiple distinct sources as in a proto-cluster, or simply extended Galactic cirrus.
We do the same for the 545 GHz \textit{Planck} sources and the \textit{Herschel} S$_{500}$ flux densities, $\sigma_{500}$.
We show these in Fig. \ref{Fig:CountsHerschel} for all 342 \textit{Planck} objects, as well as the result when taking 1000 random positions, and calculating the Log $\sigma_{350}$ in each case as a comparison.
Any source with fewer than two \textit{Herschel} sources is not included in our analysis.
There are 28 sources with 2,3 or 4 associated 350 $\mu$m detections, so the vast majority have reasonable samples from which to calculate $\sigma_{350}$.
The distribution appears bi-modal, with two distinct regions below and above Log$_{10}$($\sigma) \approx $ 1.65.
This bi-modality is not seen when examining 1,000 random positions. 
We designate these two regions as ``diffuse'' (Log$_{10}$($\sigma) < $ 1.65) and ``dominated''  (Log$_{10}$($\sigma) > $ 1.65), indicating that flux from these sources appears to be from extended diffuse / multiple faint source emission or dominated by a single source respectively.

For the 857 GHz \textit{Planck} sources, of the 299 sources not near the edge and with more than 1 associated \textit{Herschel} sources, 155 sources are identified as ``Dominated'' and 144 identified as "Diffuse".
In the 545 GHz catalogue, of the 109 sources not near the edge and with more than 1 associated \textit{Herschel} sources, 44 are ``Dominated'' and 65 are ``Diffuse''.
Overall this resulted in 159 unique ``dominated'' sources and 186 unique ``diffuse'' sources, with 9 sources having only 1 counterpart or lying near the edge of the \textit{Herschel} map.
All the cirrus sources, all the proto-cluster candidates and all but one of the not assigned sources are identified as being ``diffuse''. 
The other 156 ``dominated'' sources are all identified with local galaxies, lensed candidates, the QSO or stars. 
Of the 186 total diffuse sources, 41 are associated with local galaxies, usually because of extended emission or several bright neighbours.
We also find that four of the lens sources are diffuse, though they lie on the border between diffuse and dominated.

\begin{figure}
   \centering
   \includegraphics[width=\linewidth]{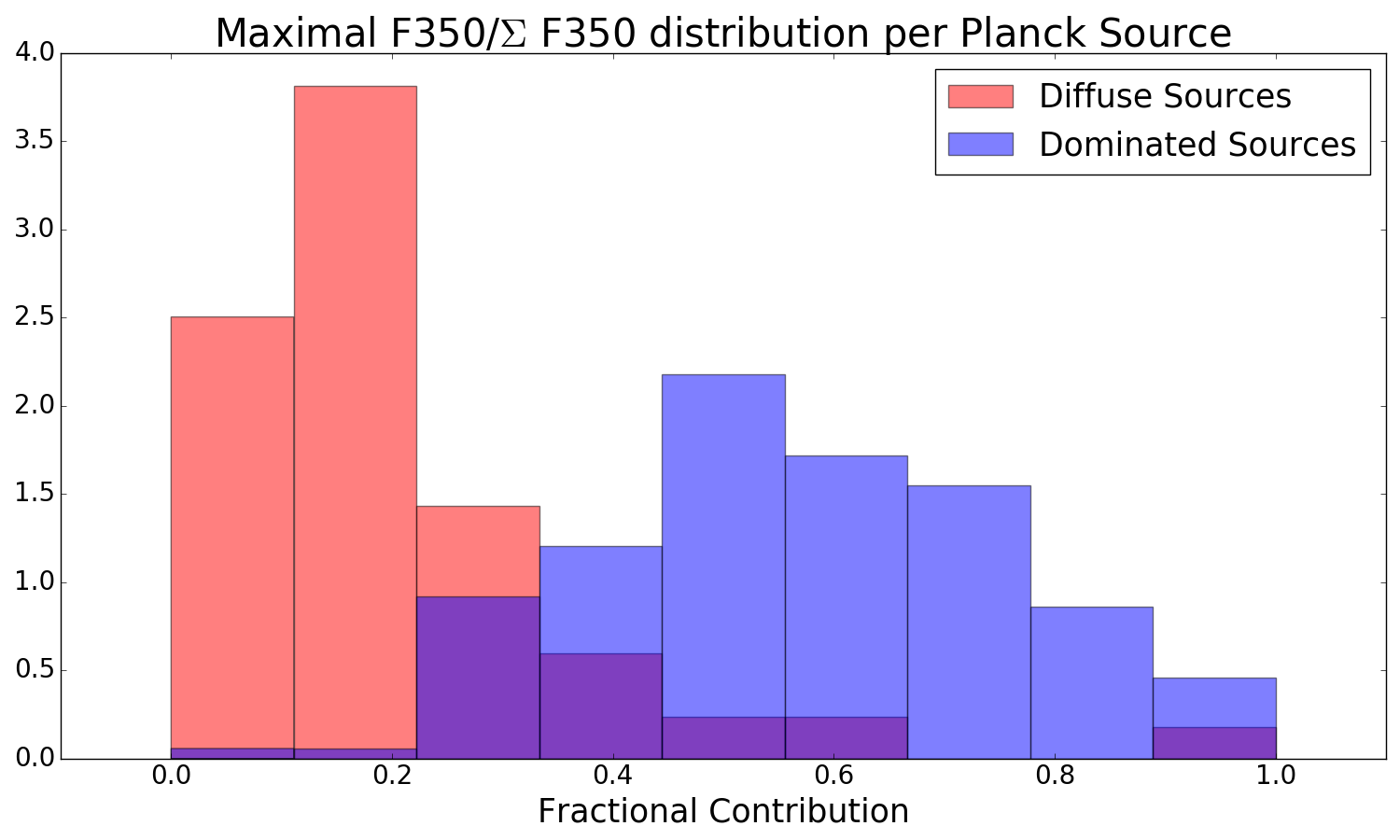}
   \caption{Fractional contribution of the brightest \textit{Herschel} source in each \textit{Planck} source to the total \textit{Herschel} 350 $\mu$m flux density from all the sources associated with each \textit{Planck} object. ``diffuse'' sources (red) and "dominated" sources (blue) are plotted separately.} 
\label{Fig:ContribHerschel}
\end{figure}

In Figure \ref{Fig:ContribHerschel} we plot the distribution of the fractional contribution from the brightest 350 $\mu$m \textit{Herschel} source to each \textit{Planck} 857 GHz source, divided by whether a \textit{Planck} source is ``diffuse'' or ``dominated''. 
This independently shows that our intuitive explanation for the division seen in the $\sigma_{350}$ seems to be the correct one; ``dominated'' objects tend to have one bright source dominating the flux whereas the ``diffuse'' objects individually have a relatively low contribution to the total flux.
A similar relationship is seen in the 545 GHz data.

The clear divide in both Fig. \ref{Fig:CountsHerschel}  and Fig. \ref{Fig:ContribHerschel} indicate that only around 60\% of the \textit{Planck} compact sources are actually compact on scales reasonably smaller than the \textit{Planck} beam.
Both figures also show that the \textit{Planck} maps are well suited for detecting extended emission from sources such as proto-clusters of DSFGs.

\subsubsection{Variations between the ERCSC, PCCS and PCCS2}
The key difference between the \textit{Planck} compact source catalogues is the use of $SExtractor$ for the ERCSC and a Mexican-hat wavelet for the detection pipeline in the PCCS and PCCS2.
This latter approach was designed to suppress emission on large scales, in order to reduce cirrus contamination in the catalogues, and simulations of its effectiveness were run on point sources \citep{Lopez-Caniego2006}.
However, its effect on extended, non-cirrus sources is unclear.

\begin{table}
\centering
\caption{Fractional make up of the three \textit{Planck} catalogues of compact sources at 857 GHz}
\begin{tabular}{l|lll}
\toprule
Source Type & ERCSC [\%] & PCCS1[\%] & PCCS2[\%] \\
\midrule
Local Galaxies & 56.0 & 61.0 & 80.0 \\
Galactic Cirrus & 16.7 & 8.8 & 5.0\\
Cluster Candidates & 9.5 & 4.6 & 1.1 \\
No Assignment Given & 11.9 & 16.1 & 5.6\\
Lenses & 1.2 & 3.8 & 3.3 \\
QSO & 0.0 & 0.8 & 0.5 \\
Stars & 0.6 & 1.1 & 1.7 \\
\bottomrule
\label{Table:Frac}
\end{tabular}
\end{table}

In Table \ref{Table:Frac}, we provide the fractional composition of each 857 GHz catalogue.
Though from the ERCSC to the PCCS2, the cirrus contamination of the catalogues has reduced from 16.7\% to 5.0\%, the fraction of proto-cluster candidates has been also reduced from 9.5\%  to 1.1\%. 
Put another way, the fraction of ``diffuse'' sources has decreased from $\sim$ 47\% in the ERCSC to 28\% in the PCCS2.
Though these proto-cluster candidates may not be real, may be line of sight effects, or potentially cirrus contamination, recent work has shown that several of these candidates are consistent with their being clusters in formation at z $\sim2$ \citep[][Cheng et al. in preparation]{Herranz2013, Clements2014, Clements2016}.
The inclusion of the Mexican-hat wavelet for source detection potentially suppresses the detection of these proto-cluster candidates, as the Bootes, EGS, Lockman and CDFS proto-cluster candidates revealed by \citet{Clements2014} do not appear in the PCCS1 or PCCS2.

\section{Photometry}

\label{Sec:Photometry}

Having identified our 27 proto-cluster candidates, alongside numerous other source types, we now examine the photometry associated with these sources.
\textit{Planck} have previously compared their photometry against \textit{Herschel} in order to verify that the two photometry measurements agree \citep{Bertincourt2016}.
In this section, we extend this analysis to checking whether summing our selected 350 $\mu$m \textit{Herschel} sources (i.e. S$_{350} > 25.4$ mJy) alone can adequately match the \textit{Planck} flux densities seen in all the \textit{Planck} compact sources.

As the band passes are well matched, a direct comparison between the 857 GHz \textit{Planck} band and the 350 $\mu$m SPIRE band can be performed with few assumptions. 
Here, we follow the same procedure set out in appendix A.1. of the PCCS1 for estimating the aperture photometry, but use the \textit{Herschel} maps instead of the \textit{Planck} maps.
We took the background subtracted maps of all of the \textit{Herschel} fields, and integrated the SPIRE 350 $\mu$m flux density over a \textit{Planck} 857 GHz beam by summing all the pixels that fell within 1 FWHM of the nominal \textit{Planck} source position.
The assumed FWHM was 4.63 arcminutes.
Once again, this was varied between 4.0 and 5.0 arcminutes to check for consistency in the results, finding similar results. 
A background annulus of inner radius 1 $\times$ FWHM and outer radius of 2 $\times$ FWHM was used to estimate the median background value and this was removed from the aperture flux estimate.
Any sources that fell on the edge of the map or contained null pixels within the primary or background aperture had a flux density assigned to them of zero to prevent edge effects contaminating our sample.
Errors were estimated from a combination of SPIRE instrumental noise, SPIRE calibration error, and a constant confusion noise conservatively estimated at 7 mJy per SPIRE-beam, all added in quadrature.
The results of this analysis, for both diffuse and dominated sources, are shown in Figure \ref{Fig:Photometry}.




\begin{figure}

\centering
\includegraphics[width=\linewidth]{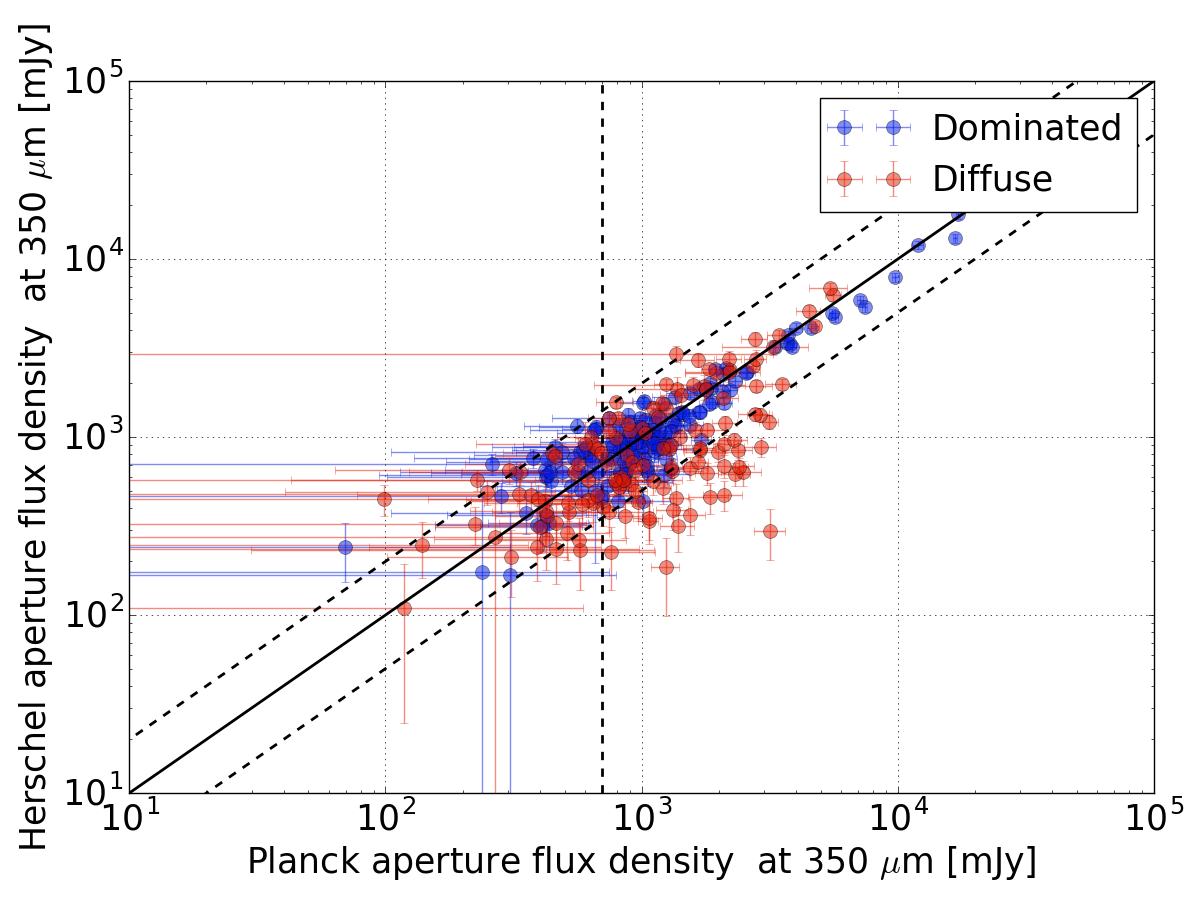}
\caption{Comparison between the \textit{Planck} aperture flux density and the \textit{Herschel} aperture flux density, as calculated in the text. The red points are the those sources considered to be diffuse, and the blue those considered dominated by a single source. The solid black line shows the 1:1 ratio. The diagonal dashed lines show the limits where the \textit{Herschel} flux is half/double that of the \textit{Planck} flux, and the vertical dashed line shows the PCCS 90\% completeness limit.} 
\label{Fig:Photometry}

\end{figure}

We then use the absolute relative flux density difference, defined as
\begin{equation}
\label{Eq:ARF}
\eta = | 100 \times \frac{S_{SPIRE} - S_{Planck}}{S_{SPIRE}} |,
\end{equation}
and use the weighted average of the \textit{Planck} and \textit{Herschel} aperture photometry, finding an absolute relative flux density difference between \textit{Planck} and \textit{Herschel} of only 4.9\%, comparable to the 1 to 5\% uncertainty found in \citet{Bertincourt2016}.

The absolute relative flux density difference is, however, not the same for the dominated (1.8\%) and diffuse (11.4\%) sources.
Given we are using background subtracted maps in each case, we repeat our analysis using the raw H-ATLAS maps that are publically avaliable.
These \textit{Herschel} maps have not had any background subtraction applied to them, and therefore could contain the flux that appears to be missing in several of our diffuse sources for \textit{Herschel}.
We found that absolute relative flux density difference for our dominated and diffuse sources changed to 4.8\% and 3.8\% respectively when we used the raw maps, both well within the \textit{Planck} calibration uncertainty. 
This indicates that the missing flux from our sources, especially diffuse sources, is being removed during the background removal process on the \textit{Herschel} maps.

The \textit{Planck} and \textit{Herschel} aperture photometry are generally in agreement for \textit{Planck} objects dominated by a single \textit{Herschel} source. 
Given roughly 40\% of all \textit{Planck} compact objects are expected to be diffuse in nature when examined at \textit{Herschel} resolutions, we consider whether the detected sources alone can account for the total \textit{Planck} flux, or whether an extended diffuse emission component is required. 

In Fig. \ref{Fig:Photometry2}, we plot the \textit{Planck} aperture flux densities and the summed 350 $\mu$m fluxes from the detected \textit{Herschel} sources, coloured by their source classification type. 
We find a 5\% absolute relative flux density difference between the summed fluxes and the aperture flux for non cirrus sources, but an 77\% relative flux difference for sources we have identified with galactic cirrus.
Several local galaxies, with emission extended well beyond the scale of the \textit{Herschel} beam, are poorly fit in the \textit{Herschel} catalogues and therefore have a smaller summed-\textit{Herschel} flux compared to the \textit{Planck} flux.

\begin{figure*}
\centering
\includegraphics[width=\linewidth]{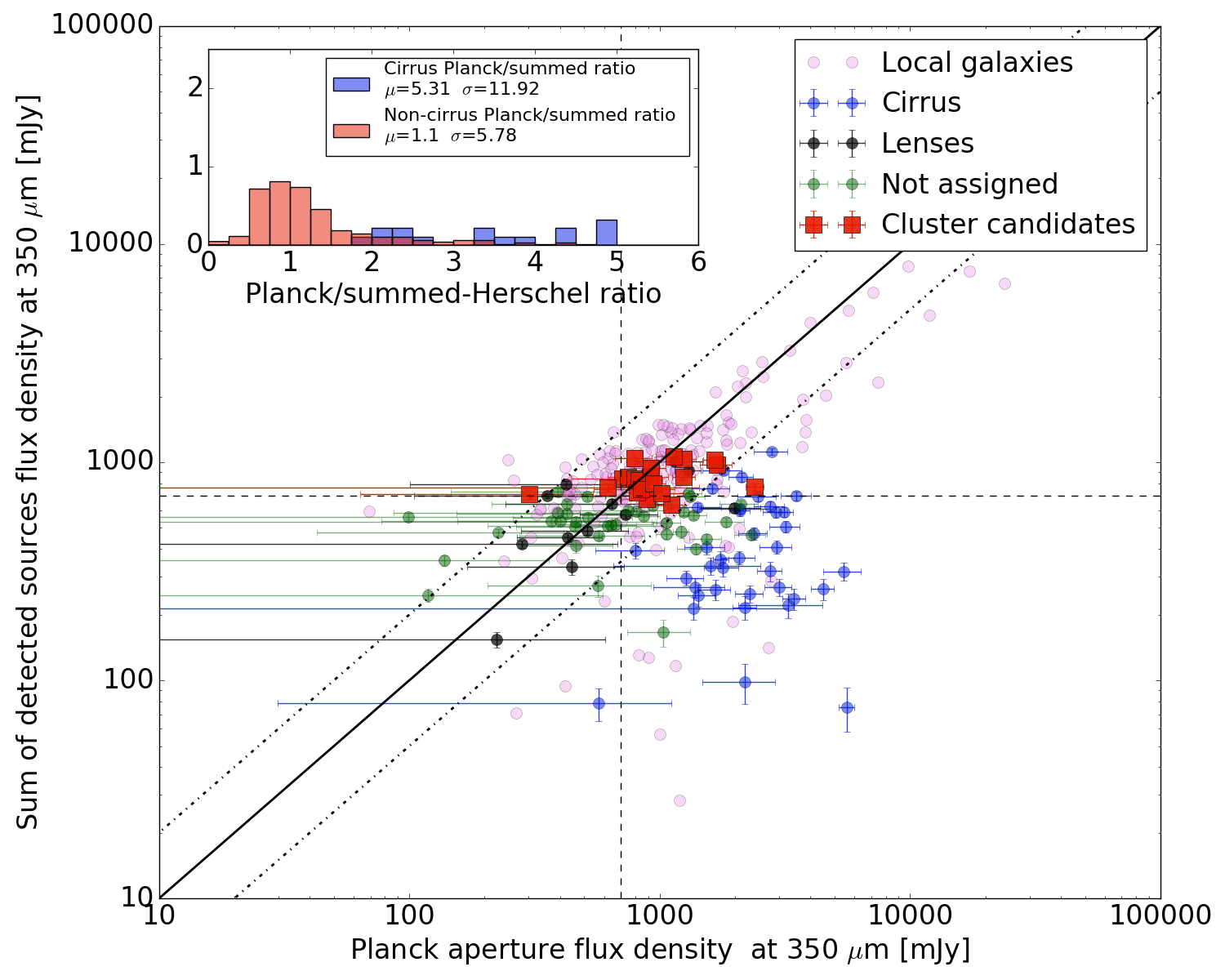}
\caption{Comparison between the \textit{Planck} aperture flux density and summing up the 350 $\mu$m flux density from the detected \textit{Herschel} sources. The light pink points are local galaxies, the blue are cirrus, the red are proto-cluster candidates, the black are lens candidates and the green are those points not assigned an identification. The solid black line shows the 1:1 ratio, whereas the dot-dashed lines show where the \textit{Planck} aperture flux is double or half the summed detected sources. The vertical and horizontal dashed lines show the nominal \textit{Planck} 90\% completeness levels from the PCCS1.  Error-bars are not shown for the local galaxies to aid in clarity, but are comparable to other sources at all fluxes. The histogram in the top left corner shows the \textit{Herschel} \textit{Planck} ratio, with cirrus sources indicated in blue, and non-cirrus sources indicated in red, as well as the mean and standard deviation. The histogram has been truncated to a maximum ratio of 6 for clarity, with 19 cirrus sources with ratios beyond this.} 
\label{Fig:Photometry2}
\end{figure*}

When summing up detected \textit{Herschel} sources, proto-cluster candidates are well matched to \textit{Planck} but Galactic cirrus sources are not, suggesting that our selection of Cirrus sources in Section \ref{Sec:SelectionMethods} was successful. 
This also implies that estimates of the physical properties of these proto-cluster candidates can be derived from the \textit{Planck} flux density alone, as it represents the summed total of the individual sources that make up the proto-cluster and no diffuse emission is needed to account for the \textit{Planck} flux.

Fig. \ref{Fig:Photometry2} also shows that the proto-cluster candidates mostly lie near to the \textit{Planck} detection limits, with a median \textit{Planck} aperture flux of 886 mJy.
Only eight of our 21 candidate proto-clusters detected at 857 GHz have an 857 GHz flux density $> 1$ Jy. 
For the unassigned objects, 14 of 63 have \textit{Planck} 857 GHz flux densities $> 1$ Jy, and these brighter sources we often find are not well matched between \textit{Planck} and \textit{Herschel}; Only two of these unassigned sources have \textit{Herschel} aperture flux densities $> 1$ Jy, and none of the unassigned sources have a 857 GHz flux density $> 1$ Jy when summing detected \textit{Herschel} sources.
Given also that Fig. \ref{Fig:DistofFluxes} indicates the ERCSC, which appears to be best at detecting these proto-cluster candidates, is limited to sources with flux density $> 750$ mJy, it is possible that the candidates we are selecting here are the bright tail of the DSFG proto-cluster population, and there could be many more proto-clusters that lie below this limit.

\section{Colours}

\label{Sec:Colours}
With only a maximum of 3 photometric points from SPIRE available, any photometric redshift attempt will have large uncertainties ($\Delta z = \pm 1$) associated with it.
However, the sub-mm colours of \textit{Herschel} sources have often been used as a proxy to give a useful indication of their redshifts \citep{Clements2014, Dowell2014, Dannerbauer2014, Asboth2016, Rowan-Robinson2016, Ivison2016}.
Therefore, in this section we set out to examine the \textit{Planck} colours of our sources, and compare them to the selection used by the PHZ in their search for high-$z$ sources, as well as using the \textit{Herschel} colours to give an indication of the likely redshifts of our \textit{Planck} sources.
We leave a more accurate determination of the redshift to a future paper that contains additional follow up observations (Cheng et al. in preparation).

\subsection{\textit{Planck} Colours}


\begin{figure*}
   \centering
  \includegraphics[width=\linewidth]{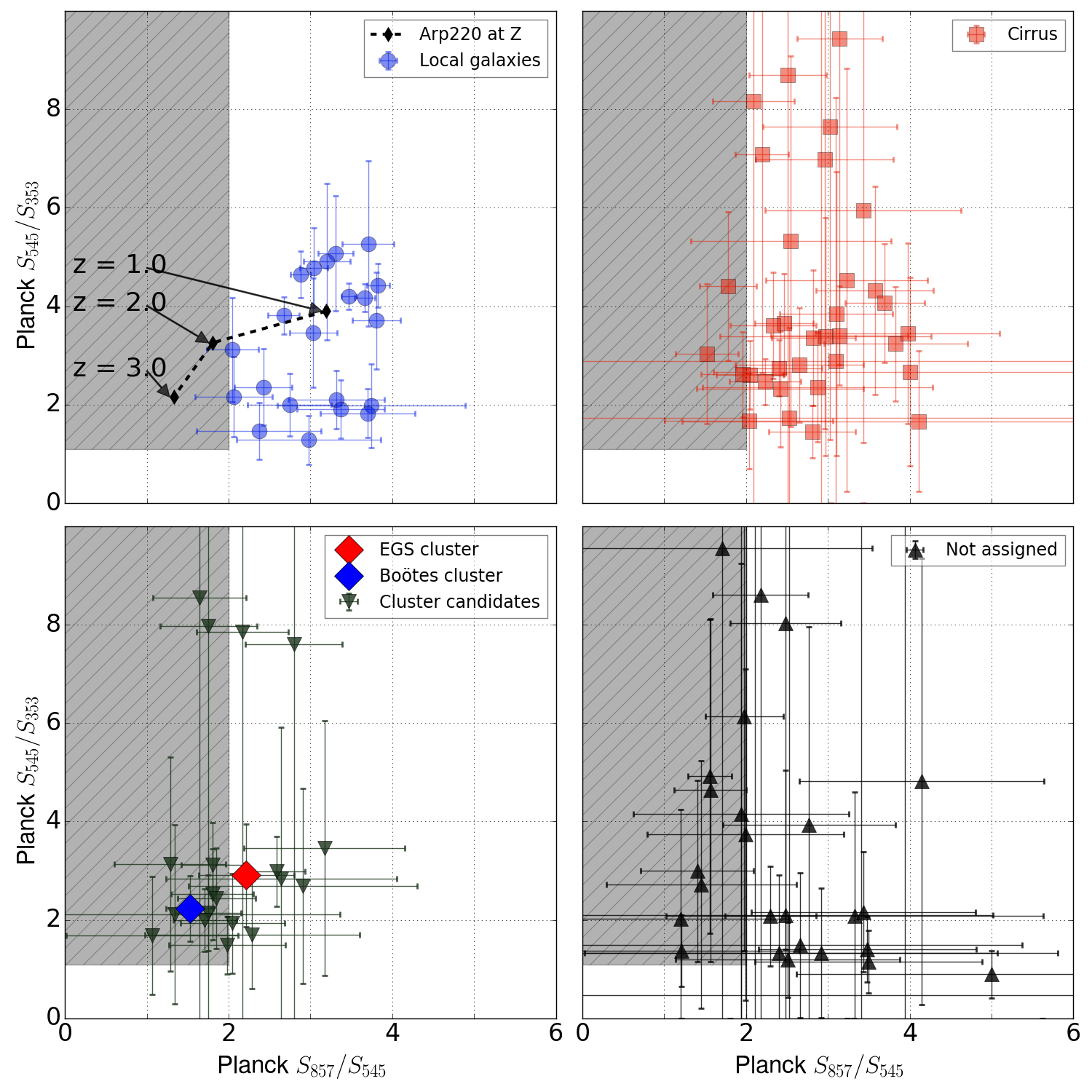}
   \centering{\caption{\textit{Planck} 857/545 GHz and 545/353 GHz colours for the categories of source we identify as local galaxies (top left), cirrus sources (top right), cluster candidates (bottom left) and unassigned sources (bottom right) . The grey shaded region represents the selection criteria used in \citet{PlanckCollaboration2015c} for their selection of high redshift source candidates. The black line in the top left plot shows the \textit{Planck} colours of Arp 220 as it would appear at z$=$1,2 and 3, and the blue and red diamonds in the proto-cluster candidates plot show, respectively, the Bootes and EGS proto-cluster candidates identified in \citet{Clements2014}. \label{Fig:PlanckColours}}}
\end{figure*}

In Figure \ref{Fig:PlanckColours}, we plot the \textit{Planck} 857/545 GHz (350/550 $\mu$m) and 545/353 GHz (550/850 $\mu$m) colours for the major populations identified in Section \ref{Sec:Ancil}.
We only plot sources from the 857 GHz selected catalogue, since it is the only catalogue which additionally provides aperture flux estimates at 545 and 353 GHz at the position of the \textit{Planck} source.
\citet{PlanckCollaboration2015c}, in their selection of high-z candidates from the \textit{Planck} maps, used a criterion with \textit{Planck} colours of 857/545 GHz $<$ 2 and 545/353 GHz $>$ 1 to search for candidate high redshift galaxies/clusters of galaxies.
We mark their selection area as the gray hashed region.
For clarity the local galaxies that are detected at 3$\sigma$ in all three of the the 857, 545 and 353 GHz bands are plotted.
We also plot two of the proto-clusters detected by \citet{Clements2014} in the Bootes and EGS fields  to demonstrate their colours (both of which are also detected in our analysis).

We note that many of our proto-cluster candidates fall outside the \textit{Planck} selection region. 
For our identified candidate proto-clusters, 21 are included in the 857GHz \textit{Planck} catalogue, and so are considered here.
Of these 21, only twelve lie within the \textit{Planck} selection region, with a mean $S_{857}/S_{545}$ ratio of $2.0 \pm 0.5$.
As the only constraint we impose upon our sources is that they are detected as a \textit{Planck} compact source, and lie in one of the major \textit{Herschel} fields, we could be selecting a population of lower redshift or warmer clusters / proto-clusters than found by \citet{PlanckCollaboration2015c}.

Local galaxies and cirrus have mean 857/545 colours of $3.0 \pm 1.0$ and $2.8 \pm 0.7$ respectively, whereas the unassigned sources have a colour of $2.5 \pm 1.0$.
For the unassigned sources, nine of the 35 have colours that would have been selected in \citet{PlanckCollaboration2015c} as potentially high redshift.
It is therefore not unreasonable to suggest that unassigned sources with both red colours and a large, but not overdense, number of \textit{Herschel} sources could also be high-redshift proto-clusters of \textit{Herschel} sources.
Our lens candidates have a median 857/545 GHz colour of of 1.8 $\pm$ 0.5, and our QSO has 857/545 GHz colour of 0.8 $\pm$ 0.4 at a redshift of 2.099.
The three stars have a mean  857/545 GHz colour of 3.0 $\pm$ 0.4. 
As expected, the stars, local galaxies and cirrus all have 857/545 colours that indicate that they are at redshifts $\ll 1$, whereas the redshift 2.099 QSO, lens candidates and our proto-cluster candidates have colours that indicate they lie at redshifts $> 1$.

The total colour from a candidate proto-cluster will be a combination of foreground/background sources and sources associated with the proto- cluster.
This is especially important, considering that overdensities of \textit{Herschel} sources have been argued to be due to line of sight effects from multiple clusters, both theoretically \citep{Negrello2017} and observationally \citep{Flores-Cacho2016}.
In order to assess the contribution from foreground sources to the colour of a \textit{Planck} source, we simulated the \textit{Planck} colours of a region of sky containing a proto-cluster.
Our simulated proto-clusters have, on average, 11 members which would be selected by our flux cutoff, and we include contribution from sources not associated with the proto-cluster by adding in, on average, 9 sources which would be selected by our flux cutoff randomly distributed between redshifts 1 and 3.
The total number of detected sources in then around 20, which is just high enough to be selected as a candidate proto-cluster for our sample.
For all sources, we drew samples from a single temperature modified blackbody function
\begin{equation}
\label{Eq:BB}
S_\nu \propto \nu^{\beta} B_\nu(T),
\end{equation}
where $\nu^{\beta}$ modifies the emissivity function of the dust and $B_\nu(T)$ is the Planck function at temperature $T$.
The temperature was fixed at 29 K and $\beta$ was fixed at 2, so that the background sources have an average $S_{857}/S_{545}$ flux density ratio that matches that seen in the \textit{Herschel} maps, in this case $S_{857}/S_{545} = 1.87$.
The fluxes of each source are drawn from an exponential distribution, which roughly matches the distribution of fluxes we see in our catalogues of 350 $\mu$m detected \textit{Herschel} sources, and our 350 $\mu$m flux is then normalised to this value.
We simulate 4 proto-clusters in total, at redshifts 1, 2, 3 and 4, and for each redshift we draw 100proto- clusters using the method described above.
We determined the total colour by summing the total 857 GHz flux density and dividing by the total 545 GHz flux density from all sources.
The results of this are shown in Fig. \ref{Fig:SimClusterPlanck}.

We find that when there are few proto-cluster sources compared to background/foreground sources, the colours tend to the average colours of the foreground/background sources, as expected, and in this case with an average of $S_{857}/S_{545} = 1.87$.
Once there are roughly equal number of proto-cluster sources and background/foreground sources however, the proto-cluster tends to dominate the colour of the source.
However, that colour is dominated by the redshift of the source, with proto-clusters at redshift 3 and 4 having a lower $S_{857}/S_{545}$ flux density ratio, proto-clusters at redshift 1 having a higher $S_{857}/S_{545}$ flux density ratio.
If a proto-cluster is roughly at the same redshift as the average redshift of the background/foreground sources, then there is no obvious difference in its colour compared to a patch of sky where there is no proto-cluster.  
This provides a simple explanation for the ``warm'' proto-cluster candidates, that they are lower redshift clusters / proto-clusters compared to the likely high-z clusters detected in the PHZ \citep{PlanckCollaboration2015c}.
However, we note in particular that our results are very sensitive to the assumption that all our galaxies are the same temperature; even if we allow the temperature to vary by $\pm 5$ K, the standard deviation in the $S_{857}/S_{545}$ flux density ratios of proto-clusters can double from 0.1 to 0.2 for a proto-cluster at redshift 2.
This further suggests that there can be significant boosting both into and out of the selection region used by the PHZ, though the general trend remains that higher redshift proto-clusters tend to have lower $S_{857}/S_{545}$ flux density ratios.
The major benefit used in this paper compared to the PHZ is that we do not make any colour selection, and are therefore sensitive to clusters / proto-clusters at all redshifts where we would detect them by our flux cut.

\begin{figure}
   \centering
  \includegraphics[width=\linewidth]{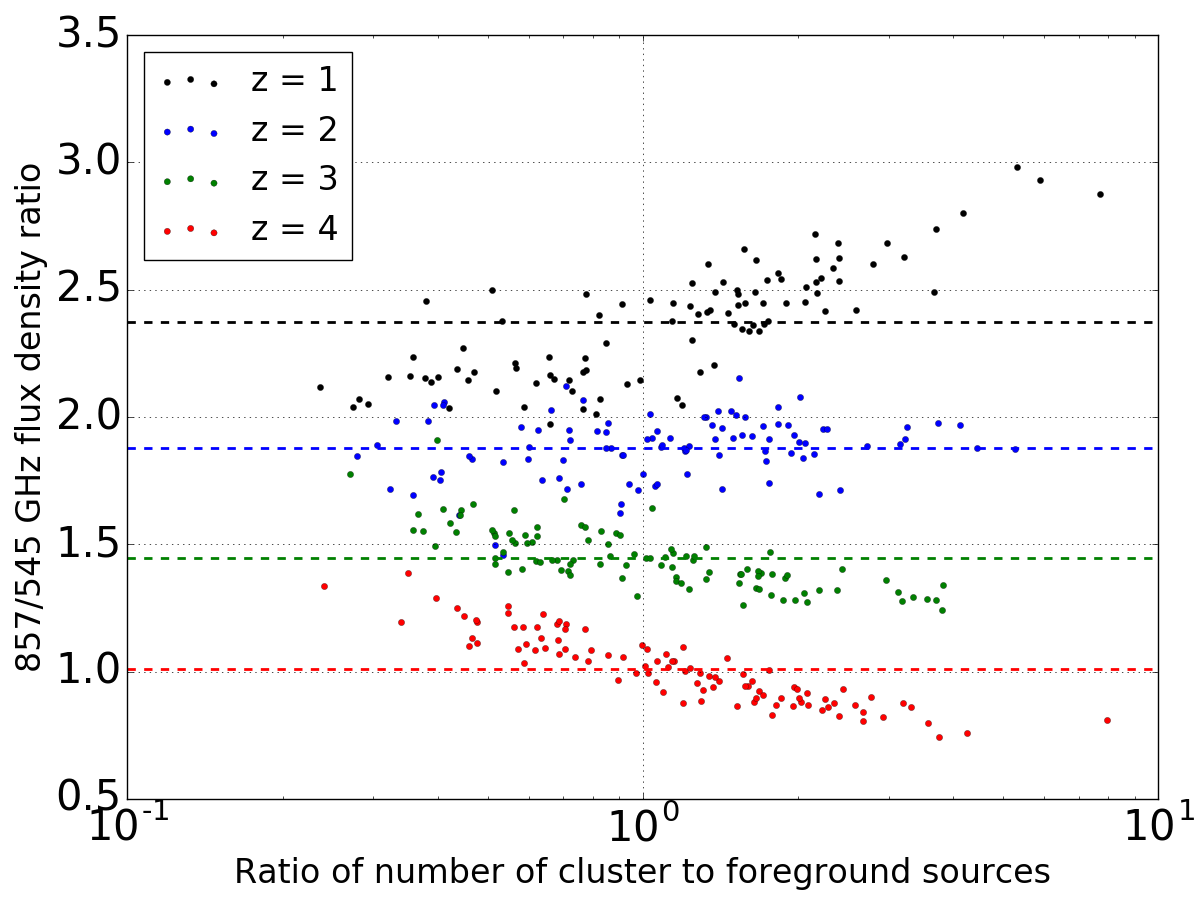}
   \centering{\caption{Estimated \textit{Planck} 857/545 GHz flux density ratio of 400 proto-clusters, as a ratio of the number of proto-cluster to background/foreground sources. Points in black are proto-clusters at a redshift of one, in blue at a redshift of two, in green at a redshift of three and in red at a redshift of four. The dashed lines show the average colour of the 100 proto-clusters at each redshift. Large symbols show proto-clusters which would be selected by our 3$\sigma$ overdensity requirement, with small labels showing proto-clusters that would not be. \label{Fig:SimClusterPlanck}}}
\end{figure}


\subsection{\textit{Herschel} Colours}

The use of \textit{Herschel}-SPIRE colour-colour diagrams to separate sources of different redshifts is well established \citep[e.g. ][]{Herranz2013, Noble2013, Clements2014, Ivison2016, Negrello2017}, though the precise interpretation of the results are uncertain. 
Typically, sources whose SED peak at longer wavelengths tend to lie at higher redshifts \citep{Casey2014, Dowell2014, Asboth2016, Ivison2016}, and therefore sources whose SED peak at 250, 350 and 500 $\mu$m likely indicate progressively higher redshifts.

\begin{figure*}
   \centering
  \includegraphics[width=1\linewidth]{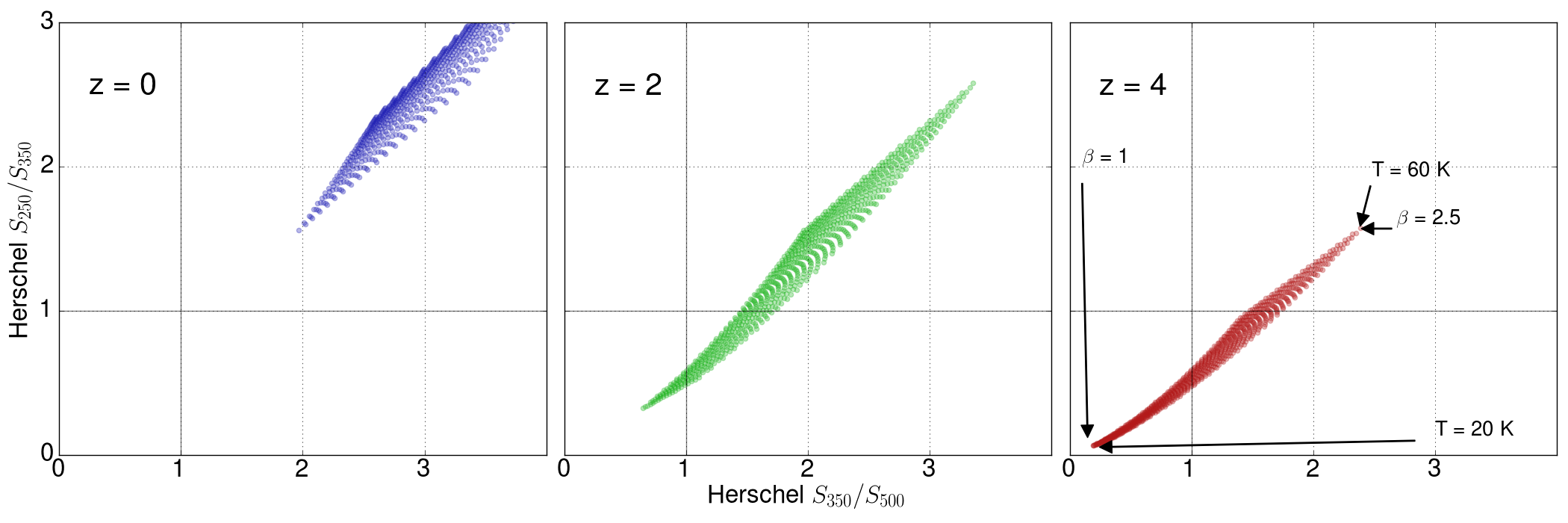}
   \centering{\caption{The \textit{Herschel}  $S_{250}/S_{350}$ vs  $S_{350}/S_{500}$ colours for modified blackbodies (see text for more detail) at redshifts of 0, 2 and 4, allowing $T$ and $\beta$ to vary between 20 and 60 K and 1 and 2.5, respectively. In the far right plot, the regions of maximum and minimum $T$ and $\beta$ are indicated for clarity.}  \label{Fig:SimulationVariation}}
\end{figure*}

In Fig. \ref{Fig:SimulationVariation}, we simulate the \textit{Herschel} colours, again using a single temperature modified blackbody function, in an attempt to show the rough redshift a source is likely to have, given its \textit{Herschel} colours.
We fix the redshifts at 0, 2 and 4, where we expect our sources to approximately peak in the 3 SPIRE bands, and uniformly distribute the temperatures and $\beta$ values between 20 and 60 K and 1 and 2.5, respectively.
Figure \ref{Fig:SimulationVariation} shows that the \textit{Herschel} colours of a source can provide a good proxy for the redshift of that source.

To compare to our simulation, in Fig. \ref{Figure:HerschelColours2}, we plot the individual $S_{250}/S_{350}$ and $S_{350}/S_{500}$ \textit{Herschel} colours for the local galaxies and proto-cluster candidate \textit{Planck} sources.
Any local galaxy extended on arcminute scales, or where extraction on the \textit{Herschel} map has clearly divided the source into multiple sources were removed.
For the 250/350 $\mu$m and 350/500 $\mu$m colours of the local galaxies, we find a mean of $2.05 \pm 0.43$ and $2.60 \pm 0.74$, respectively, whereas for the proto-cluster candidates these values are $1.13 \pm 0.47$ and $1.57 \pm 0.49$.
Fig. \ref{Figure:HerschelColours2} clearly divides into two regions, one bluer region associated with the local galaxies, and one red region where the bulk of the \textit{Herschel} detected proto-cluster candidates lie.
Similar to the \textit{Planck} colours, the \textit{Herschel} colours of the proto-cluster candidates are on average redder than for the local galaxies.
At the same time, we take a template starburst galaxy, Arp 220 \citep{Donley2007}, and examine the \textit{Herschel} colours as it would appear at various redshifts.
Direct comparison suggests the proto-cluster candidates lie at a redshift of $\simeq$ 2.
This is also in good agreement with our estimates of single temperature blackbody fits in Fig. \ref{Fig:SimulationVariation}, with local galaxies inhabiting the low redshift region and proto-cluster candidates inhabiting the region suggested for redshifts between 2 and 4.
However, estimating the redshift of sources from its Herschel colours alone can be difficult; often the errors are large, and the variation seen in Figures \ref{Fig:SimulationVariation} and \ref{Figure:HerschelColours2} alone is enough to make the true redshift of a source uncertain.
Given the simulations, observed error, and variation we see here, we can therefore reasonably say these sources likely lie at z $>$ 1, but little more until future follow up work can constrain these sources further.


\begin{figure}
   \centering
  \includegraphics[width=1\linewidth]{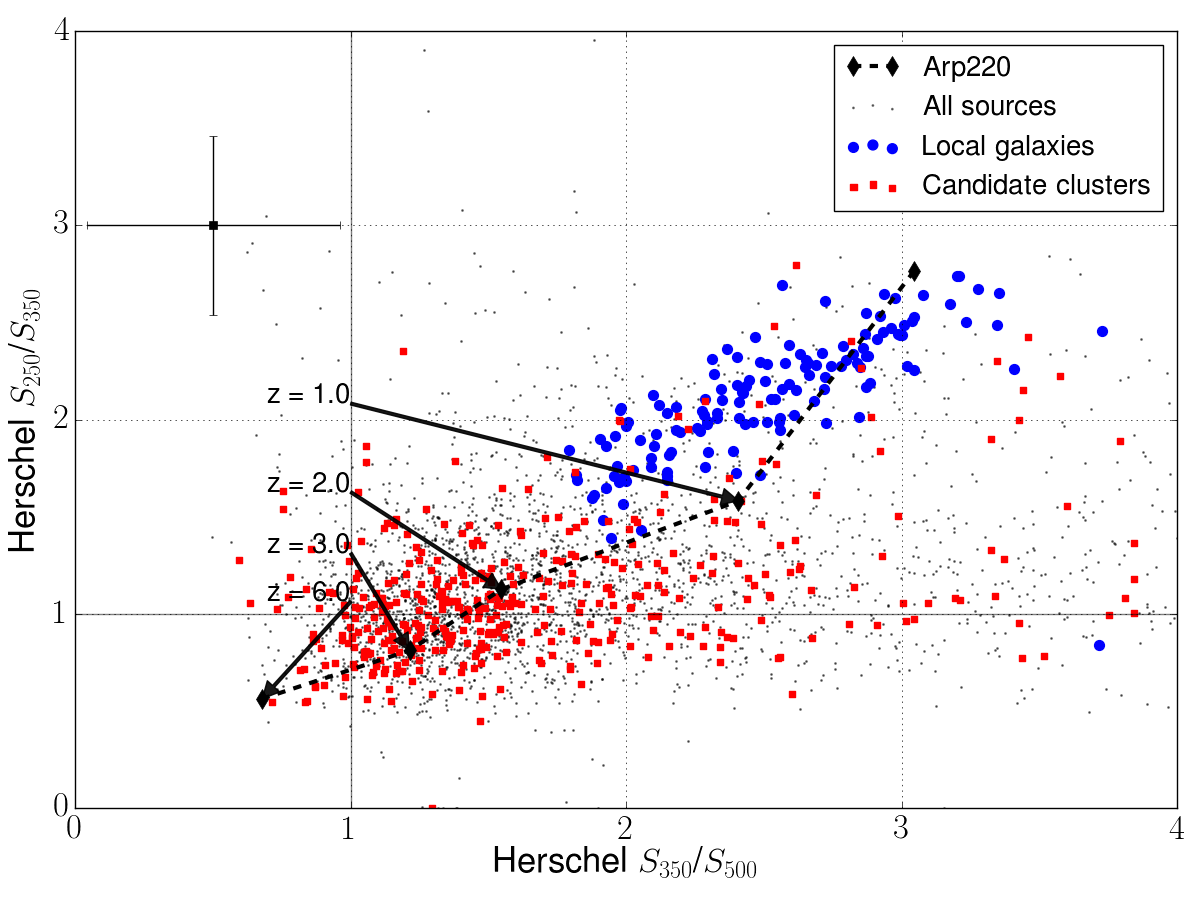}
   \centering{\caption{\textit{Herschel} $S_{250}/S_{350}$ and $S_{350}/S_{500}$ colours of local galaxies (blue circles) or all the Herschel sources associated with the proto-cluster candidates (red squares). The small black circles include all \textit{Herschel} sources detected for all of our \textit{Planck} sources. Typical errors are given on the left (black square).  The dashed black line with the black diamonds shows the \textit{Herschel} colours of the local ULIRG Arp 220, as it would appear at z = 2,4 and 6.}  \label{Figure:HerschelColours2}}
\end{figure}

\section{The candidate clusters}

\label{Sec:Clusters}

Out of the 279 unique \textit{Planck} sources we have identified, 27 appear to be $>3\sigma$ overdensities of \textit{Herschel} sources.
The photometry of these objects indicates that the flux density comes from a number of discrete, individual sources, and their colours indicate that they likely lie at $z \sim 2$. 
These observations could correspond to a physical cluster of DSFGs, a series of line of sight sources stretching from $z \sim 2$ to $\sim 4$, or multiple clusters / proto-clusters along the line of sight.
In this section, we attempt to quantify these proto-cluster candidates further, and examine whether the large area surveyed can explain these sources through fluctuations in the number counts alone.

\subsection{Probability of observing $>N$ sources by chance}

\begin{figure*}
   \centering
  \includegraphics[width=1\linewidth]{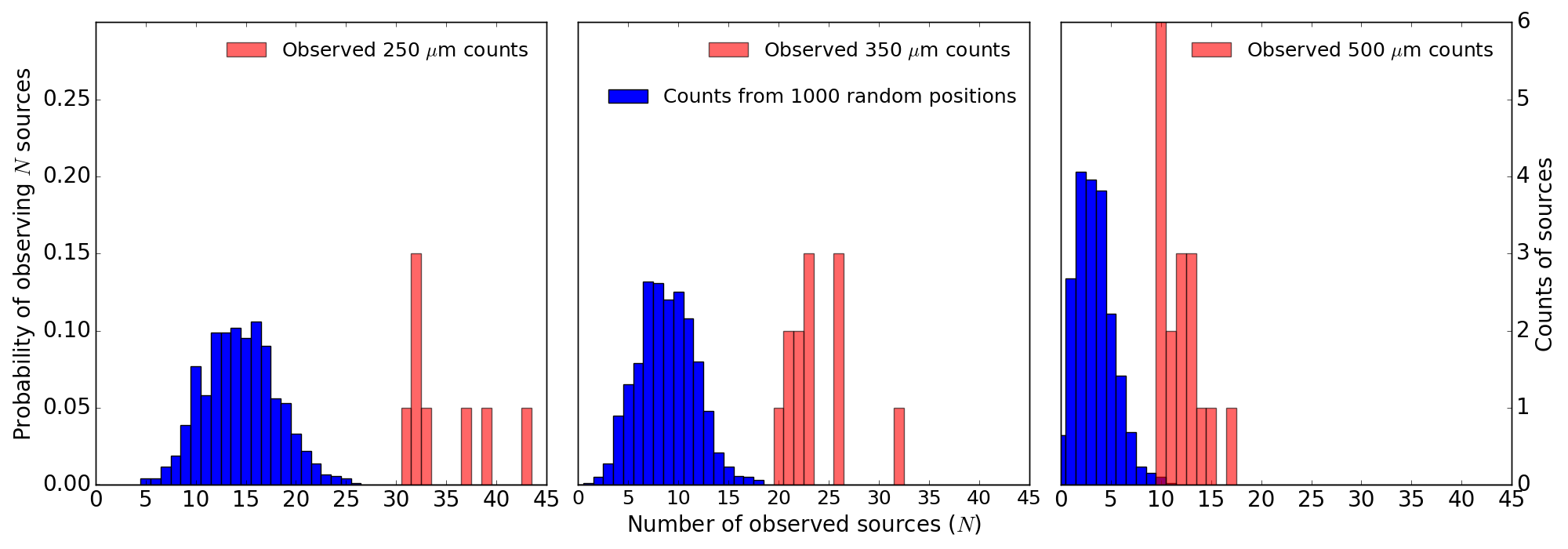}
   \centering{\caption{(Blue) histograms of the result when 1,000 random \textit{Planck} beams are placed on the NGP map and the number of sources with S$_{350} $, S$_{250}$, or S$_{500} > 24.5$ mJy are counted for: (left) The 250 $\mu$m band, (middle) the 350 $\mu$m band and (right) the 500 $\mu$m band. (Red) histograms of the observed numbers of 250, 350 and 500 $\mu$m sources for our candidate proto-clusters, which are considered overdense in their respective bands.\label{Fig:ProbN}}}
\end{figure*}

If our candidate proto-clusters are actually only line of sight or number count fluctuations, then it should be possible to model the probability of finding one using Poisson statistics. 
In Fig. \ref{Fig:ProbN}, we sample the NGP field with 1,000 random \textit{Planck} beams of radius 4.63 arcminutes, and count the number of 250, 350, and 500$\mu$m sources with fluxes greater than 25.4mJy in each of the three respective bands.
We then plot the normalised version of this sample, as well as histograms of the numbers of \textit{Herschel} sources associated with each of our candidate proto-clusters from the 857 GHz band.
Our candidate proto-clusters are clearly overdense with respect to our random samples of 1,000 positions.
The mean number of associated \textit{Herschel} sources for our proto-cluster candidates is 29.1, 20.6 and 10.7 for the 250, 350 and 500 $\mu$m bands respectively, corresponding to a 2.9, 3.5 and 4.0$\sigma$ overdensity respectively \footnote{These probabilities have been converted to their corresponding $\sigma$ value in the Normal distribution.}. 

Given we here examine roughly 800 deg$^2$ of sky, and according to Poisson statistics, we may expect to find around 89.0 patches where there are 26 or more 250 $\mu$m sources, 33 regions where there are 10.2 or more 350 $\mu$m sources, and 1.3 regions where there are 11 or more 500 $\mu$m sources.
If all our proto-clusters were only this overdense, this might explain our results, however, many of our proto-clusters host far stronger overdensities, with 14 of our proto-cluster candidates containing $\geq$ 36, 23 or 12 250, 350 and 500 $\mu$m sources respectively (with maximal numbers of associated \textit{Herschel} sources of 43, 32 and 17 for the three bands).
Over 800 deg$^2$ of sky, we would therefore expect to see 0.5, 1.5 and 0.3 patches containing $\geq$ 36, 23 or 12 250, 350 and 500 $\mu$m sources, if they were Poisson distributed.
We in fact see 4 patches at least this overdense in the 250 $\mu$m band, 8 at least this overdense in the 350 $\mu$m band, and 10 at least this overdense in the 500 $\mu$m band, which cannot be explained solely by the large area surveyed in this paper.
Our candidate proto-clusters are therefore likely to be physically associated or be the product of several clusters / proto-clusters or overdensities along the line of sight.

We would still expect some level of contaimination from unassociated sources.
Under the assumption that the \textit{Herschel} sources are a mix of proto-cluster members and Poisson distributed unassociated sources, for an expected $\mu$ sources, the probability that there are $N$ proto-cluster sources out of $M$ detected sources is given by:
\begin{equation}
p(N|M, \mu) = \frac{\big[\frac{\mu^{(M-N)}}{(M-N)!}\big]}{\Sigma_{i = 0}^{i = M} \big[\frac{\mu^{(i)}}{(i)!}\big]},
\end{equation}
the derivation of which is given in Appendix \ref{apen:2}.
For our mean of 20.6 350 $\mu$m \textit{Herschel} sources associated with each cluster, this suggests that on average, around 11 of the sources would be associated with the proto-cluster, with only a 0.7\% chance of having 3 or fewer proto-cluster members.

Though we do not have accurate redshifts for our sources, we can get some idea if they lie at similar redshifts by examining where the individual \textit{Herschel} sources for a single proto-cluster candidate lie in colour-colour space.
In Fig. \ref{Fig:RealCluster}, we plot the \textit{Herschel} colours for the \textit{Herschel} components of three of our \textit{Planck} sources; the Bootes proto-cluster identified by \citet{Clements2014}, a candidate proto-cluster PCCS1 857 G085.48+43.36 identified in this work, and a cirrus source.
The Bootes proto-cluster and the ELAIS-N1 proto-cluster show clear clustering in the colour colour plot, whereas the cirrus source shows a much larger spread.
From Fig. \ref{Fig:SimulationVariation}, this clustering in colour-colour space indicates it is likely these sources are physically associated, but the uncertainties are large enough that we cannot rule out the possibility that these multiple clusters / proto-clusters along the line of sight.

\begin{figure*}
   \centering
  \includegraphics[width=\linewidth]{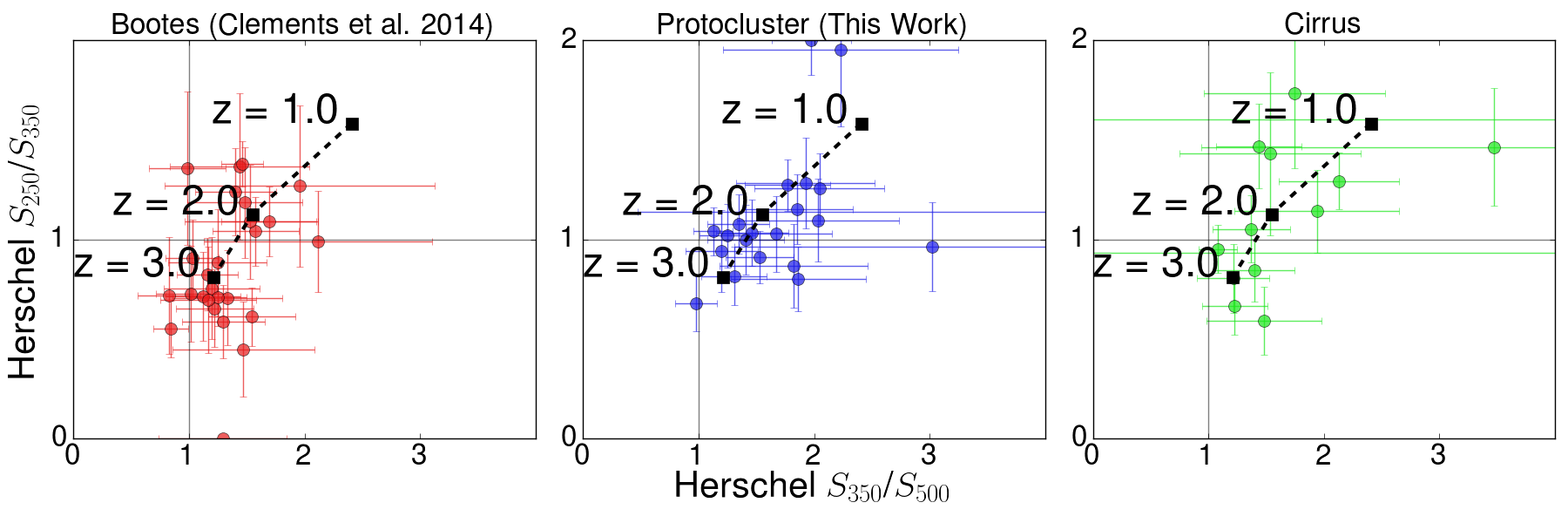}
   \centering{\caption{\textit{Herschel} S$_{350}$/S$_{500}$ S$_{250}$/S$_{350}$ plot showing the colours for three \textit{Planck} sources; The Bootes clump identified by \citet{Clements2014} on the left; a candidate cluster seen in the ELAIS-N1 field in the centre, and a source identified with Galactic cirrus  on the right. The black dashed line and squares indicate the \textit{Herschel} S$_{350}$/S$_{500}$ and S$_{250}$/S$_{350}$ of the local ULIRG Arp 220, as it would appear at z =  1.0,  2.0 and 3.0. \label{Fig:RealCluster}}}
\end{figure*}

\subsection{Properties of the proto-cluster candidates}

Given our previous analysis, in the following sections we assume that all 27 of our candidate proto-clusters are physically assocated proto-clusters or multiple clusters / proto-clusters along the line as sight, as opposed to chance overdensities along the line of sight. 
We find a surface density of candidate proto-clusters of ($3.3 \pm 0.7) \times10^{-2}$ sources deg$^{-2}$.
In their assessment of the number of \textit{Planck} detectable clusters, \citet{Clements2014} find a surface density of ($4.4 \pm 2.2) \times10^{-2}$ sources deg$^{-2}$, in good agreement with our results here. 
\citet{PlanckCollaboration2015c} in the PHZ, searched directly on the \textit{Planck} maps, discovering a total of 2,151 candidate high-z sources across around 10,000 deg$^2$ of the cleanest part of the sky, with intial follow up suggesting 94\% of these are overdensities of sources \citep{PlanckCollaboration2015d}.
Given the different selection functions used in the PHZ and this paper, it is difficult to make a direct comparison, but this would correspond to a approximate surface density of $(0.18 \pm 0.01)$ sources deg$^{-2}$, roughly 5 times larger than found here.
This can be somewhat offset if we include our sources where do not not assign a classification, as our surface density rises to ($0.11 \pm 0.02$) sources deg$^{-2}$, in closer agreement with the PHZ.
Further follow up of the PHZ sources, especially at the fainter end, is needed to investigate the discrepancies.

The number counts within individual fields mostly agree with the estimated number counts given here, with 10 out of an expected 11 from the SGP, seven out of an expected six for the NGP, zero out of four for HERS (which has large amounts of Galactic cirrus), and roughly one in each of the smaller HerMES fields.
The GAMA fields are lacking in sources, with no proto-cluster candidates detected in any of them.
The lack of objects in GAMA could be due to the large amount of foreground cirrus present in GAMA09 and GAMA15, which could obscure a number of candidate proto-clusters.

Many confirmed proto-clusters are found to be extended on scales of tens of arcminutes \citep{Casey2016}.
The smaller \textit{Planck} beam implies that we are detecting highly compact systems of DSFGs, compared to generic proto-clusters which tend to show less of a density contrast with respect to the background \citep{Casey2016}.
For instance, the Bootes proto-cluster candidate appears to be at a redshift of z $\sim$ 2.3. 
\citet{Pearson2013} estimate the redshift distribution of sources in the phase 1 release of H-ATLAS, where they find there should be roughly 10-100 \textit{Herschel} sources per square degree with $F_{350} > 35$ mJy at a redshift $\sim$ 2, or roughly 0.2-1.5 sources per \textit{Planck} beam.
Using the definition of \citet{Chiang2013} of density contrast: 
\begin{equation}
\delta_{gal}(\textbf{x}) = \frac{n_{gal}(\textbf{x}) - <n_{gal}>}{<n_{gal}>}
\end{equation}  
and a simple photo-z fitter, which fits our \textit{Herschel} sources to a SED template of Arp 220, we find 12 sources with $F_{350} > 35$ whose photo-z is consistent within 1$\sigma$ of z$=2.3$, giving a density contrast between $\delta(12) = 7 - 60$, depending on whether one uses a low or high estimate of the density of \textit{Herschel} sources at z $= 2.3$.
The low density contrast estimate is still consistent with these sources being proto-clusters, but for density contrasts of $>10$ this becomes more difficult to understand; the large density contrasts imply that these are systems which are well on their way to collapse and virialization.
However two of our candidate proto-clusters appear to be associated with known galaxy clusters; PCCS1 545 G058.72+82.59 (PCCS1 857 G058.53+82.57) lies 4.3 arcminutes away from the core of galaxy cluster GHO 1319+3023 \citep{Gunn1986} at a redshift of 0.4, PCCS1 545 G027.38+84.85 (PLCKERC857 G027.36+84.83) is associated with the redshift 0.43 galaxy cluster GMBCG J198.59994+26.5688 \citep{Hao2010} and PCCS1 545 G084.40+81.05 is associated with the estimated redshift 0.43 galaxy cluster NSCS J131812+335831 \citep{Lopes2004}.
Given our earlier estimates on the redshift of our souces being at $z>1$, it is possible that our cluster of DSFGs is being lensed by a foreground cluster, rather than that they are physically associated with the foreground cluster.
Three of our proto-clusters, PLCKERC857 G017.86-68.67, PLCKERC857 G149.81+50.11 and  PLCKERC857 G095.44+58.94, also appear to host QSOs that are mostly, not emitting in the FIR.
Again, whether or not these QSOs are associated with the cluster of DSFGs is uncertain, but they they're redshifts are typically between $z=1$ to $2$, so could be signposting the true redshifts of our proto-clusters.

\subsection{Simulations of DSFGs in clusters}
\label{sec:sim}

\citet{Granato2015} simulate the FIR/sub-mm properties of high-redshift clusters and proto-clusters by combining hydrodynamical simulations with GRASIL-3D, a radiative transfer code that accounts for dust reprocessing in arbitrary geometries.
In Fig. \ref{Figure:Granato} we comapre the number counts of clusters of DSFGs from the Herschel data with the predicted number counts obtained by \citet{Granato2015}, assuming their 24 simulated clusters, which all had a final virial mass at z = 0 above $1 \times 10^{15} h^{-1} M_{\odot}$, are representative of the cluster population we detect here.
We impose a $3\sigma$ S/N cut for each band considered, and use the aperture photometry estimate from \textit{Planck}.
Again, we assume that all our 27 candidate proto-clusters are actual physical clusters of sources.
Our observations indicate that our detected clumps are more numerous, or are brighter, than predicted from these simulations.
The flux density from our proto-clusters appear to be on average $\sim$ 5 times greater than predicted. 

\begin{figure}
   \centering
  \includegraphics[width=1\linewidth]{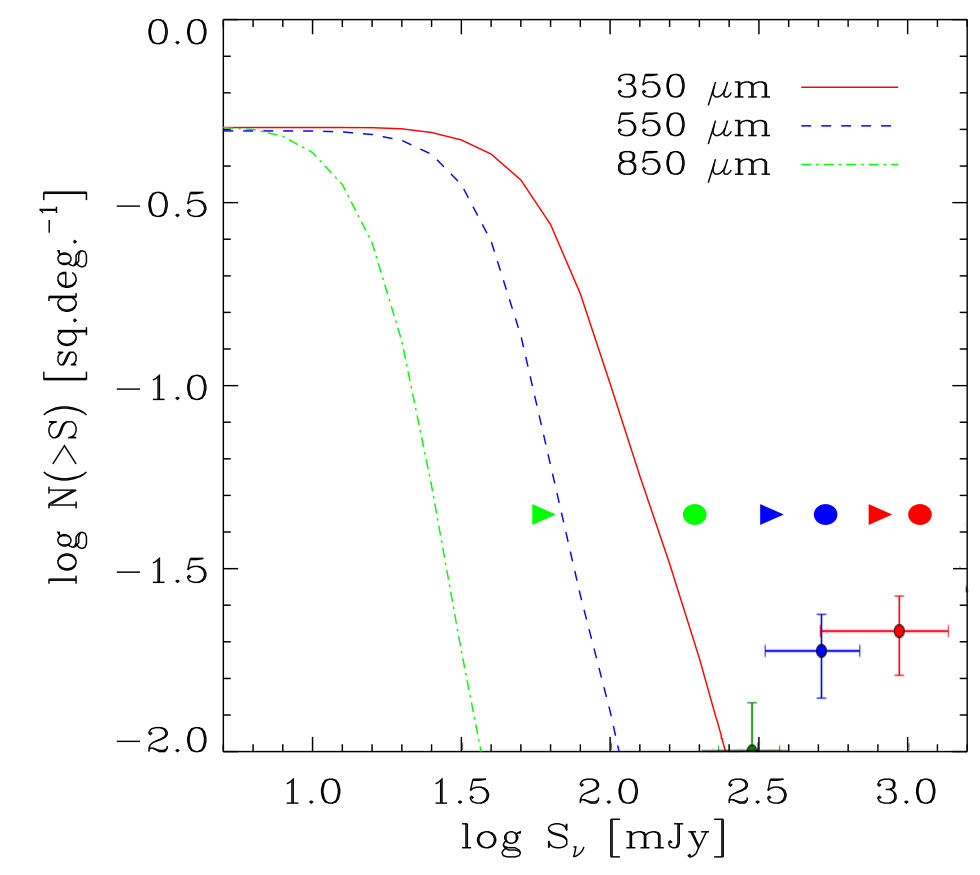}
   \centering{\caption{Expected cumulative number counts of clusters reproduced from \citet{Granato2015}. The solid, dashed and dot-dashed lines show the predicted number counts of sources at 850, 550 and 350 $\mu$m. The circles and right pointing triangles are the results from \citet{Clements2014}, and points with errorbars are the results from this work.} \label{Figure:Granato}}
\end{figure}

In Figure \ref{Figure:Granato2}, we show that this is likely due to our observed sources being brighter than expected in simulations, by reproducing the histogram of expected 350, 550 and 850 $\mu$m flux densities from  \citet{Granato2015}, and comparing the distribution of the 350 $\mu$m flux densities of the proto-clusters identified in this work.
Since some of the flux from our proto-cluster candidates will come from sources not associated with the proto-cluster, we attempt to remove this foreground contribution.
We place 1000 \textit{Planck} beams at random positions on each of the \textit{Herschel} maps, calculate the total flux density in those beams following the same prescription in section \ref{Sec:Photometry}, and take the median value of the aperture fluxes over those 1000 beams as the typical foreground contamination.
The median value varies between maps, but is usually of the order of 100 to 300 mJy.
\begin{figure}
   \centering
  \includegraphics[width=1\linewidth]{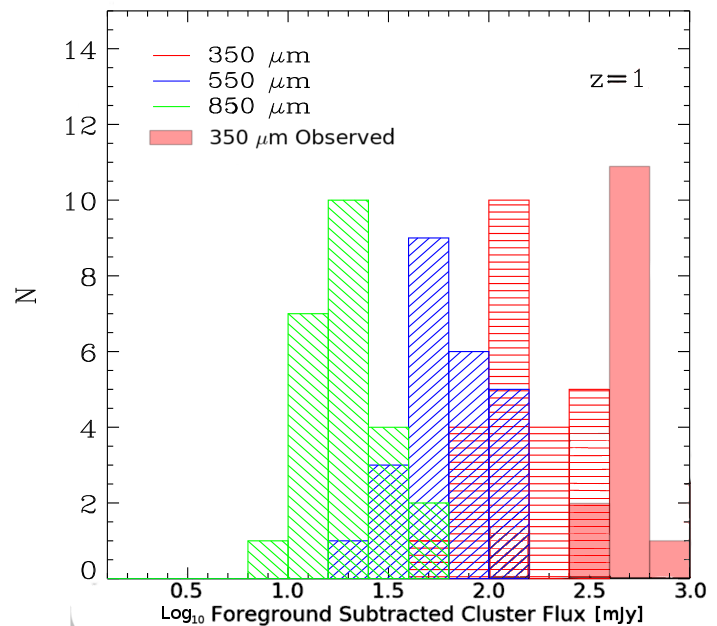}
   \centering{\caption{ A histogram of the estimated flux densities of clusters at z = 1 reproduced from \citet{Granato2015}. The red, green and blue hashed bins represent the histograms from the simulation of clusters as they would appear in the \textit{Planck} HFI bands. The solid red histogram gives the foreground subtracted candidate proto-cluster flux densities from this work if placed at z = 1.} \label{Figure:Granato2}}
\end{figure}
These values are then removed from the \textit{Herschel} aperture flux densities for each of the proto-clusters, and the results plotted in Fig. \ref{Figure:Granato2}.
The difference between our observed flux densities and the simulated clusters is exacerbated at higher redshifts, as the simulated flux densities tend to decrease \citep{Granato2015}.
The original plots in \citet{Granato2015} split the data into three separate redshift bins at z = 1,2 and 3, with the z = 1 flux densities generally being the greatest.
Therefore, to be conservative, we compare our results to those at redshift z = 1, under the extreme assumption that all our candidate proto-clusters exist at this redshift.
Even in the extreme case that all our candidate proto-clusters lie at z $\simeq$ 1, the observed flux densities appear systematically higher than the simulated flux densities, with a median flux density of proto-cluster candidates of 500 mJy at 350 $\mu$m observed compared to 100 mJy simulated. 

It is difficult to match the observed proto-cluster flux densities to the simulated.
We earlier demonstrated that the flux from these sources comes almost entirely from multiple, detected, discrete sources rather than cirrus or fainter sources. 
Additionally, we remove any foreground or background contaminent and compare our sources to those simulated clusters with the highest flux densities.
Even with these constrains, we still find our proto-clusters are around a factor of 5 $\times$ brighter in comparison to the simulations.
If these proto-clusters are confirmed to be real, physical associations, then these results demonstrate that current models of cluster formation struggle to reproduce the FIR/sub-mm flux densities seen in observations by a factor of 5, and likely underestimate the SFR in clusters / proto-clusters during their formation.
One possible explination is that these DSFGs are not tracing only the most massive clusters, and that clusters with lower final virial masses could match our observations, but redshift and mass confirmations would be required before this can be tested.


\subsection{Evolution of large scale structure}

\begin{figure}
   \centering
  \includegraphics[width=1\linewidth]{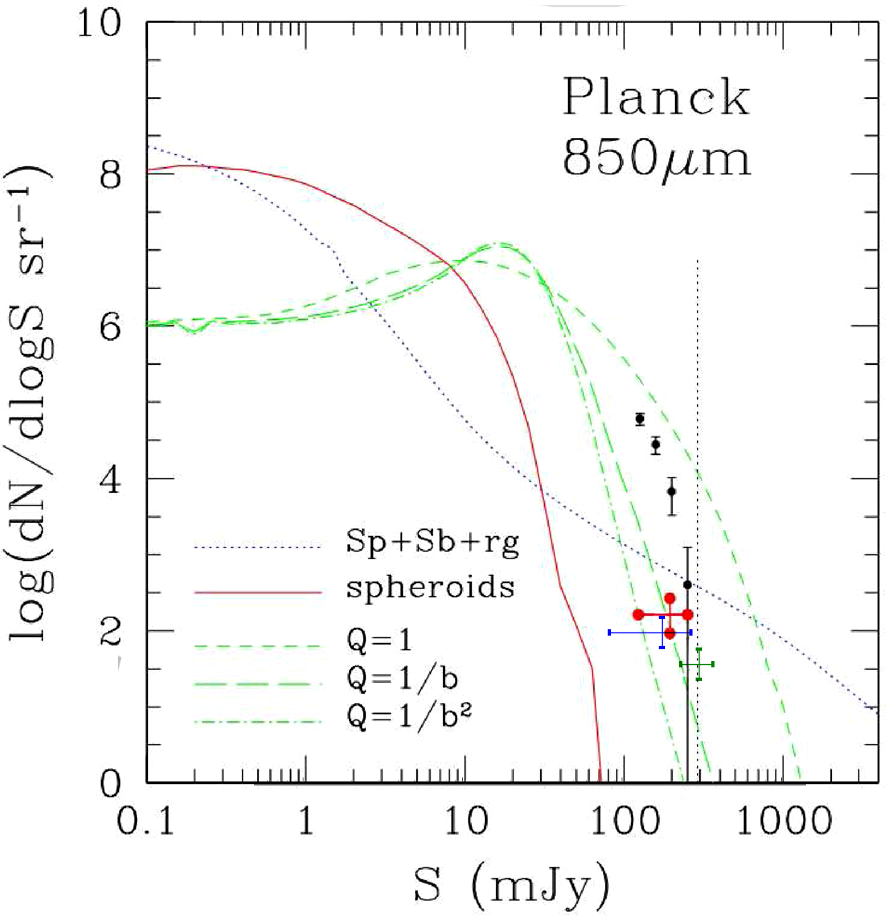}
   \centering{\caption{Predicted number counts for over-densities of 850 $\mu$m sources taken from \citet{Negrello2005}. The dashed green lines gives the predictions for the number counts as the three point correlation function evolves according to Q = 1 (no evolution) Q = $\frac{1}{b}$ and Q = $\frac{1}{b^2}$, where b is the linear bias factor between galaxies and dark matter. The black points are from the simulations from \citet{Negrello2005}, the red point gives the results from \citet{Clements2014} and the blue point gives the results from this paper using all our proto-clusters. The green point is our result if we restrict ourselves only to proto-clusters detected at the $>3\sigma$ level.} \label{Figure:Negrello}}
\end{figure}

According to the formalism of \citet{Negrello2005}, the number counts of clusters should be sensitive to the evolution of the amplitude Q of the three-point correlation function.
Under the tentative assumption that the galaxies in our proto-cluster candidates are in fact all physically associated, we compare the counts of our candidate proto-clusters to the predictions made by \citet{Negrello2005} in Figure \ref{Figure:Negrello}.
We plot both our result when no restrictions are imposed (in blue) and when we impose the constraint that only the 7 proto-cluster candidates detected to at least $3\sigma$ in the 353 GHz channel (850 $\mu$m) are included (in green). 
Without constraints, we find a median flux for our proto-clusters of (169 $\pm$ 95) mJy, and a number density of 135 $\pm$ 24 sources per steradian.
With constraints, we find a median flux for our proto-clusters of (294 $\pm$ 65) mJy, and a number density of 39.4 $\pm$ 13 sources per steradian.

Both \citet{Clements2014} and this work strongly exclude the Q $=$ 1 analytical model, which corresponds to the amplitude Q of the three-point correlation not evolving with redshift.
Using 3$\sigma$ detected proto-clusters at 353 GHz suggests that the Q $\propto$ $\frac{1}{b^2}$ model is also incorrect.
However, caution should be taken as these are only candidate proto-clusters rather than confirmed, and with only 7 proto-clusters detected to 3$\sigma$ the number counts remain low.
Addionally, flux contamination from unassociated sources (which we account for in Section \ref{sec:sim} but cannot account for here) is likely to play a role. 
If we assume that roughly 25\% of the flux can be from unassociated sources, similar to what was found in Section \ref{sec:sim}, our conclusions remain similar, with a better agreement between the $Q = \frac{1}{b}$ model and our $>3\sigma$ detected proto-clusters.

\section{Discussion}

\label{Sec:Discussion}

We have identified 27 candidate proto-clusters from a cross-match of \textit{Planck} compact source catalogues and \textit{Herschel} maps.
The numbers of sources are difficult to explain if none of them are associated with each other, and their colours indicate they all likely like at z $> 1$.
We have also found several proto-cluster candidates with lower $S_{857}/S_{545}$ flux ratios than expected. 
We have shown this could be from a large number of foreground contaminants, but it is also possible that there exists a warmer population of clusters / proto-clusters of DSFGs.
In this section, we discuss these results in the context of the literature, as well as briefly discussing the natures of the other \textit{Planck} compact sources we have identified.

\subsection{The HeLMS field}

The HerMES Large Mode Survey (HeLMS, P.I. Marco Viero) is a shallow 280 deg$^2$ field  imaged with \textit{Herschel}-SPIRE at 250, 350 and 500 $\mu$m.
In comparison to the other extra-galactic fields under consideration here, it is highly contaminated by cirrus.
No publicly released, verified catalogue exists for this complex field, but using a private catalogue (Marco Viero, Private Communication) we find 130 857 GHz and 40 545 GHz \textit{Planck} compact sources in this field, and 137 unique sources.
The maps are highly cirrus contaminated, with 64 (46\%) of sources being identified with Galactic cirrus, 61 (45\%) local galaxies, 2 QSOs (LBQS 0106+0119 and CRATES J2323-0316), 1 candidate proto-cluster, and 9 sources with no clear identification.
Even in the PCCS2, where we found that only 2.5\% of sources were associated with Galactic cirrus, we find  27\% of PCCS2 sources in the HeLMS field are associated with Galactic cirrus.


Overall, given the reasonably small differences found between the other H-ATLAS, HerMES and Hers fields, and the large differences found between them and HeLMS, we ascribe the differences in our results to the complex nature of the HeLMS field, and the preliminary nature of the catalogues currently available.

\subsection{The nature of the \textit{Planck} compact sources}

Fig. \ref{Fig:CountsHerschel}, \ref{Fig:ContribHerschel} and Table \ref{Table:Id} all indicate that the \textit{Planck} compact sources resolve into a range of different phenomena.
Furthermore, almost half of the \textit{Planck} compact sources are actually extended on the scale of \textit{Herschel}, so filters designed purely for point like sources can miss a range of sources, as shown in Table \ref{Table:Frac}.
Given this, the ERCSC, PCCS and PCCS2 should not simply be considered deeper versions of the same catalogue, but catalogues that specifically probe different source types owing to the different filters and extraction methods used in their creation.

The \textit{Planck} compact source catalogues appear to host several stars. 
Two of the three stars present in our catalogue, Mira (also known as $\omicron$ Ceti), and R Sculptoris, are both Asymptotic Giant Branch (AGB) stars, known to produce large amounts of dust \citep{Mayer2011, Mayer2014}, whilst the third, $\alpha$ PsA o, is known to host a dusty debris disk \citep{Acke2012}.

One of our lens candidates, like our proto-cluster candidates, appear to host an overdensity of \textit{Herschel} sources.
These could indicate the presence of a physical cluster or proto-cluster. 
PLCKERC857 G047.32+82.53 \citep[H-ATLAS J132426.9+284452, ][]{Negrello2017} at a redshift of 1.676 \citep{George2013, Bussmann2013, Timmons2015} has a 3.1$\sigma$ over-density of 500 $\mu$m sources.
We also note that PCCS2 857 G270.56+58.54 \citep[H12-00, ][]{Herranz2013, Fu2012, Clements2016} hosts 2.8$\sigma$ over-density of 350 $\mu$m sources and \citet{Clements2016} find an overdensity of SCUBA-2 850 $\mu$m sources associated with H12-00, and is unclassified in this work, though it is selected as a candidate proto-cluster if we use a slightly smaller beamsize of 4.33 arcminutes.
Furthermore, H12-00 is also independently selected in \citet{Canameras2015}, where they specifically search for and follow up the brightest gravitational lensed sources discovered with \textit{Planck}.
Whether DSFGs are good tracers of the most massive dark matter overdensities at $z > 2$ continues to be widely debated \citep{Blain2004, Chapman2009, Dannerbauer2014, Miller2015, Casey2015, Hung2016}, but if they do, then these lensed sources could make excellent signposts for the locations of further clusters / proto-clusters.

H12-00 \citep{Clements2016} does not qualify as a proto-cluster candidate using our criterion in Section \ref{Sec:SelectionMethods}.
Given that follow up work on H12-00 demonstrates its likely cluster nature \citep{Clements2016}, the large number of ``red'' unassigned sources, and a number of sources that are on the edge of being selected as candidate proto-clusters, it is entirely possible that \textit{Planck} is detecting a far larger number of protocluster sources, but that the specific quantifiable criteria used here mean that they are not assigned as such during the selection process.
An examination of the unassigned sources reveals almost half (28 of 61) have a $>2\sigma$ in the 250, 350 or 500 $\mu$m bands, but are not 3$\sigma$ overdense.
Additionally, three of the four matches we found with the PHZ were unclassified.
The final source, PHz G160.57-56.79 / PCCS1 857 G160.59-56.74, we identify as the local galaxy 2MASX J02094125+0015587 at a redshift of $z=0.2020$.
This is somewhat surprising, and could hint that our selection probes a different population of proto-clusters / overdensities of sources.
It is possible that several of the ``red'' unassigned sources could be due to CIB fluctuations which, due to clustering, have a strong super-Gaussian tail so can appear as high S/N sources (See Figure 10 of De Zotti et al. 2015).
However, Fig. \ref{Fig:Photometry2} shows that the flux density of many of the unassigned sources is entirely accounted for by discrete, detected sources. 
These could still be line of sight chance alignments, but it does show that these are unlikely to be fluctuations in the background sources too faint to be detected by \textit{Herschel} and are, at best, fluctuations in the number counts of bright ($S_{350} > 25.4$ mJy) \textit{Herschel} sources.  

\subsection{The nature of our proto-cluster candidates}


DSFGs have now been found in a range of cluster environments, from extremely large proto-clusters on angular scales $> 10$ arcminutes \citep{Dannerbauer2014, Casey2015, Casey2016}, to scales similar to those of the \textit{Planck} HFI beam \citep[][This work]{Herranz2013, Clements2014, PlanckCollaboration2015c, PlanckCollaboration2015d}, to $> 10$ sources on $\sim 20$ arcsecond scales \citep[][]{Oteo2017}.
The existence of many physically associated DSFGs is surprising; simulations expect these sources to be physically unassociated \citep{Hayward2013, Cowley2014}, and without a mechanism for triggering several DSFGs simultaneously or a longer duty cycle \citep{Emonts2016, Dannerbauer2017, Oteo2017c}, we would not expect to observe several physically associated DSFGs at once \citep{Casey2016}.
Similar to the PHZ, we can be confident but not certain that the compact candidate proto-clusters we have detected are physically associated or multiple clusters.
However follow up of other apparent overdensities of DSFGs \citep[][]{Flores-Cacho2016, Wang2016, Oteo2017}, suggests many of these objects are indeed physically associated. 
We leave redshift estimates and therefore SFR estimates for our proto-clusters to a future paper (Cheng et al. in preparation), but if the DSFG members of these proto-cluster candidates are similar to other DSFGs, their likely SFR will be of the order of $100$ M$_{\odot}$ yr$^{-1}$, and a likely total cluster SFR of up to several thousands of M$_{\odot}$ yr$^{-1}$ \cite{Dannerbauer2014, Casey2016, Oteo2017}. 
Both \citet{Scoville2013} and \citet{Darvish2016} find that below z $\sim$ 1, SFR is efficiently quenched in denser environments, but the mechanicsms for this quenching remain uncertain, and Figure 2 of \citet{Casey2016} shows that there is clearly a downturn below z $\sim$ 1 between the theoeretical and observed SFR density of clusters.
In this paper, we have found DSFGs, with elevated associated SFRs, in clusters / proto-clusters over a range of scales, from the arcsecond to the arcminute.
This indicates that it is unlikely that it is simply the scale or size of the structure that determines its SFR density.
Given we do not see DSFGs in local clusters, it could be that is it the virialisation state or the presence of a evolved intracluster medium which determines whether the presence of multiple DSFGs is likely to occur.
No clusters containing significant numbers of DSFGs have so far been confirmed to be viralised, though the Spiderweb Galaxy structure may contain DSFGs in a viralised sub-halo \citep{Dannerbauer2014}.
If it is the virialisation or presence of an evolved intracluster medium that prevents or quenches DSFGs, it would suggest that none of the proto-cluster candidates detected here, and indeed none of the confirmed clusters containing significant numbers of DSFGs, are yet viralised or posses an evolved intracluster medium.
Finding clusters / proto-clusters of DSFGs, and determining particularly the viralisation and environmental state around and within them, may therefore be key to understanding the mechanisms behind the quenching of galaxies in different environments.

However, it should be stressed that the candidate proto-clusters detected in this work remain candidates, and not only is there a need to confirm that the DSFGs detected lie at the same redshift, but further work should be undertaken to confirm that there is also an optical/NIR overdensity at these positions, confirming that this is indeed a cluster rather than associated sources that come from looking down a filament.
Additionally, work should be done to characterise these clusters, particularly at what evolutionary stage they are at (i.e. have they virialised?).
Both the PHZ and this work provide complementary targets for these systems on scales around a few arcminutes, with the PHZ generally selecting the brightest and reddest, and this work selecting both fainter and warmer candidates.
Additional sources of cluster / proto-cluster candidates can come from selection directly on the \textit{Herschel} maps \citep{Valtchanov2013a}.


The diversity of DSFGs in clusters is further suggested by the difference we find with the PHZ; we only find four sources in common, none of which we identify as a candidate cluster.
When applying their flux and colour cuts directly on our catalogue, we only find 15 objects, 3 (20\%)  are cluster candidates (including the Bootes cluster candidates identified by \citet{Clements2014}), 3 (20\%) are cirrus, 4 (26\%) are local galaxies (UGC 09215, UGC 08017, NGC 5056 and CGCG 160-170), 1 (6.6\%) is a lens candidate (H-ATLAS J132426.9+284452), and 4 (26\%) we were not able to assign a identity. 
Of these 15, 11 are detected only in the ERCSC.

Similar work has also been undertaken by \citet{Baes2014} in the 84 deg$^2$ of the \textit{Herschel} Virgo Cluster Survey, where they find that most \textit{Planck} compact sources are dominated by local late type galaxies, with few sources being classified as galactic cirrus, spurious detections, and no sources classified as candidate clusters. 
This can largely be explained by the small areas examined; As Baes et al. note, they are directly examining a local cluster, whereas our total area surveyed is $\sim$ 10$\times$ larger across most of the accessible extragalactic sky.
Given also that they use the PCCS1, which both they and we note is devoid of sources without a bright local counterpart (see Table \ref{Table:Frac}), we therefore conclude that sky variance and use of PCCS1 can explain the apparently low numbers of cluster candidates in the \textit{Herschel} Virgo Cluster Survey.

\section{Conclusions}

\label{Sec:Conclusion}

Through a cross-match of the \textit{Planck} compact source catalogues, and 808.4 deg$^2$ of \textit{Herschel} fields from H-ATLAS, HerMES and Hers, we have identified 27 proto-clusters of DSFGs that are at least 3$\sigma$ overdense in either 250, 350 or 500 $\mu$m sources.
Additionally, we have identified, 192 local galaxies, 43 regions of galactic cirrus, 12 candidate lensed sources, 3 stars and 2 QSOs which also make up the \textit{Planck} compact source catalogues.
A further 61 sources we are unable to assign a classification, but many host a large number of \textit{Herschel} sources ($>2\sigma$ in the 250, 350 or 500 $\mu$m bands), and other have colours indicative of a high redshift origin. 
It is possible that many of these unassigned sources are also proto-clusters of DSFGs, though it is more difficult to rule out fluctuations in the number counts as an explanation.

We find there is significant differences between the three released versions of the catalogues, with the ERCSC hosting a larger fraction of candidate proto-clusters than the PCCS or PCCS.
We ascribe this to the filters used in the creation of the three catalogues, with the PCCS and PCCS2 Mexican-hat wavelet filter likely suppressing extended emission from both Galactic cirrus and candidate proto-clusters.

We verified that there is good agreement between \textit{Planck} and \textit{Herschel} aperture photometry for all sources, and further find that with the exception of Galactic cirrus, simply summing up the detected \textit{Herschel} source with $S_{350}$ or $S_{500} > 25.4$ mJy results in a good match between \textit{Planck} and \textit{Herschel} photometry. 

The \textit{Planck} colours of our proto-cluster candidates indicate that a selection criteria of $S_{857}/S_{545} < 2$ performs well for selecting out candidate proto-clusters.
However, we have also found a number of warmer proto-cluster candidates, which would be missed by such a selection, though we have shown this can be also explained by a significant contamination of low redshift $z < 1$ DSFGs.
The \textit{Herschel} colours of our sources indicate they all likely lie at z $> 2$, and the small scatter of points in the \textit{Herschel} colour-colour plots can indicate a physical cluster / proto-cluster, though the uncertainties are large.

We find a surface density of candidate proto-clusters of ($3.3 \pm 0.7) \times10^{-2}$ sources deg$^{-2}$, in good agreement with previous similar studies. 
Crossmatching our catalogue with the PHZ, we find only four matches, none of which we identify as a candidate proto-cluster.

Finally, we compare our results to simulations, finding both that our proto-clusters are a factor of 5 times brighter at 353 GHz than expected from simulations, even in the most conservative estimates, and that the amplitude of the three-point correlation function $Q$ likely evolves with $Q$ $\propto$ $\frac{1}{b}$.

Without redshift confirmation, there remains the possibility that none of these objects are physical clusters / proto-clusters. 
However, given the number we have found alongside other groups, if they are clusters / proto-clusters it is a challenge to explain how groups of $>20$ associated DSFGs exist, given their expected lifetimes of $\sim 100$ Myrs.
Such proto-clusters of DSFGs are being found from arcminute to arcsecond scales, yet we do not see this in the local Universe, indicating that star formation is quenched in clusters at low redshifts, but does take place in clusters / proto-clusters at higher redshifts, possibly due to a clusters virilisation state.
Since we do not know if these sources are virialisaed, further characterisation, particularly of the environment and state of virialisation, should be a key focus for follow up observations.
Given also that we expect DSFGs such as these to evolve into the brightest cluster members at the cores of galaxy clusters, they likely play a vital role in the earliest stages of cluster formation and evolution.

\section*{Acknowledgements}
We thank the anonymous referee for their insightful and useful comments.
Based on observations obtained with Planck (http://www.esa.int/Planck), an ESA science mission with instruments and contributions directly funded by ESA Member States, NASA, and Canada.
The Herschel-ATLAS is a project with Herschel, which is an ESA space observatory with science instruments provided by European-led Principal Investigator consortia and with important participation from NASA. The H-ATLAS website is \url{http://www.h-atlas.org/} U.S. participants in Herschel ATLAS acknowledge support provided by NASA through a contract issued from JPL.
This research has made use of data from HerMES project (\url{http://hermes.sussex.ac.uk/}). HerMES is a Herschel Key Programme utilising
Guaranteed Time from the SPIRE instrument team, ESAC scientists and a mission
scientist.
The HerMES data was accessed through the Herschel Database in Marseille
(HeDaM - http://hedam.lam.fr) operated by CeSAM and hosted by the Laboratoire
d'Astrophysique de Marseille.
HerMES DR3 was made possible through support of the Herschel Extragalactic
Legacy Project, HELP (http://herschel.sussex.ac.uk).
This work made extensive use of the Starlink Table/VOTable Processing Software, TOPCAT \citep{Taylor2005}.
GAMA is a joint European-Australasian project based around a spectroscopic campaign using the Anglo-Australian Telescope. The GAMA input catalogue is based on data taken from the Sloan Digital Sky Survey and the UKIRT Infrared Deep Sky Survey. Complementary imaging of the GAMA regions is being obtained by a number of independent survey programmes including GALEX MIS, VST KiDS, VISTA VIKING, WISE, Herschel-ATLAS, GMRT and ASKAP providing UV to radio coverage. GAMA is funded by the STFC (UK), the ARC (Australia), the AAO, and the participating institutions. The GAMA website is http://www.gama-survey.org/ . 
This research made use of Astropy, a community-developed core Python package for Astronomy (Astropy Collaboration, 2013) \citep{Astropy2013}.
DLC and JG acknowledge support from STFC, in part through grant numbers ST/N000838/1 and ST/K001051/1.
H.D. acknowledges financial support from the Spanish Ministry of Economy and Competitiveness (MINECO) under the 2014 Ram\'oy Cajal program MINECO RYC-2014-15686.
This research has made use of the NASA/IPAC Extragalactic Database (NED) which is operated by the Jet Propulsion Laboratory, California Institute of Technology, under contract with the National Aeronautics and Space Administration. 
This research has made use of the SIMBAD database,
operated at CDS, Strasbourg, France.
M.J.M.~acknowledges the support of  the National Science Centre, Poland through the POLONEZ grant 2015/19/P/ST9/04010.
This project has received funding from the European Union's Horizon 2020 research and innovation programme under the Marie Sk{\l}odowska-Curie grant agreement No. 665778.
LD acknowledge support from the European Research Council in the form of Advanced Investigator grant COSMICISM and also the Consolidator Grant CosmicDust.




\bibliographystyle{mnras}
\bibliography{./library} 

\begin{thebibliography}{}

\bibitem[Ade et~al., 2014]{Ade2014}
Ade, P. A.~R., Aghanim, N., Arg{\"{u}}eso, F., Armitage-Caplan, C., Arnaud, M.,
  Ashdown, M., Atrio-Barandela, F., Aumont, J., Baccigalupi, C., Banday, A.~J.,
  Barreiro, R.~B., Bartlett, J.~G., Battaner, E., Beelen, A., Benabed, K.,
  Beno{\^{i}}t, A., Benoit-L{\'{e}}vy, A., Bernard, J. J.-P., Bersanelli, M.,
  Bielewicz, P., Bobin, J., Bock, J.~J., Bonaldi, A., Bonavera, L., Bond,
  J.~R., Borrill, J., Bouchet, F.~R., Bridges, M., Bucher, M., Burigana, C.,
  Butler, R.~C., Cardoso, J.-F.~J., Carvalho, P., Catalano, A., Challinor, A.,
  Chamballu, A., Chen, X., Chiang, H.~C., Chiang, L. L.-Y., Christensen, P.~R.,
  Church, S., Clemens, M., Clements, D.~L., Colombi, S., Colombo, L. P.~L.,
  Couchot, F., Coulais, A., Crill, B.~P., Curto, A., Cuttaia, F., Danese, L.,
  Davies, R.~D., Davis, R.~J., de~Bernardis, P., de~Rosa, A., de~Zotti, G.,
  Delabrouille, J., Delouis, J. J.-M., D{\'{e}}sert, F. F.-X., Dickinson, C.,
  Diego, J.~M., Dole, H., Donzelli, S., Dor{\'{e}}, O., Douspis, M., Dupac, X.,
  Efstathiou, G., En{\ss}lin, T.~A., Eriksen, H.~K., Finelli, F., Forni, O.,
  Frailis, M., Franceschi, E., Galeotta, S., Ganga, K., Giard, M., Giardino,
  G., Giraud-H{\'{e}}raud, Y., Gonz{\'{a}}lez-Nuevo, J., G{\'{o}}rski, K.~M.,
  Gratton, S., Gregorio, A., Gruppuso, A., Hansen, F.~K., Hanson, D., Harrison,
  D.~L., Henrot-Versill{\'{e}}, S., Hern{\'{a}}ndez-Monteagudo, C., Herranz,
  D., Hildebrandt, S.~R., Hivon, E., Hobson, M., Holmes, W.~A., Hornstrup, A.,
  Hovest, W., Huffenberger, K.~M., Jaffe, A.~H., Jaffe, T.~R., Jones, W.~C.,
  Juvela, M., Keih{\"{a}}nen, E., Keskitalo, R., Kisner, T.~S., Kneissl, R.,
  Knoche, J., Knox, L., Kunz, M., Kurki-Suonio, H., Lagache, G.,
  L{\"{a}}hteenm{\"{a}}ki, A., Lamarre, J.-M.~M., Lasenby, A., Laureijs, R.~J.,
  Lawrence, C.~R., Leahy, J.~P., Leonardi, R., Le{\'{o}}n-Tavares, J., Leroy,
  C., Lesgourgues, J., Liguori, M., Lilje, P.~B., Linden-V{\o}rnle, M.,
  L{\'{o}}pez-Caniego, M., Lubin, P.~M., Mac{\'{i}}as-P{\'{e}}rez, J.~F.,
  Maffei, B., Maino, D., Mandolesi, N., Maris, M., Marshall, D.~J., Martin,
  P.~G., Mart{\'{i}}nez-Gonz{\'{a}}lez, E., Masi, S., Massardi, M., Matarrese,
  S., Matthai, F., Mazzotta, P., McGehee, P., Meinhold, P.~R., Melchiorri, A.,
  Mendes, L., Mennella, A., Migliaccio, M., Mitra, S., Miville-Desch{\^{e}}nes,
  M.-A.~A., Moneti, A., Montier, L., Morgante, G., Mortlock, D., Munshi, D.,
  Murphy, J.~A., Naselsky, P., Nati, F., Natoli, P., Negrello, M., Netterfield,
  C.~B., N{\o}rgaard-Nielsen, H.~U., Noviello, F., Novikov, D., Novikov, I.,
  O'Dwyer, I.~J., Osborne, S., Oxborrow, C.~A., Paci, F., Pagano, L., Pajot,
  F., Paladini, R., Paoletti, D., Partridge, B., Pasian, F., Patanchon, G.,
  Pearson, T.~J., Perdereau, O., Perotto, L., Perrotta, F., Piacentini, F.,
  Piat, M., Pierpaoli, E., Pietrobon, D., Plaszczynski, S., Pointecouteau, E.,
  Polenta, G., Ponthieu, N., Popa, L., Poutanen, T., Pratt, G.~W.,
  Pr{\'{e}}zeau, G., Prunet, S., Puget, J.-L.~L., Rachen, J.~P., Reach, W.~T.,
  Rebolo, R., Reinecke, M., Remazeilles, M., Renault, C., Ricciardi, S.,
  Riller, T., Ristorcelli, I., Rocha, G., Rosset, C., Roudier, G.,
  Rowan-Robinson, M., Rubi{\~{n}}o-Mart{\'{i}}n, J.~A., Rusholme, B., Sandri,
  M., Santos, D., Savini, G., Scott, D., Seiffert, M.~D., Shellard, E. P.~S.,
  Spencer, L.~D., Starck, J.-L.~L., Stolyarov, V., Stompor, R., Sudiwala, R.,
  Sunyaev, R., Sureau, F., Sutton, D., Suur-Uski, A.-S.~S., Sygnet, J.-F.~F.,
  Tauber, J.~A., Tavagnacco, D., Terenzi, L., Toffolatti, L., Tomasi, M.,
  Tristram, M., Tucci, M., Tuovinen, J., T{\"{u}}rler, M., Umana, G.,
  Valenziano, L., Valiviita, J., {Van Tent}, B., Varis, J., Vielva, P., Villa,
  F., Vittorio, N., Wade, L.~A., Walter, B., Wandelt, B.~D., Yvon, D., Zacchei,
  A., Zonca, A., {Armitage Caplan}, C., Arnaud, M., Ashdown, M., {Atrio
  Barandela}, F., Aumont, J., Baccigalupi, C., Banday, A.~J., Barreiro, R.~B.,
  Bartlett, J.~G., Battaner, E., Beelen, A., Benabed, K., Beno{\^{i}}t, A.,
  {Benoit L{\'{e}}vy}, A., Bernard, J. J.-P., Bersanelli, M., Bielewicz, P.,
  Bobin, J., Bock, J.~J., Bonaldi, A., Bonavera, L., Bond, J.~R., Borrill, J.,
  Bouchet, F.~R., Bridges, M., Bucher, M., Burigana, C., Butler, R.~C.,
  Cardoso, J.-F.~J., Carvalho, P., Catalano, A., Challinor, A., Chamballu, A.,
  Chen, X., Chiang, H.~C., Chiang, L. L.-Y., Christensen, P.~R., Church, S.,
  Clemens, M., Clements, D.~L., Colombi, S., Colombo, L. P.~L., Couchot, F.,
  Coulais, A., Crill, B.~P., Curto, A., Cuttaia, F., Danese, L., Davies, R.~D.,
  Davis, R.~J., de~Bernardis, P., de~Rosa, A., de~Zotti, G., Delabrouille, J.,
  Delouis, J. J.-M., D{\'{e}}sert, F. F.-X., Dickinson, C., Diego, J.~M., Dole,
  H., Donzelli, S., Dor{\'{e}}, O., Douspis, M., Dupac, X., Efstathiou, G.,
  En{\ss}lin, T.~A., Eriksen, H.~K., Finelli, F., Forni, O., Frailis, M.,
  Franceschi, E., Galeotta, S., Ganga, K., Giard, M., Giardino, G., {Giraud
  H{\'{e}}raud}, Y., {Gonz{\'{a}}lez Nuevo}, J., G{\'{o}}rski, K.~M., Gratton,
  S., Gregorio, A., Gruppuso, A., Hansen, F.~K., Hanson, D., Harrison, D.~L.,
  {Henrot Versill{\'{e}}}, S., {Hern{\'{a}}ndez Monteagudo}, C. (2014).
\newblock {Planck 2013 results. XXVIII. The Planck Catalogue of Compact
  Sources}.
\newblock {\em Astronomy {\&} Astrophysics}, 571:A28.

\bibitem[Ade et~al., 2011]{Ade2011}
Ade, P. A.~R., Aghanim, N., Arnaud, M., Ashdown, M., Aumont, J., Baccigalupi,
  C., Balbi, A., Banday, A.~J., Barreiro, R.~B., Bartlett, J.~G., Battaner, E.,
  Benabed, K., Beno{\^{i}}t, A., Bernard, J.-P., Bersanelli, M., Bhatia, R.,
  Bonaldi, A., Bonavera, L., Bond, J.~R., Borrill, J., Bouchet, F.~R., Bucher,
  M., Burigana, C., Butler, R.~C., Cabella, P., Cantalupo, C.~M., Cappellini,
  B., Cardoso, J.-F., Carvalho, P., Catalano, A., Cay{\'{o}}n, L., Challinor,
  A., Chamballu, A., Chary, R.-R., Chen, X., Chiang, L.-Y., Chiang, C.,
  Christensen, P.~R., Clements, D.~L., Colombi, S., Couchot, F., Coulais, A.,
  Crill, B.~P., Cuttaia, F., Danese, L., Davis, R.~J., de~Bernardis, P.,
  de~Rosa, A., de~Zotti, G., Delabrouille, J., Delouis, J.-M., D{\'{e}}sert,
  F.-X., Dickinson, C., Diego, J.~M., Dolag, K., Dole, H., Donzelli, S.,
  Dor{\'{e}}, O., D{\"{o}}rl, U., Douspis, M., Dupac, X., Efstathiou, G.,
  En{\ss}lin, T.~A., Eriksen, H.~K., Finelli, F., Forni, O., Fosalba, P.,
  Frailis, M., Franceschi, E., Galeotta, S., Ganga, K., Giard, M.,
  Giraud-H{\'{e}}raud, Y., Gonz{\'{a}}lez-Nuevo, J., G{\'{o}}rski, K.~M.,
  Gratton, S., Gregorio, A., Gruppuso, A., Haissinski, J., Hansen, F.~K.,
  Harrison, D., Helou, G., Henrot-Versill{\'{e}}, S.,
  Hern{\'{a}}ndez-Monteagudo, C., Herranz, D., Hildebrandt, S.~R., Hivon, E.,
  Hobson, M., Holmes, W.~A., Hornstrup, A., Hovest, W., Hoyland, R.~J.,
  Huffenberger, K.~M., Huynh, M., Jaffe, A.~H., Jones, W.~C., Juvela, M.,
  Keih{\"{a}}nen, E., Keskitalo, R., Kisner, T.~S., Kneissl, R., Knox, L.,
  Kurki-Suonio, H., Lagache, G., L{\"{a}}hteenm{\"{a}}ki, A., Lamarre, J.-M.,
  Lasenby, A., Laureijs, R.~J., Lawrence, C.~R., Leach, S., Leahy, J.~P.,
  Leonardi, R., Le{\'{o}}n-Tavares, J., Leroy, C., Lilje, P.~B.,
  Linden-V{\o}rnle, M., L{\'{o}}pez-Caniego, M., Lubin, P.~M.,
  Mac{\'{i}}as-P{\'{e}}rez, J.~F., MacTavish, C.~J., Maffei, B., Maggio, G.,
  Maino, D., Mandolesi, N., Mann, R., Maris, M., Marleau, F., Marshall, D.~J.,
  Mart{\'{i}}nez-Gonz{\'{a}}lez, E., Masi, S., Massardi, M., Matarrese, S.,
  Matthai, F., Mazzotta, P., McGehee, P., Meinhold, P.~R., Melchiorri, A.,
  Melin, J.-B., Mendes, L., Mennella, A., Mitra, S., Miville-Desch{\^{e}}nes,
  M.-A., Moneti, A., Montier, L., Morgante, G., Mortlock, D., Munshi, D.,
  Murphy, A., Naselsky, P., Natoli, P., Netterfield, C.~B.,
  N{\o}rgaard-Nielsen, H.~U., Noviello, F., Novikov, D., Novikov, I., O'Dwyer,
  I.~J., Osborne, S., Pajot, F., Paladini, R., Partridge, B., Pasian, F.,
  Patanchon, G., Pearson, T.~J., Perdereau, O., Perotto, L., Perrotta, F.,
  Piacentini, F., Piat, M., Piffaretti, R., Plaszczynski, S., Platania, P.,
  Pointecouteau, E., Polenta, G., Ponthieu, N., Poutanen, T., Pratt, G.~W.,
  Pr{\'{e}}zeau, G., Prunet, S., Puget, J.-L., Rachen, J.~P., Reach, W.~T.,
  Rebolo, R., Reinecke, M., Renault, C., Ricciardi, S., Riller, T.,
  Ristorcelli, I., Rocha, G., Rosset, C., Rowan-Robinson, M.,
  Rubi{\~{n}}o-Mart{\'{i}}n, J.~A., Rusholme, B., Sajina, A., Sandri, M.,
  Santos, D., Savini, G., Schaefer, B.~M., Scott, D., Seiffert, M.~D.,
  Shellard, P., Smoot, G.~F., Starck, J.-L., Stivoli, F., Stolyarov, V.,
  Sudiwala, R., Sunyaev, R., Sygnet, J.-F., Tauber, J.~A., Tavagnacco, D.,
  Terenzi, L., Toffolatti, L., Tomasi, M., Torre, J.-P., Tristram, M.,
  Tuovinen, J., T{\"{u}}rler, M., Umana, G., Valenziano, L., Valiviita, J.,
  Varis, J., Vielva, P., Villa, F., Vittorio, N., Wade, L.~A., Wandelt, B.~D.,
  White, S. D.~M., Wilkinson, A., Yvon, D., Zacchei, A., and Zonca, A. (2011).
\newblock {Planck early results. VII. The Early Release Compact Source
  Catalogue}.
\newblock {\em Astronomy {\&} Astrophysics}, 536:A7.

\bibitem[Clements et~al., 2014]{Clements2014}
Clements, D.~L., Braglia, F.~G., Hyde, A.~K., Perez-Fournon, I., Bock, J.,
  Cava, A., Chapman, S., Conley, A., Cooray, A., Farrah, D., Solares, E. A.~G.,
  Marchetti, L., Marsden, G., Oliver, S.~J., Roseboom, I.~G., Schulz, B.,
  Smith, A.~J., Vaccari, M., Vieira, J., Viero, M., Wang, L., Wardlow, J.,
  Zemcov, M., and de~Zotti, G. (2014).
\newblock {Herschel Multitiered Extragalactic Survey: clusters of dusty
  galaxies uncovered by Herschel and Planck}.
\newblock {\em Monthly Notices of the Royal Astronomical Society},
  439(2):1193--1211.

\bibitem[Herranz et~al., 2013]{Herranz2013}
Herranz, D., Gonz{\'{a}}lez-Nuevo, J., Clements, D.~L., {De Zotti}, G.,
  Lopez-Caniego, M., Lapi, A., Rodighiero, G., Danese, L., Fu, H., Cooray, A.,
  Baes, M., Bendo, G.~J., Bonavera, L., Carrera, F.~J., Dole, H., Eales, S.,
  Ivison, R.~J., Jarvis, M., Lagache, G., Massardi, M., Michalowski, M.~J.,
  Negrello, M., Rigby, E., Scott, D., Valiante, E., Valtchanov, I., {Van der
  Werf}, P., Auld, R., Buttiglione, S., Dariush, A., Dunne, L., Hopwood, R.,
  Hoyos, C., Ibar, E., and Maddox, S. (2013).
\newblock {Herschel ATLAS: Planck sources in the phase 1 fields}.
\newblock {\em Astronomy {\&} Astrophysics}, 549.

\bibitem[{Miville Deschenes} and Lagache, 2005]{MivilleDeschenes2005}
{Miville Deschenes}, M.~A. and Lagache, G. (2005).
\newblock {IRIS: A New Generation of IRAS Maps}.
\newblock {\em The Astrophysical Journal Supplement Series}, 157(2):302:323.

\bibitem[Negrello et~al., 2005]{Negrello2005}
Negrello, M., Gonzalez-Nuevo, J., Magliocchetti, M., Moscardini, L., {De
  Zotti}, G., Toffolatti, L., and Danese, L. (2005).
\newblock {Effect of clustering on extragalactic source counts with
  low-resolution instruments}.
\newblock {\em Monthly Notices of the Royal Astronomical Society},
  358(3):869--874.

\bibitem[{Planck Collaboration} et~al., 2015]{PlanckCollaboration2015b}
{Planck Collaboration}, Ade, P. A.~R., Aghanim, N., Arg{\"{u}}eso, F., Arnaud,
  M., Ashdown, M., Aumont, J., Baccigalupi, C., Banday, A.~J., Barreiro, R.~B.,
  Bartolo, N., Battaner, E., Beichman, C., Benabed, K., Beno{\^{i}}t, A.,
  Benoit-L{\'{e}}vy, A., Bernard, J.~P., Bersanelli, M., Bielewicz, P., Bock,
  J.~J., B{\"{o}}hringer, H., Bonaldi, A., Bonavera, L., Bond, J.~R., Borrill,
  J., Bouchet, F.~R., Boulanger, F., Bucher, M., Burigana, C., Butler, R.~C.,
  Calabrese, E., Cardoso, J.~F., Carvalho, P., Catalano, A., Challinor, A.,
  Chamballu, A., Chary, R.~R., Chiang, H.~C., Christensen, P.~R., Clemens, M.,
  Clements, D.~L., Colombi, S., Colombo, L. P.~L., Combet, C., Couchot, F.,
  Coulais, A., Crill, B.~P., Curto, A., Cuttaia, F., Danese, L., Davies, R.~D.,
  Davis, R.~J., de~Bernardis, P., de~Rosa, A., de~Zotti, G., Delabrouille, J.,
  D{\'{e}}sert, F.~X., Dickinson, C., Diego, J.~M., Dole, H., Donzelli, S.,
  Dor{\'{e}}, O., Douspis, M., Ducout, A., Dupac, X., Efstathiou, G., Elsner,
  F., En{\ss}lin, T.~A., Eriksen, H.~K., Falgarone, E., Fergusson, J., Finelli,
  F., Forni, O., Frailis, M., Fraisse, A.~A., Franceschi, E., Frejsel, A.,
  Galeotta, S., Galli, S., Ganga, K., Giard, M., Giraud-H{\'{e}}raud, Y.,
  Gjerl{\o}w, E., Gonz{\'{a}}lez-Nuevo, J., G{\'{o}}rski, K.~M., Gratton, S.,
  Gregorio, A., Gruppuso, A., Gudmundsson, J.~E., Hansen, F.~K., Hanson, D.,
  Harrison, D.~L., Helou, G., Henrot-Versill{\'{e}}, S.,
  Hern{\'{a}}ndez-Monteagudo, C., Herranz, D., Hildebrandt, S.~R., Hivon, E.,
  Hobson, M., Holmes, W.~A., Hornstrup, A., Hovest, W., Huffenberger, K.~M.,
  Hurier, G., Jaffe, A.~H., Jaffe, T.~R., Jones, W.~C., Juvela, M.,
  Keih{\"{a}}nen, E., Keskitalo, R., Kisner, T.~S., Kneissl, R., Knoche, J.,
  Kunz, M., Kurki-Suonio, H., Lagache, G., L{\"{a}}hteenm{\"{a}}ki, A.,
  Lamarre, J.~M., Lasenby, A., Lattanzi, M., Lawrence, C.~R., Leahy, J.~P.,
  Leonardi, R., Le{\'{o}}n-Tavares, J., Lesgourgues, J., Levrier, F., Liguori,
  M., Lilje, P.~B., Linden-V{\o}rnle, M., L{\'{o}}pez-Caniego, M., Lubin,
  P.~M., Mac{\'{i}}as-P{\'{e}}rez, J.~F., Maggio, G., Maino, D., Mandolesi, N.,
  Mangilli, A., Maris, M., Marshall, D.~J., Martin, P.~G.,
  Mart{\'{i}}nez-Gonz{\'{a}}lez, E., Masi, S., Matarrese, S., McGehee, P.,
  Meinhold, P.~R., Melchiorri, A., Mendes, L., Mennella, A., Migliaccio, M.,
  Mitra, S., Miville-Desch{\^{e}}nes, M.~A., Moneti, A., Montier, L., Morgante,
  G., Mortlock, D., Moss, A., Munshi, D., Murphy, J.~A., Naselsky, P., Nati,
  F., Natoli, P., Negrello, M., Netterfield, C.~B., N{\o}rgaard-Nielsen, H.~U.,
  Noviello, F., Novikov, D., Novikov, I., Oxborrow, C.~A., Paci, F., Pagano,
  L., Pajot, F., Paladini, R., Paoletti, D., Partridge, B., Pasian, F.,
  Patanchon, G., Pearson, T.~J., Perdereau, O., Perotto, L., Perrotta, F.,
  Pettorino, V., Piacentini, F., Piat, M., Pierpaoli, E., Pietrobon, D.,
  Plaszczynski, S., Pointecouteau, E., Polenta, G., Pratt, G.~W.,
  Pr{\'{e}}zeau, G., Prunet, S., Puget, J.~L., Rachen, J.~P., Reach, W.~T.,
  Rebolo, R., Reinecke, M., Remazeilles, M., Renault, C., Renzi, A.,
  Ristorcelli, I., Rocha, G., Rosset, C., Rossetti, M., Roudier, G.,
  Rowan-Robinson, M., Rubi{\~{n}}o-Mart{\'{i}}n, J.~A., Rusholme, B., Sandri,
  M., Sanghera, H.~S., Santos, D., Savelainen, M., Savini, G., Scott, D.,
  Seiffert, M.~D., Shellard, E. P.~S., Spencer, L.~D., Stolyarov, V., Sudiwala,
  R., Sunyaev, R., Sutton, D., Suur-Uski, A.~S., Sygnet, J.~F., Tauber, J.~A.,
  Terenzi, L., Toffolatti, L., Tomasi, M., Tornikoski, M., Tristram, M., Tucci,
  M., Tuovinen, J., T{\"{u}}rler, M., Umana, G., Valenziano, L., Valiviita, J.,
  {Van Tent}, B., Vielva, P., Villa, F., Wade, L.~A., Walter, B., Wandelt,
  B.~D., Wehus, I.~K., Yvon, D., Zacchei, A., and Zonca, A. (2015).
\newblock {Planck 2015 results. XXVI. The Second Planck Catalogue of Compact
  Sources}.
\newblock {\em in Press 2015}.

\bibitem[Viero et~al., 2014]{Viero2014}
Viero, M.~P., Asboth, V., Roseboom, I.~G., Moncelsi, L., Marsden, G., {Mentuch
  Cooper}, E., Zemcov, M., Addison, G., Baker, A.~J., Beelen, A., Bock, J.,
  Bridge, C., Conley, A., Devlin, M.~J., Dor{\'{e}}, O., Farrah, D.,
  Finkelstein, S., Font-Ribera, A., Geach, J.~E., Gebhardt, K., Gill, A.,
  Glenn, J., Hajian, A., Halpern, M., Jogee, S., Kurczynski, P., Lapi, A.,
  Negrello, M., Oliver, S.~J., Papovich, C., Quadri, R., Ross, N., Scott, D.,
  Schulz, B., Somerville, R., Spergel, D.~N., Vieira, J.~D., Wang, L., and
  Wechsler, R. (2014).
\newblock {THE HERSCHEL STRIPE 82 SURVEY (HerS): MAPS AND EARLY CATALOG}.
\newblock {\em The Astrophysical Journal Supplement Series}, 210(2):22.

\end{thebibliography}


\begin{thebibliography}{}
\makeatletter
\relax
\def\mn@urlcharsother{\let\do\@makeother \do\$\do\&\do\#\do\^\do\_\do\%\do\~}
\def\mn@doi{\begingroup\mn@urlcharsother \@ifnextchar [ {\mn@doi@}
  {\mn@doi@[]}}
\def\mn@doi@[#1]#2{\def\@tempa{#1}\ifx\@tempa\@empty \href
  {http://dx.doi.org/#2} {doi:#2}\else \href {http://dx.doi.org/#2} {#1}\fi
  \endgroup}
\def\mn@eprint#1#2{\mn@eprint@#1:#2::\@nil}
\def\mn@eprint@arXiv#1{\href {http://arxiv.org/abs/#1} {{\tt arXiv:#1}}}
\def\mn@eprint@dblp#1{\href {http://dblp.uni-trier.de/rec/bibtex/#1.xml}
  {dblp:#1}}
\def\mn@eprint@#1:#2:#3:#4\@nil{\def\@tempa {#1}\def\@tempb {#2}\def\@tempc
  {#3}\ifx \@tempc \@empty \let \@tempc \@tempb \let \@tempb \@tempa \fi \ifx
  \@tempb \@empty \def\@tempb {arXiv}\fi \@ifundefined
  {mn@eprint@\@tempb}{\@tempb:\@tempc}{\expandafter \expandafter \csname
  mn@eprint@\@tempb\endcsname \expandafter{\@tempc}}}

\bibitem[\protect\citeauthoryear{Acke et~al.,}{Acke et~al.}{2012}]{Acke2012}
Acke B.,  et~al., 2012, \mn@doi [Astronomy {\&} Astrophysics, Volume 540,
  id.A125, 9 pp.] {10.1051/0004-6361/201118581}, 540

\bibitem[\protect\citeauthoryear{Asboth et~al.,}{Asboth
  et~al.}{2016}]{Asboth2016}
Asboth V.,  et~al., 2016, \mn@doi [Monthly Notices of the Royal Astronomical
  Society, Volume 462, Issue 2, p.1989-2000] {10.1093/mnras/stw1769}, 462, 1989

\bibitem[\protect\citeauthoryear{Baes et~al.,}{Baes et~al.}{2014}]{Baes2014}
Baes M.,  et~al., 2014, \mn@doi [Astronomy {\&} Astrophysics, Volume 562,
  id.A106, 16 pp.] {10.1051/0004-6361/201322384}, 562

\bibitem[\protect\citeauthoryear{Bertin \& Arnouts}{Bertin \&
  Arnouts}{1996}]{Bertin1996}
Bertin E.,  Arnouts S.,  1996, \mn@doi [Astronomy and Astrophysics Supplement
  Series] {10.1051/aas:1996164}, 117, 393

\bibitem[\protect\citeauthoryear{Bertincourt et~al.,}{Bertincourt
  et~al.}{2016}]{Bertincourt2016}
Bertincourt B.,  et~al., 2016, \mn@doi [Astronomy {\&} Astrophysics]
  {10.1051/0004-6361/201527313}, 588, A107

\bibitem[\protect\citeauthoryear{Blain}{Blain}{2002}]{Blain2002}
Blain A.,  2002, \mn@doi [Physics Reports] {10.1016/S0370-1573(02)00134-5},
  369, 111

\bibitem[\protect\citeauthoryear{Blain, Chapman, Smail  \& Ivison}{Blain
  et~al.}{2004}]{Blain2004}
Blain A.~W.,  Chapman S.~C.,  Smail I.,   Ivison R.,  2004, \mn@doi [The
  Astrophysical Journal] {10.1086/422026}, 611, 52

\bibitem[\protect\citeauthoryear{Bussmann et~al.,}{Bussmann
  et~al.}{2013}]{Bussmann2013}
Bussmann R.~S.,  et~al., 2013, \mn@doi [The Astrophysical Journal, Volume 779,
  Issue 1, article id. 25, 26 pp. (2013).] {10.1088/0004-637X/779/1/25}, 779

\bibitem[\protect\citeauthoryear{Cai et~al.,}{Cai et~al.}{2013}]{Cai2013}
Cai Z.-Y.,  et~al., 2013, \mn@doi [The Astrophysical Journal, Volume 768, Issue
  1, article id. 21, 24 pp. (2013).] {10.1088/0004-637X/768/1/21}, 768

\bibitem[\protect\citeauthoryear{Canameras et~al.,}{Canameras
  et~al.}{2015}]{Canameras2015}
Canameras R.,  et~al., 2015, \mn@doi [Astronomy {\&} Astrophysics, Volume 581,
  id.A105, 18 pp.] {10.1051/0004-6361/201425128}, 581

\bibitem[\protect\citeauthoryear{Capak et~al.,}{Capak et~al.}{2011}]{Capak2011}
Capak P.~L.,  et~al., 2011, \mn@doi [Nature] {10.1038/nature09681}, 470, 233

\bibitem[\protect\citeauthoryear{Casey}{Casey}{2016}]{Casey2016}
Casey C.~M.,  2016, \mn@doi [The Astrophysical Journal, Volume 824, Issue 1,
  article id. 36, pp. (2016).] {10.3847/0004-637X/824/1/36}, 824

\bibitem[\protect\citeauthoryear{Casey, Narayanan  \& Cooray}{Casey
  et~al.}{2014}]{Casey2014}
Casey C.~M.,  Narayanan D.,   Cooray A.,  2014, \mn@doi [Physics Reports]
  {10.1016/j.physrep.2014.02.009}, 541, 45161

\bibitem[\protect\citeauthoryear{Casey et~al.,}{Casey et~al.}{2015}]{Casey2015}
Casey C.~M.,  et~al., 2015, \mn@doi [The Astrophysical Journal]
  {10.1088/2041-8205/808/2/L33}, 808, L33

\bibitem[\protect\citeauthoryear{Chapman, Blain, Ibata, Ivison, Smail  \&
  Morrison}{Chapman et~al.}{2009}]{Chapman2009}
Chapman S.~C.,  Blain A.,  Ibata R.,  Ivison R.~J.,  Smail I.,   Morrison G.,
  2009, \mn@doi [The Astrophysical Journal] {10.1088/0004-637X/691/1/560}, 691,
  560

\bibitem[\protect\citeauthoryear{Chiang, Overzier  \& Gebhardt}{Chiang
  et~al.}{2013}]{Chiang2013}
Chiang Y.-K.,  Overzier R.,   Gebhardt K.,  2013, \mn@doi [The Astrophysical
  Journal, Volume 779, Issue 2, article id. 127, 16 pp. (2013).]
  {10.1088/0004-637X/779/2/127}, 779

\bibitem[\protect\citeauthoryear{Clements et~al.,}{Clements
  et~al.}{2010}]{Clements2010}
Clements D.~L.,  et~al., 2010, \mn@doi [Astronomy and Astrophysics]
  {10.1051/0004-6361/201014581}, 518, L8

\bibitem[\protect\citeauthoryear{Clements et~al.,}{Clements
  et~al.}{2014}]{Clements2014}
Clements D.~L.,  et~al., 2014, \mn@doi [Monthly Notices of the Royal
  Astronomical Society] {10.1093/mnras/stt2253}, 439, 1193

\bibitem[\protect\citeauthoryear{Clements et~al.,}{Clements
  et~al.}{2016}]{Clements2016}
Clements D.~L.,  et~al., 2016, \mn@doi [Monthly Notices of the Royal
  Astronomical Society] {10.1093/mnras/stw1224}, 461

\bibitem[\protect\citeauthoryear{Cowley, Lacey, Baugh  \& Cole}{Cowley
  et~al.}{2014}]{Cowley2014}
Cowley W.~I.,  Lacey C.~G.,  Baugh C.~M.,   Cole S.,  2014, \mn@doi [Monthly
  Notices of the Royal Astronomical Society] {10.1093/mnras/stu2179}, 446, 1784

\bibitem[\protect\citeauthoryear{Daddi et~al.,}{Daddi et~al.}{2008}]{Daddi2009}
Daddi E.,  et~al., 2008, \mn@doi [The Astrophysical Journal, Volume 694, Issue
  2, pp. 1517-1538 (2009).] {10.1088/0004-637X/694/2/1517}, 694, 1517

\bibitem[\protect\citeauthoryear{Dannerbauer et~al.,}{Dannerbauer
  et~al.}{2014}]{Dannerbauer2014}
Dannerbauer H.,  et~al., 2014, \mn@doi [Astronomy {\&} Astrophysics]
  {10.1051/0004-6361/201423771}, 570, A55

\bibitem[\protect\citeauthoryear{Dannerbauer et~al.,}{Dannerbauer
  et~al.}{2017}]{Dannerbauer2017}
Dannerbauer H.,  et~al., 2017, \mn@doi [eprint arXiv:1701.05250]
  {10.1051/0004-6361/201730449}

\bibitem[\protect\citeauthoryear{Darvish, Mobasher, Sobral, Rettura, Scoville,
  Faisst  \& Capak}{Darvish et~al.}{2016}]{Darvish2016}
Darvish B.,  Mobasher B.,  Sobral D.,  Rettura A.,  Scoville N.,  Faisst A.,
  Capak P.,  2016, \mn@doi [The Astrophysical Journal, Volume 825, Issue 2,
  article id. 113, 16 pp. (2016).] {10.3847/0004-637X/825/2/113}, 825

\bibitem[\protect\citeauthoryear{Donley, Rieke, Perez-Gonzalez, Rigby  \&
  Alonso-Herrero}{Donley et~al.}{2007}]{Donley2007}
Donley J.~L.,  Rieke G.~H.,  Perez-Gonzalez P.~G.,  Rigby J.~R.,
  Alonso-Herrero A.,  2007, \mn@doi [The Astrophysical Journal, Volume 660,
  Issue 1, pp. 167-190.] {10.1086/512798}, 660, 167

\bibitem[\protect\citeauthoryear{Dowell et~al.,}{Dowell
  et~al.}{2014}]{Dowell2014}
Dowell C.~D.,  et~al., 2014, \mn@doi [The Astrophysical Journal]
  {10.1088/0004-637X/780/1/75}, 780, 75

\bibitem[\protect\citeauthoryear{Driver et~al.,}{Driver
  et~al.}{2011}]{Driver2011}
Driver S.~P.,  et~al., 2011, \mn@doi [Monthly Notices of the Royal Astronomical
  Society] {10.1111/j.1365-2966.2010.18188.x}, 413, 971

\bibitem[\protect\citeauthoryear{Eales et~al.,}{Eales et~al.}{2010}]{Eales2010}
Eales S.,  et~al., 2010, \mn@doi [Publications of the Astronomical Society of
  the Pacific] {10.1086/653086}, 122, 499

\bibitem[\protect\citeauthoryear{Eisenhardt et~al.,}{Eisenhardt
  et~al.}{2008}]{Eisenhardt2008}
Eisenhardt P. R.~M.,  et~al., 2008, \mn@doi [The Astrophysical Journal]
  {10.1086/590105}, 684, 905

\bibitem[\protect\citeauthoryear{Emonts et~al.,}{Emonts
  et~al.}{2016}]{Emonts2016}
Emonts B. H.~C.,  et~al., 2016, \mn@doi [Science, Volume 354, Issue 6316, pp.
  1128-1130 (2016).] {10.1126/science.aag0512}, 354, 1128

\bibitem[\protect\citeauthoryear{Falgarone et~al.,}{Falgarone
  et~al.}{2017}]{Falgarone2017}
Falgarone E.,  et~al., 2017, \mn@doi [Nature, Volume 548, Issue 7668, pp.
  430-433 (2017).] {10.1038/nature23298}, 548, 430

\bibitem[\protect\citeauthoryear{Flores-Cacho et~al.,}{Flores-Cacho
  et~al.}{2016}]{Flores-Cacho2016}
Flores-Cacho I.,  et~al., 2016, \mn@doi [Astronomy {\&} Astrophysics]
  {10.1051/0004-6361/201425226}, 585, A54

\bibitem[\protect\citeauthoryear{Fu et~al.,}{Fu et~al.}{2012}]{Fu2012}
Fu H.,  et~al., 2012, \mn@doi [The Astrophysical Journal]
  {10.1088/0004-637X/753/2/134}, 753, 134

\bibitem[\protect\citeauthoryear{George et~al.,}{George
  et~al.}{2013}]{George2013}
George R.~D.,  et~al., 2013, \mn@doi [Monthly Notices of the Royal Astronomical
  Society: Letters, Volume 436, Issue 1, p.L99-L103] {10.1093/mnrasl/slt122},
  436, L99

\bibitem[\protect\citeauthoryear{Granato, {De Zotti}, Silva, Bressan  \&
  Danese}{Granato et~al.}{2004}]{Granato2004}
Granato G.~L.,  {De Zotti} G.,  Silva L.,  Bressan A.,   Danese L.,  2004,
  \mn@doi [The Astrophysical Journal] {10.1086/379875}, 600, 580

\bibitem[\protect\citeauthoryear{Granato, Ragone-Figueroa, Dominguez-Tenreiro,
  Obreja, Borgani, {De Lucia}  \& Murante}{Granato et~al.}{2015}]{Granato2015}
Granato G.~L.,  Ragone-Figueroa C.,  Dominguez-Tenreiro R.,  Obreja A.,
  Borgani S.,  {De Lucia} G.,   Murante G.,  2015, \mn@doi [Monthly Notices of
  the Royal Astronomical Society] {10.1093/mnras/stv676}, 450, 1320

\bibitem[\protect\citeauthoryear{Griffin et~al.,}{Griffin
  et~al.}{2010}]{Griffin2010}
Griffin M.~J.,  et~al., 2010, \mn@doi [Astronomy and Astrophysics]
  {10.1051/0004-6361/201014519}, 518, L3

\bibitem[\protect\citeauthoryear{Gunn, Hoessel  \& Oke}{Gunn
  et~al.}{1986}]{Gunn1986}
Gunn J.~E.,  Hoessel J.~G.,   Oke J.~B.,  1986, \mn@doi [The Astrophysical
  Journal] {10.1086/164317}, 306, 30

\bibitem[\protect\citeauthoryear{Hao et~al.,}{Hao et~al.}{2010}]{Hao2010}
Hao J.,  et~al., 2010, \mn@doi [The Astrophysical Journal Supplement, Volume
  191, Issue 2, pp. 254-274 (2010).] {10.1088/0067-0049/191/2/254}, 191, 254

\bibitem[\protect\citeauthoryear{Harrison \& Coles}{Harrison \&
  Coles}{2011}]{Harrison2011}
Harrison I.,  Coles P.,  2011, \mn@doi [Monthly Notices of the Royal
  Astronomical Society] {10.1111/j.1745-3933.2011.01198.x}, 421, L19

\bibitem[\protect\citeauthoryear{Hayward, Behroozi, Somerville, Primack, Moreno
   \& Wechsler}{Hayward et~al.}{2013}]{Hayward2013}
Hayward C.~C.,  Behroozi P.~S.,  Somerville R.~S.,  Primack J.~R.,  Moreno J.,
   Wechsler R.~H.,  2013, \mn@doi [Monthly Notices of the Royal Astronomical
  Society, Volume 434, Issue 3, p.2572-2581] {10.1093/mnras/stt1202}, 434, 2572

\bibitem[\protect\citeauthoryear{Herranz et~al.,}{Herranz
  et~al.}{2013}]{Herranz2013}
Herranz D.,  et~al., 2013, Astronomy {\&} Astrophysics, 549

\bibitem[\protect\citeauthoryear{Hickox et~al.,}{Hickox
  et~al.}{2012}]{Hickox2012}
Hickox R.~C.,  et~al., 2012, \mn@doi [Monthly Notices of the Royal Astronomical
  Society] {10.1111/j.1365-2966.2011.20303.x}, 421

\bibitem[\protect\citeauthoryear{Hopkins, Hernquist, Cox  \&
  Kere{\v{s}}}{Hopkins et~al.}{2008}]{Hopkins2008}
Hopkins P.~F.,  Hernquist L.,  Cox T.~J.,   Kere{\v{s}} D.,  2008, \mn@doi [The
  Astrophysical Journal Supplement Series] {10.1086/524362}, 175, 356

\bibitem[\protect\citeauthoryear{Hung et~al.,}{Hung et~al.}{2016}]{Hung2016}
Hung C.-L.,  et~al., 2016, \mn@doi [The Astrophysical Journal, Volume 826,
  Issue 2, article id. 130, 10 pp. (2016).] {10.3847/0004-637X/826/2/130}, 826

\bibitem[\protect\citeauthoryear{Ivison et~al.,}{Ivison
  et~al.}{2013}]{Ivison2013}
Ivison R.~J.,  et~al., 2013, \mn@doi [The Astrophysical Journal, Volume 772,
  Issue 2, article id. 137, 15 pp. (2013).] {10.1088/0004-637X/772/2/137}, 772

\bibitem[\protect\citeauthoryear{Ivison et~al.,}{Ivison
  et~al.}{2016}]{Ivison2016}
Ivison R.~J.,  et~al., 2016, \mn@doi [APJ] {10.3847/0004-637X/832/1/78}, 832

\bibitem[\protect\citeauthoryear{Kravtsov \& Borgani}{Kravtsov \&
  Borgani}{2012}]{Kravatsov2012}
Kravtsov A.,  Borgani S.,  2012, \mn@doi [Annual Review of Astronomy and
  Astrophysics] {10.1146/annurev-astro-081811-125502}, 50, 353

\bibitem[\protect\citeauthoryear{Lapi et~al.,}{Lapi et~al.}{2011}]{Lapi2011}
Lapi A.,  et~al., 2011, \mn@doi [The Astrophysical Journal, Volume 742, Issue
  1, article id. 24, 21 pp. (2011).] {10.1088/0004-637X/742/1/24}, 742

\bibitem[\protect\citeauthoryear{Levenson et~al.,}{Levenson
  et~al.}{2010}]{Levenson2010}
Levenson L.,  et~al., 2010, \mn@doi [Monthly Notices of the Royal Astronomical
  Society] {10.1111/j.1365-2966.2010.17771.x}, 409, 83

\bibitem[\protect\citeauthoryear{Lopes, de Carvalho, Gal, Djorgovski, Odewahn,
  Mahabal  \& Brunner}{Lopes et~al.}{2004}]{Lopes2004}
Lopes P. A.~A.,  de Carvalho R.~R.,  Gal R.~R.,  Djorgovski S.~G.,  Odewahn
  S.~C.,  Mahabal A.~A.,   Brunner R.~J.,  2004, \mn@doi [The Astronomical
  Journal] {10.1086/423038}, 128, 1017

\bibitem[\protect\citeauthoryear{L{\'{o}}pez-Caniego, Herranz,
  Gonz{\'{a}}lez-Nuevo, Sanz, Barreiro, Vielva, Arg{\"{u}}eso  \&
  Toffolatti}{L{\'{o}}pez-Caniego et~al.}{2006}]{Lopez-Caniego2006}
L{\'{o}}pez-Caniego M.,  Herranz D.,  Gonz{\'{a}}lez-Nuevo J.,  Sanz J.~L.,
  Barreiro R.~B.,  Vielva P.,  Arg{\"{u}}eso F.,   Toffolatti L.,  2006,
  \mn@doi [Monthly Notices of the Royal Astronomical Society]
  {10.1111/j.1365-2966.2006.10639.x}, 370, 2047

\bibitem[\protect\citeauthoryear{Ma et~al.,}{Ma et~al.}{2015}]{Ma2015}
Ma C.-J.,  et~al., 2015, \mn@doi [The Astrophysical Journal]
  {10.1088/0004-637X/806/2/257}, 806, 257

\bibitem[\protect\citeauthoryear{MacKenzie et~al.,}{MacKenzie
  et~al.}{2017}]{MacKenzie2017}
MacKenzie T.~P.,  et~al., 2017, \mn@doi [Monthly Notices of the Royal
  Astronomical Society, Volume 468, Issue 4, p.4006-4017]
  {10.1093/mnras/stx512}, 468, 4006

\bibitem[\protect\citeauthoryear{Mayer et~al.,}{Mayer et~al.}{2011}]{Mayer2011}
Mayer A.,  et~al., 2011, \mn@doi [Astronomy {\&} Astrophysics, Volume 531,
  id.L4, 4 pp.] {10.1051/0004-6361/201117203}, 531

\bibitem[\protect\citeauthoryear{Mayer et~al.,}{Mayer et~al.}{2014}]{Mayer2014}
Mayer A.,  et~al., 2014, Asymmetrical Planetary Nebulae VI conference,
  Proceedings of the conference held 4-8 November, 2013. Edited by C. Morisset,
  G. Delgado-Inglada and S. Torres-Peimbert. Online at
  http://www.astroscu.unam.mx/apn6/PROCEEDINGS/, id.59

\bibitem[\protect\citeauthoryear{Micha{\l}owski, Hjorth  \&
  Watson}{Micha{\l}owski et~al.}{2010}]{Michalowski2009}
Micha{\l}owski M.,  Hjorth J.,   Watson D.,  2010, \mn@doi [Astronomy and
  Astrophysics] {10.1051/0004-6361/200913634}, 514, A67

\bibitem[\protect\citeauthoryear{Miller, Hayward, Chapman  \& Behroozi}{Miller
  et~al.}{2015}]{Miller2015}
Miller T.~B.,  Hayward C.~C.,  Chapman S.~C.,   Behroozi P.~S.,  2015, \mn@doi
  [Monthly Notices of the Royal Astronomical Society] {10.1093/mnras/stv1267},
  452, 878

\bibitem[\protect\citeauthoryear{{Miville Deschenes} \& Lagache}{{Miville
  Deschenes} \& Lagache}{2005}]{MivilleDeschenes2005}
{Miville Deschenes} M.~A.,  Lagache G.,  2005, \mn@doi [The Astrophysical
  Journal Supplement Series] {10.1086/427938}, 157, 302:323

\bibitem[\protect\citeauthoryear{Negrello, Gonzalez-Nuevo, Magliocchetti,
  Moscardini, {De Zotti}, Toffolatti  \& Danese}{Negrello
  et~al.}{2005}]{Negrello2005}
Negrello M.,  Gonzalez-Nuevo J.,  Magliocchetti M.,  Moscardini L.,  {De Zotti}
  G.,  Toffolatti L.,   Danese L.,  2005, \mn@doi [Monthly Notices of the Royal
  Astronomical Society] {10.1111/j.1365-2966.2005.08783.x}, 358, 869

\bibitem[\protect\citeauthoryear{Negrello et~al.,}{Negrello
  et~al.}{2016}]{Negrello2017}
Negrello M.,  et~al., 2016, \mn@doi [Monthly Notices of the Royal Astronomical
  Society, Volume 465, Issue 3, p.3558-3580] {10.1093/mnras/stw2911}, 465, 3558

\bibitem[\protect\citeauthoryear{Nguyen et~al.,}{Nguyen
  et~al.}{2010}]{Nguyen2010}
Nguyen H.~T.,  et~al., 2010, \mn@doi [Astronomy and Astrophysics, Volume 518,
  id.L5, 5 pp.] {10.1051/0004-6361/201014680}, 518

\bibitem[\protect\citeauthoryear{Noble et~al.,}{Noble et~al.}{2013}]{Noble2013}
Noble A.~G.,  et~al., 2013, \mn@doi [Monthly Notices of the Royal Astronomical
  Society: Letters] {10.1093/mnrasl/slt108}, 436, L40

\bibitem[\protect\citeauthoryear{Oliver et~al.,}{Oliver
  et~al.}{2010}]{Oliver2010}
Oliver S.~J.,  et~al., 2010, \mn@doi [Astronomy and Astrophysics]
  {10.1051/0004-6361/201014697}, 518, L21

\bibitem[\protect\citeauthoryear{Oliver et~al.,}{Oliver
  et~al.}{2012}]{Oliver2012}
Oliver S.~J.,  et~al., 2012, \mn@doi [Monthly Notices of the Royal Astronomical
  Society] {10.1111/j.1365-2966.2012.20912.x}, 424, 1614

\bibitem[\protect\citeauthoryear{Oteo, Zwaan, Ivison, Smail  \& Biggs}{Oteo
  et~al.}{2016}]{Oteo2017b}
Oteo I.,  Zwaan M.~A.,  Ivison R.~J.,  Smail I.,   Biggs A.~D.,  2016, \mn@doi
  [The Astrophysical Journal, Volume 837, Issue 2, article id. 182, 9 pp.
  (2017).] {10.3847/1538-4357/aa5da4}, 837

\bibitem[\protect\citeauthoryear{Oteo et~al.,}{Oteo et~al.}{2017a}]{Oteo2017}
Oteo I.,  et~al., 2017a, eprint arXiv:1709.02809

\bibitem[\protect\citeauthoryear{Oteo et~al.,}{Oteo et~al.}{2017b}]{Oteo2017c}
Oteo I.,  et~al., 2017b, eprint arXiv:1701.05901

\bibitem[\protect\citeauthoryear{Ott, Centre  \& Agency}{Ott
  et~al.}{2010}]{Ott2010}
Ott S.,  Centre H.~S.,   Agency E.~S.,  2010, Astronomical Data Analysis
  Software and Systems XIX, 434, 139

\bibitem[\protect\citeauthoryear{Pearson et~al.,}{Pearson
  et~al.}{2013}]{Pearson2013}
Pearson E.~A.,  et~al., 2013, \mn@doi [Monthly Notices of the Royal
  Astronomical Society, Volume 435, Issue 4, p.2753-2763]
  {10.1093/mnras/stt1369}, 435, 2753

\bibitem[\protect\citeauthoryear{Petty et~al.,}{Petty et~al.}{2013}]{Petty2013}
Petty S.~M.,  et~al., 2013, \mn@doi [The Astronomical Journal, Volume 146,
  Issue 4, article id. 77, 17 pp. (2013).] {10.1088/0004-6256/146/4/77}, 146

\bibitem[\protect\citeauthoryear{Pilbratt et~al.,}{Pilbratt
  et~al.}{2010}]{Pilbratt2010}
Pilbratt G.~L.,  et~al., 2010, \mn@doi [Astronomy and Astrophysics]
  {10.1051/0004-6361/201014759}, 518, L1

\bibitem[\protect\citeauthoryear{{Planck Collaboration} et~al.,}{{Planck
  Collaboration} et~al.}{2011a}]{PlanckCollaboration2011}
{Planck Collaboration} P.,  et~al., 2011a, \mn@doi [Astronomy {\&}
  Astrophysics, Volume 536, id.A1, 16 pp.] {10.1051/0004-6361/201116464}, 536

\bibitem[\protect\citeauthoryear{{Planck Collaboration} et~al.,}{{Planck
  Collaboration} et~al.}{2011b}]{Ade2011}
{Planck Collaboration} P.,  et~al., 2011b, \mn@doi [Astronomy {\&}
  Astrophysics] {10.1051/0004-6361/201116474}, 536, A7

\bibitem[\protect\citeauthoryear{{Planck Collaboration} et~al.,}{{Planck
  Collaboration} et~al.}{2014}]{Ade2014}
{Planck Collaboration} P.,  et~al., 2014, \mn@doi [Astronomy {\&} Astrophysics]
  {10.1051/0004-6361/201321524}, 571, A28

\bibitem[\protect\citeauthoryear{{Planck Collaboration} et~al.,}{{Planck
  Collaboration} et~al.}{2015a}]{PlanckCollaboration2015d}
{Planck Collaboration} P.,  et~al., 2015a, \mn@doi [Astronomy {\&}
  Astrophysics, Volume 582, id.A30, 29 pp.] {10.1051/0004-6361/201424790}, 582

\bibitem[\protect\citeauthoryear{{Planck Collaboration} et~al.,}{{Planck
  Collaboration} et~al.}{2015b}]{PlanckCollaboration2015b}
{Planck Collaboration} P.,  et~al., 2015b, \mn@doi [Astronomy {\&}
  Astrophysics, Volume 594, id.A26, 39 pp.] {10.1051/0004-6361/201526914}, 594

\bibitem[\protect\citeauthoryear{{Planck Collaboration} et~al.,}{{Planck
  Collaboration} et~al.}{2015c}]{PlanckCollaboration2015a}
{Planck Collaboration} P.,  et~al., 2015c, \mn@doi [Astronomy {\&}
  Astrophysics, Volume 594, id.A27, 38 pp.] {10.1051/0004-6361/201525823}, 594

\bibitem[\protect\citeauthoryear{{Planck Collaboration} et~al.,}{{Planck
  Collaboration} et~al.}{2015d}]{PlanckCollaboration2015e}
{Planck Collaboration} P.,  et~al., 2015d, \mn@doi [Astronomy {\&}
  Astrophysics, Volume 594, id.A28, 28 pp.] {10.1051/0004-6361/201525819}, 594

\bibitem[\protect\citeauthoryear{{Planck Collaboration} et~al.,}{{Planck
  Collaboration} et~al.}{2016}]{PlanckCollaboration2015c}
{Planck Collaboration} P.,  et~al., 2016, \mn@doi [Astronomy {\&} Astrophysics,
  Volume 596, id.A100, 28 pp.] {10.1051/0004-6361/201527206}, 596, A100

\bibitem[\protect\citeauthoryear{{Planck HFI Team} et~al.,}{{Planck HFI Team}
  et~al.}{2010}]{PlanckHFITeam2010}
{Planck HFI Team} P.,  et~al., 2010, \mn@doi [Astronomy and Astrophysics]
  {10.1051/0004-6361/200912975}, 520, A9

\bibitem[\protect\citeauthoryear{Rowan-Robinson et~al.,}{Rowan-Robinson
  et~al.}{2016}]{Rowan-Robinson2016}
Rowan-Robinson M.,  et~al., 2016, \mn@doi [Monthly Notices of the Royal
  Astronomical Society, Volume 461, Issue 1, p.1100-1111]
  {10.1093/mnras/stw1169}, 461, 1100

\bibitem[\protect\citeauthoryear{Scoville et~al.,}{Scoville
  et~al.}{2013}]{Scoville2013}
Scoville N.,  et~al., 2013, \mn@doi [The Astrophysical Journal Supplement,
  Volume 206, Issue 1, article id. 3, 26 pp. (2013).]
  {10.1088/0067-0049/206/1/3}, 206

\bibitem[\protect\citeauthoryear{Simpson et~al.,}{Simpson
  et~al.}{2014}]{Simpson2014a}
Simpson J.~M.,  et~al., 2014, \mn@doi [The Astrophysical Journal]
  {10.1088/0004-637X/788/2/125}, 788, 125

\bibitem[\protect\citeauthoryear{Stevens, Jarvis, Coppin, Page, Greve, Carrera
  \& Ivison}{Stevens et~al.}{2010}]{Stevens2010}
Stevens J.~A.,  Jarvis M.~J.,  Coppin K. E.~K.,  Page M.~J.,  Greve T.~R.,
  Carrera F.~J.,   Ivison R.~J.,  2010, \mn@doi [Monthly Notices of the Royal
  Astronomical Society] {10.1111/j.1365-2966.2010.16641.x}, 405, no

\bibitem[\protect\citeauthoryear{Swinbank, Chapman, Smail, Lindner, Borys,
  Blain, Ivison  \& Lewis}{Swinbank et~al.}{2006}]{Swinbank2006}
Swinbank A.~M.,  Chapman S.~C.,  Smail I.,  Lindner C.,  Borys C.,  Blain
  A.~W.,  Ivison R.~J.,   Lewis G.~F.,  2006, \mn@doi [Monthly Notices of the
  Royal Astronomical Society] {10.1111/j.1365-2966.2006.10673.x}, 371, 465

\bibitem[\protect\citeauthoryear{Tacconi et~al.,}{Tacconi
  et~al.}{2008}]{Tacconi2008}
Tacconi L.~J.,  et~al., 2008, \mn@doi [The Astrophysical Journal]
  {10.1086/587168}, 680, 246

\bibitem[\protect\citeauthoryear{Tauber et~al.,}{Tauber
  et~al.}{2010}]{Tauber2010}
Tauber J.~A.,  et~al., 2010, \mn@doi [Astronomy and Astrophysics]
  {10.1051/0004-6361/200912983}, 520, A1

\bibitem[\protect\citeauthoryear{Taylor}{Taylor}{2005}]{Taylor2005}
Taylor M.~B.,  2005, Astronomical Data Analysis Software and Systems XIV - ASP
  Conference Series, 347, 29

\bibitem[\protect\citeauthoryear{{The Astropy Collaboration} et~al.,}{{The
  Astropy Collaboration} et~al.}{2013}]{Astropy2013}
{The Astropy Collaboration} A.,  et~al., 2013, \mn@doi [Astronomy {\&}
  Astrophysics, Volume 558, id.A33, 9 pp.] {10.1051/0004-6361/201322068}, 558

\bibitem[\protect\citeauthoryear{Timmons et~al.,}{Timmons
  et~al.}{2015}]{Timmons2015}
Timmons N.,  et~al., 2015, \mn@doi [The Astrophysical Journal, Volume 805,
  Issue 2, article id. 140, 6 pp. (2015).] {10.1088/0004-637X/805/2/140}, 805

\bibitem[\protect\citeauthoryear{Toft et~al.,}{Toft et~al.}{2014}]{Toft2014}
Toft S.,  et~al., 2014, \mn@doi [The Astrophysical Journal, Volume 782, Issue
  2, article id. 68, 12 pp. (2014).] {10.1088/0004-637X/782/2/68}, 782

\bibitem[\protect\citeauthoryear{Valiante et~al.,}{Valiante
  et~al.}{2016}]{Valiante2016}
Valiante E.,  et~al., 2016, \mn@doi [Monthly Notices of the Royal Astronomical
  Society, Volume 462, Issue 3, p.3146-3179] {10.1093/mnras/stw1806}, 462, 3146

\bibitem[\protect\citeauthoryear{Valtchanov et~al.,}{Valtchanov
  et~al.}{2013}]{Valtchanov2013a}
Valtchanov I.,  et~al., 2013, \mn@doi [Monthly Notices of the Royal
  Astronomical Society, Volume 436, Issue 3, p.2505-2514]
  {10.1093/mnras/stt1753}, 436, 2505

\bibitem[\protect\citeauthoryear{Viero et~al.,}{Viero et~al.}{2014}]{Viero2014}
Viero M.~P.,  et~al., 2014, \mn@doi [The Astrophysical Journal Supplement
  Series] {10.1088/0067-0049/210/2/22}, 210, 22

\bibitem[\protect\citeauthoryear{Walter et~al.,}{Walter
  et~al.}{2012}]{Walter2012}
Walter F.,  et~al., 2012, \mn@doi [Nature] {10.1038/nature11073}, 486, 233

\bibitem[\protect\citeauthoryear{Wang et~al.,}{Wang et~al.}{2013}]{Wang2014}
Wang L.,  et~al., 2013, \mn@doi [Monthly Notices of the Royal Astronomical
  Society, Volume 444, Issue 3, p.2870-2883] {10.1093/mnras/stu1569}, 444, 2870

\bibitem[\protect\citeauthoryear{Wang et~al.,}{Wang et~al.}{2016}]{Wang2016}
Wang T.,  et~al., 2016, \mn@doi [The Astrophysical Journal, Volume 828, Issue
  1, article id. 56, 15 pp. (2016).] {10.3847/0004-637X/828/1/56}, 828

\bibitem[\protect\citeauthoryear{Wenger et~al.,}{Wenger
  et~al.}{2000}]{Wenger2000}
Wenger M.,  et~al., 2000, \mn@doi [Astronomy and Astrophysics Supplement,
  v.143, p.9-22] {10.1051/aas:2000332}, 143, 9

\bibitem[\protect\citeauthoryear{Wilkinson et~al.,}{Wilkinson
  et~al.}{2016}]{Wilkinson2016}
Wilkinson A.,  et~al., 2016, \mn@doi [Monthly Notices of the Royal Astronomical
  Society, Volume 464, Issue 2, p.1380-1392] {10.1093/mnras/stw2405}, 464, 1380

\bibitem[\protect\citeauthoryear{Yuan et~al.,}{Yuan et~al.}{2014}]{Yuan2014}
Yuan T.~T.,  et~al., 2014, \mn@doi [The Astrophysical Journal Letters, Volume
  795, Issue 1, article id. L20, 6 pp. (2014).] {10.1088/2041-8205/795/1/L20},
  795

\makeatother
\end{thebibliography}




\appendix

\section{Images of the \textit{Planck} compact sources}

\label{Sec:Im}

\textit{To comply with arXivs size limits, only 1 of the 4 grids of pictures is included here. The full published article will include all of the images.}

\begin{figure*}
   \centering
  \includegraphics[width=1\linewidth]{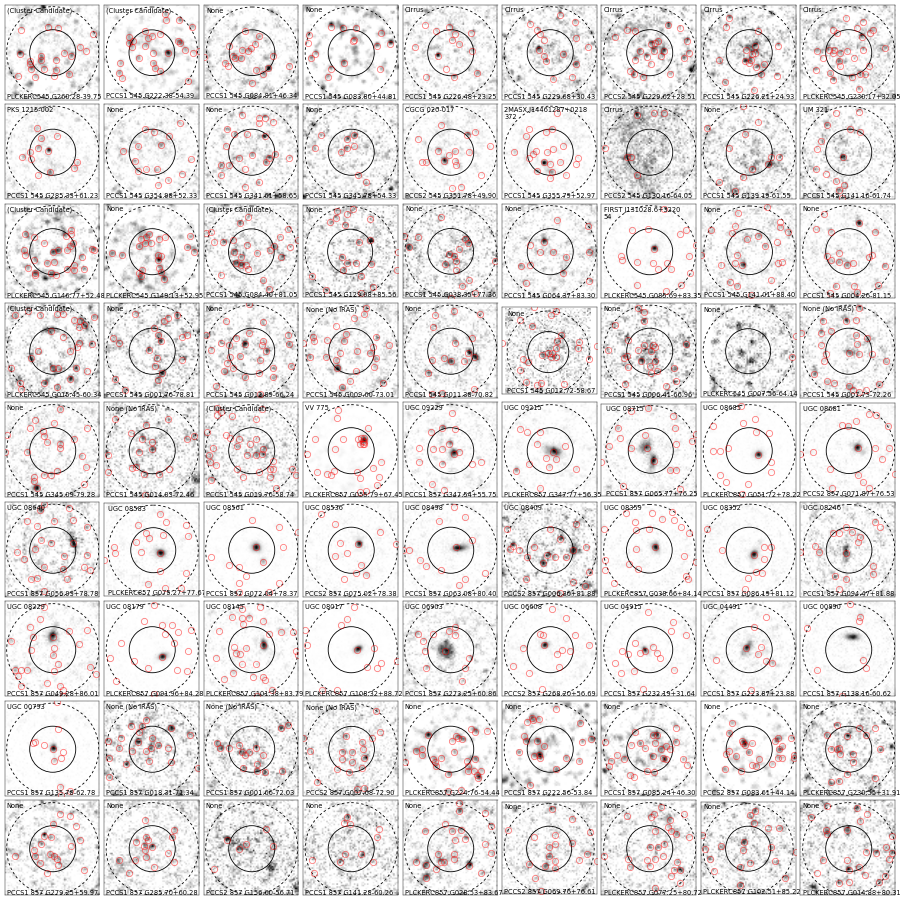}

  \caption{The 350 $\mu$m Herschel map for all of our sources, with the Planck beam in solid black circle, the aperture photometry in dashed black circle, and the red circles indicate the positions of sources which have a flux density $> 25.4$ mJy in either the 250, 350 or 500 $\mu$m bands.}
  
\end{figure*}

\section{Table of the\textit{Planck} compact sources }
\label{Sec:Apen}

Here we include a list of all our \textit{Planck} compact sources that lie on the maps

\newpage
\begin{landscape}
\begin{table}
\caption{Candidate protoclusters from the \textit{Planck} 857GHz catalogues of compact sources. The $\sigma$ values provide the strength of the overdensity at 250, 350 and 500 $\mu$m. A table containing the properties and identifications of all the \textit{Planck} compact sources is avaliable online.}
\begin{tabular}{|l|r|r|l|l|r|r|r|}

\hline
  \multicolumn{1}{|c|}{Name} &
  \multicolumn{1}{c|}{RA} &
  \multicolumn{1}{c|}{DEC} &
  \multicolumn{1}{c|}{Associations} &
  \multicolumn{1}{c|}{Planck 857 Flux [mJy]} &
  \multicolumn{1}{c|}{$\sigma_{250}$} &
  \multicolumn{1}{c|}{$\sigma_{350}$} &
  \multicolumn{1}{c|}{$\sigma_{500}$} \\
\hline

  PCCS1 857 G014.92-58.26 & 336.635 & -32.177 & Cluster Candidate & 302 $\pm$ 239 & 1.88 & 1.96 & 4.36\\
  PCCS1 857 G354.81-79.56 & 3.049 & -33.228 & Cluster Candidate & 1014 $\pm$ 233 & 1.67 & 2.23 & 3.25\\
  PLCKERC857 G257.09-87.10 & 15.233 & -29.122 & Cluster Candidate & 2403 $\pm$ 198 & 3.52 & 3.02 & 3.25\\
  PLCKERC857 G014.99-59.64 & 338.260 & -32.139 & Cluster Candidate & 943 $\pm$ 115 & 3.12 & 1.96 & 4.36\\
  PLCKERC857 G239.13-78.19 & 25.333 & -31.786 & Cluster Candidate & 1110 $\pm$ 145 & 2.72 & 1.68 & 3.25\\
  PLCKERC857 G017.86-68.67 & 348.790 & -30.591 & Cluster Candidate & 1657 $\pm$ 167 & 2.92 & 3.76 & 5.05\\
  PLCKERC857 G007.34-65.24 & 345.366 & -35.103 & Cluster Candidate & 920 $\pm$ 132 & 3.32 & 3.52 & 3.25\\
  PCCS1 857 G252.98-85.59 & 16.749 & -29.910 & Cluster Candidate & 619 $\pm$ 817 & 1.88 & 2.76 & 3.25\\
  PCCS1 857 G058.69+81.03 & 202.258 & 30.712 & Cluster Candidate & 845 $\pm$ 298 & 1.45 & 1.68 & 4.01\\
  PCCS1 857 G058.53+82.57 & 200.607 & 30.124 & Cluster Candidate & 827 $\pm$ 197 & 1.45 & 2.23 & 4.36\\
  PLCKERC857 G062.48+78.89 & 204.276 & 32.142 & Cluster Candidate & 814 $\pm$ 109 & 0.77 & 2.50 & 3.25\\
  PLCKERC857 G063.13+78.00 & 205.172 & 32.621 & Cluster Candidate & 912 $\pm$ 169 & 2.10 & 3.52 & 3.25\\
  PLCKERC857 G027.36+84.83 & 198.608 & 26.510 & Cluster Candidate & 792 $\pm$ 93 & 2.52 & 3.27 & 4.01\\
  PLCKERC857 G042.54+81.51 & 202.358 & 28.224 & Cluster Candidate & 1697 $\pm$ 251 & 1.67 & 3.76 & 4.01\\
  PLCKERC857 G149.81+50.11 & 158.364 & 59.196 & Cluster Candidate & 1249 $\pm$ 131 & 5.38 & 4.47 & 4.71\\
  PCCS1 857 G089.66+36.10 & 256.828 & 60.449 & Cluster Candidate & 886 $\pm$ 194 & 3.32 & 2.76 & 2.01\\
  PCCS1 857 G085.48+43.36 & 244.657 & 55.771 & Cluster Candidate & 748 $\pm$ 128 & 2.72 & 3.27 & 3.64\\
  PLCKERC857 G095.44+58.94 & 216.128 & 52.936 & Cluster Candidate & 1141 $\pm$ 176 & 4.65 & 5.80 & 2.44\\
  PCCS1 857 G238.20+42.27 & 150.845 & 1.469 & Cluster Candidate & 712 $\pm$ 269 & 4.28 & 3.76 & 2.01\\
  PLCKERC857 G060.37+66.55 & 218.579 & 35.559 & Cluster Candidate & 1241 $\pm$ 151 & 1.00 & 4.47 & 5.70\\
  PLCKERC857 G261.89-40.69 & 70.168 & -53.734 & Cluster Candidate & 795 $\pm$ 134 & 3.32 & 4.47 & 3.64\\

\hline

\end{tabular}
\end{table}
\end{landscape}

\newpage
\begin{landscape}
\begin{table}
\caption{Candidate protoclusters from the \textit{Planck} 545GHz catalogues of compact sources. A table containing the properties and identifications of all the \textit{Planck} compact sources is avaliable online.}
\begin{tabular}{|l|r|r|l|l|r|r|r|}

\hline
  \multicolumn{1}{|c|}{Name} &
  \multicolumn{1}{c|}{RA} &
  \multicolumn{1}{c|}{DEC} &
  \multicolumn{1}{c|}{Associations} &
  \multicolumn{1}{c|}{$S_{545}$ [mJy]} &
  \multicolumn{1}{c|}{$\sigma_{250}$} &
  \multicolumn{1}{c|}{$\sigma_{350}$} &
  \multicolumn{1}{c|}{$\sigma_{500}$} \\
\hline

  PCCS1 545 G019.76-58.74 & 337.311 & -29.670 & Cluster Candidate & 225$\pm$164 & 2.92 & 1.68 & 3.25\\
  PCCS1 545 G354.79-79.57 & 3.060 & -33.226 & Cluster Candidate & 335$\pm$158 & 1.22 & 1.39 & 3.25\\
  PLCKERC545 G015.45-60.34 & 339.084 & -31.902 & Cluster Candidate & 607$\pm$87 & 2.72 & 3.27 & 4.01\\
  PCCS1 545 G058.72+82.59 & 200.566 & 30.136 & Cluster Candidate & 349$\pm$161 & 1.22 & 1.39 & 3.64\\
  PCCS1 545 G027.38+84.85 & 198.581 & 26.515 & Cluster Candidate & 433$\pm$136 & 1.45 & 2.23 & 4.01\\
  PCCS1 545 G084.40+81.05 & 199.569 & 33.968 & Cluster Candidate & 387$\pm$230 & 1.88 & 1.39 & 3.25\\
  PLCKERC545 G146.77+52.48 & 164.280 & 59.031 & Cluster Candidate & 606$\pm$73 & 4.09 & 4.93 & 1.06\\
  PCCS1 545 G222.38-54.39 & 53.046 & -27.117 & Cluster Candidate & 182$\pm$185 & 2.10 & 3.02 & 2.01\\
  PLCKERC545 G060.36+66.56 & 218.577 & 35.554 & Cluster Candidate & 872$\pm$114 & 0.77 & 4.47 & 5.70\\
  PLCKERC545 G260.28-39.75 & 71.996 & -52.643 & Cluster Candidate & 471$\pm$46 & 3.71 & 3.27 & 3.25\\
  
\hline\end{tabular}
\end{table}
\end{landscape}


\section{Probability of N cluster galaxies in M detected sources}
\label{apen:2}
For a given proto-cluster candidate, we observe $M$ sources.
Under the assumption that some of these are physically associated with a proto-cluster, whilst some are not, $M$ is a combination of both the number of field and proto-cluster galaxies, $$M = N_{field} + N_{cluster}.$$
Furthermore, we assume that the field galaxies are distributed in a Poisson manner, and can be described by Poisson statistics.
For a Poisson process with a mean and variance of $\mu$, the probability of observing M sources is given by:
$$\big[\frac{\mu^{M}}{M!}\big]exp(-\mu)$$
If we assume that $N$ of our $M$ sources are associated with the proto-cluster, then $M-N$ sources will be associated with the field, and the probability of observing these $M-N$ field galaxies is:
$$\big[\frac{\mu^{(M-N)}}{(M-N)!}\big]exp(-\mu)$$
However, this needs to be renormalised as the maximum possible observed field galaxies is now $M$ rather than $\infty$ (It is impossible to observe $M+1$ field galaxies out of $M$ total galaxies).
This can be done using the Poisson cumulative distribution function, given by: 
$$\Sigma_{i = 0}^{i = M} \big[\frac{\mu^{i}}{i!}\big]exp(-\mu)$$
where we simply sum over all of the possible arrangements of $N_{field} + N_{cluster}$ to give a total of $M$ sources. 
This is now the correct normalisation factor, as it allows for the full range of possibilities stretching from no sources are associated with the proto-cluster, to all the sources are associated with the proto-cluster.
The full equation becomes:
$$p(N|M, \mu) = \frac{\big[\frac{\mu^{(M-N)}}{(M-N)!}\big]exp(-\mu)}{\Sigma_{i = 0}^{i = M} \big[\frac{\mu^{(i)}}{(i)!}\big]exp(-\mu)}.$$
Which can be further simplified to:
$$p(N|M, \mu) = \frac{\big[\frac{\mu^{(M-N)}}{(M-N)!}\big]}{\Sigma_{i = 0}^{i = M} \big[\frac{\mu^{(i)}}{(i)!}\big]}.$$

\bsp	
\label{lastpage}
\end{document}